\documentclass[%
 reprint,
 amsmath,amssymb,
 aps,
]{revtex4-1}

\usepackage{graphicx}% Include figure files
\usepackage{natbib}

\usepackage{dcolumn}% Align table columns on decimal point
\usepackage{bm}% bold math
\begin{document}

%\preprint{APS/123-QED}

\title{Black Hole Binaries and Microquasars}

%\begin{CJK*}{GB}{}

\author{Shuang-Nan Zhang$^{1,2}$}
\email{zhangsn@ihep.ac.cn}
\affiliation{%
 $^{1}$Laboratory for Particle Astrophysics, Institute of High Energy Physics,Beijing 100049, China. \\
 $^{2}$National Astronomical Observatories, Chinese Academy Of Sciences, Beijing 100012, China.\\}%
%\end{CJK*}

\date{\today}% It is always \today, today,

\begin{abstract}
This is a general review on the observations and physics of black hole X-ray binaries and microquasars, with the
emphasize on recent developments in the high energy regime. The focus is put on understanding the accretion
flows and measuring the parameters of black holes in them. It includes mainly two parts: (1) Brief review of
several recent review article on this subject; (2) Further development on several topics, including black hole
spin measurements, hot accretion flows, corona formation, state transitions and thermal stability of standard
think disk. This is thus not a regular bottom-up approach, which I feel not necessary at this stage. Major
effort is made in making and incorporating from many sources useful plots and illustrations, in order to make
this article more comprehensible to non-expert readers. In the end I attempt to make a unification scheme on the accretion-outflow (wind/jet) connections of all types of accreting BHs of all accretion rates and all BH mass scales, and finally provide a brief outlook.
\begin{description}
\item[PACS numbers]{}
\end{description}
\end{abstract}

\pacs{Valid PACS appear here}% PACS, the Physics and Astronomy
\maketitle

\tableofcontents

%\section{\label{sec:level1}First-level heading:\protect\\ The line
%break was forced \lowercase{via} \textbackslash\textbackslash}
\section{Synopsis}

I will start by defining what I mean by black hole binaries (BHBs) and microquasars in this article. I decide to
restrict myself to only a subclass of BHBs, namely, BH X-ray binaries (BHXBs), since these are the only class of
BHBs known observationally. I will then simply refer microquasars as BHXBs for reasons discussed in
Section~\ref{term}.

Since many excellent, comprehensive and quite up-to-date review articles on BHXBs are readily available in
literature, I feel it is not necessary to write another bottom-up and comprehensive review article on the same
subjects at this stage. I will thus take quite an unusual approach in this article. I will first give some
concise guides on several representative review articles \cite{Fender2012a,Remillard2006,Done2010,Done2007},
with some necessary updates. I will then focus on the further developments on several topics I feel deserve more
discussions, i.e., BH spin measurements (Section~\ref{BH_spin}), hot accretion flows (Section~\ref{hot_flow}),
corona formation (Section~\ref{corona}), state transitions (Section~\ref{transitions}) and thermal stability of
SSD (Section~\ref{disk}). The emphasis is thus put on understanding the accretion flows and measuring the
parameters of black holes in them. Some rather general issues on BH astrophysics, such as what astrophysical BHs
are and how to identify them observationally, are not discussed here but can be found from my recent book
chapter entitled ``Astrophysical Black Holes in the Physical Universe" \cite{Zhang2011b}.

The usual practice of writing a review article is to end by listing some outstanding issues and major unsolved
problems, and then to propose some possible approaches to them. I initially did not do this in the first draft. The history of astronomy tells us that major progress is almost always made by unexpected discoveries and research results;
unpredictability is an essential nature of astronomy. This article is not intended to be read by funding
agencies or proposal reviewers, so I thought I did not have to do it. In astronomy, knowing what has happened, but looking
and doing it differently are far more important and effective than following other people's advises. However, the editors of this book suggested me to write a brief ``outlook" in the end. I thus did it nevertheless.

\section{Acronyms and Terminology}\label{term}

In Table~\ref{Acronyms}, I list all acronyms used in this article; most of these are quite commonly used in this
community.

\begin{table}%
\small \caption{List of Acronyms.}\label{Acronyms}
\begin{tabular}{l|l}
\toprule
Acronym &Definition\\
\colrule
ADAF & advection dominated accretion flow\\
ADIOS & advection dominated accretion \\
&inflow/outflow solution\\
AGN & active galactic nuclei\\
BH & black hole\\
BHB & black hole binary\\
BHXB & black hole X-ray binary\\
BL3Q & broad-line-less luminous quasar\\
BLR & broad-line region\\
BP & Blandford-Payne\\
BZ & Blandford-Znajek\\
CDAF & convection dominated accretion flow\\
CF & continuum fitting\\
DIM & disk instability model\\
FRED & fast rise and exponential decay\\
FSRQ & flat spectrum radio quasar\\
GR & general relativity\\
GRB & gamma-ray burst\\
HAF & hot accretion flow\\
HFQPO & high frequency quasi-periodic oscillation\\
HID & Hardness-Intensity-Diagram\\
HIF & hot inner flow\\
ISCO & inner-most stable circular orbit\\
JDAF & jet dominated accretion flow\\
LFQPO & low frequency quasi-periodic oscillation\\
LHAF & luminous hot accretion flow\\
LLAGN & low-luminosity AGN\\
LMC & Large Magellanic Cloud\\
LMXB & low-mass X-ray binary\\
NDAF & neutrino dominated accretion flow\\
NS & neutron star\\
NSXB & neutron star X-ray binary\\
PDS & power density spectrum\\
PL & power-law\\
QPO & quasi-periodic oscillation\\
rms & root-mean-squares\\
RID & RMS-Intensity-Diagram\\
RQQ & radio-quiet quasar\\
SEAF & Super-Eddington accretion flow\\
SLE & Shapiro, Lightman \& Eardley\\
SPL & steep power-law\\
SSD & Shakura-Sunyaev Disk\\
SXE & soft X-ray excess\\
TID & truncated inner disk\\
ULX & ultra-luminous X-ray source\\
WD & white dwarf\\
XRB & X-ray binary\\
 \botrule
\end{tabular}
\end{table}

\begin{figure}
\center{
\includegraphics[angle=0,scale=0.30]{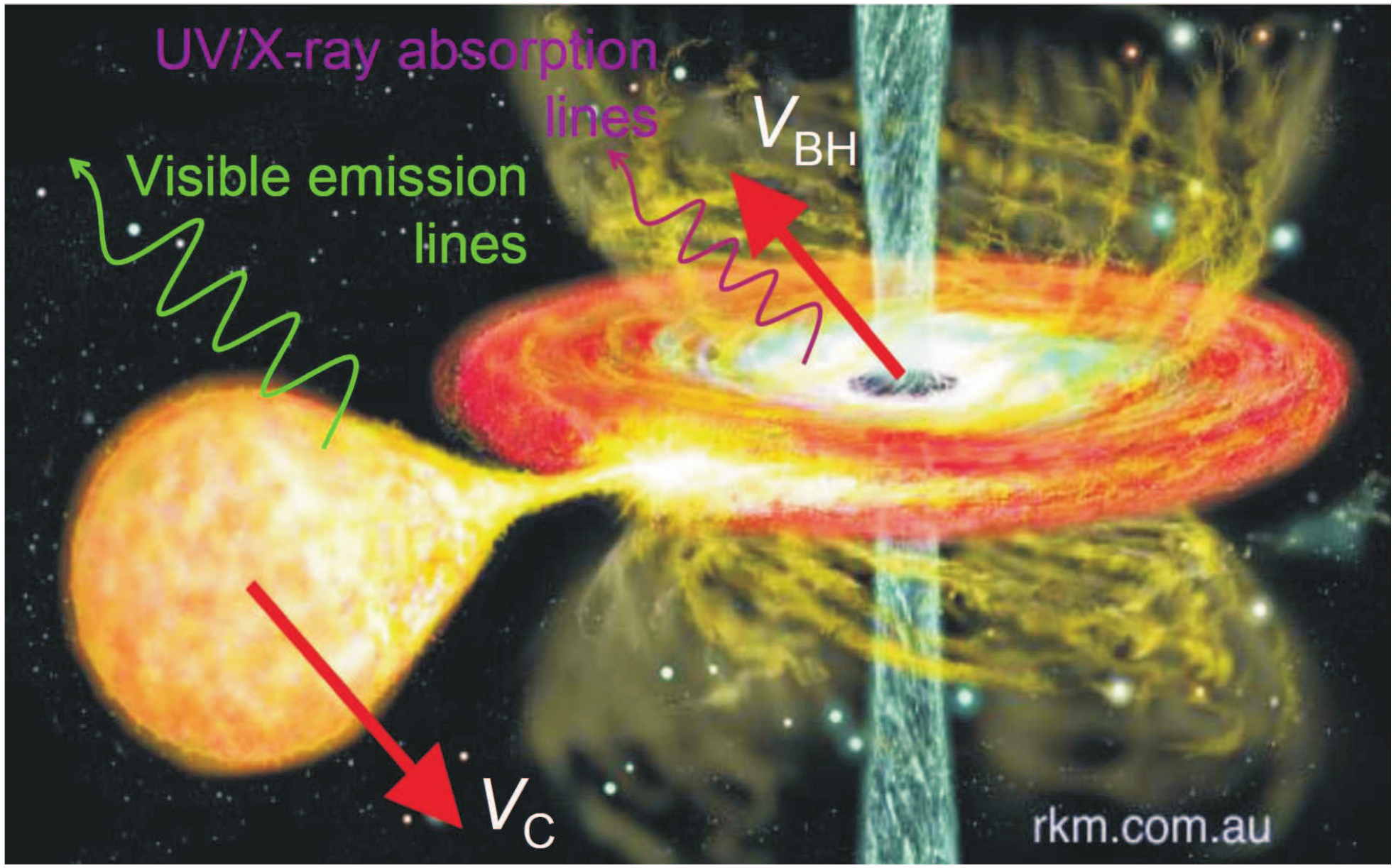}}
\caption{Illustration of a BHXB and microquasar. X-ray emission is produced from the central hot accretion disk.
A jet is normally observed in the radio band. The companion star may produce optical emission lines. The disk
wind may be observed with UV/X-ray absorption lines.} \label{BHXB}
\end{figure}

A BH binary (BHB) is a gravitationally bound binary system in which one of the objects is a stellar mass BH with
mass from several to tens of solar masses ($M_{\odot}$); the other object, i.e. its companion, can be either a
normal star, a white dwarf, or a neutron star (NS). In case a binary system consists of two BHs, it is is
referred to as a binary BH system, which is not covered in this article. When the companion in a BHB is a normal
star, the gas from the star may be accreted to the BH and X-rays are produced, as a consequence of the heating
by converting the gravitational potential energy into the kinetic energy of the gas, and a {\em BH X-ray binary}
(BHXB) is referred to as such a binary system, as shown in Figure \ref{BHXB}. The possible existence of BHXBs
was first suggested by Zel'dovich \& Novikov \cite{Zel'dovich1966,Zel'dovich1965}. The first BHXB found is
Cygnus~X--1 \cite{WEBSTER1972}, now a well-studied system among many others found subsequently in the Milky Way
and nearby galaxies.

The terminology of {\em microquasar} has some twisting in it. Historically it was first referred to the BHXB
1E1740.7$-$2942 in the Galactic center region, because a double-sided jet was detected from it, mimicking some
quasars with similar radio lobes, which have much larger scales \cite{Mirabel1992}. Soon after, superluminal
jets are observed from a BHXB GRS~1915+105 \cite{Mirabel1994}, which is also referred to as a microquasar.
Nevertheless both BHXBs are quite unusual compared to many others, thus microquasars were considered quite
unusual. However it became clear that microquasars may be quite common among BHXBs, since the discovery of a
normal BHXB GRO~J1655$-$40 \cite{Zhanga}, whose superluminal jets were observed \cite{Tingay1995} and showed
some correlations with its X-ray emission \cite{Harmon1995}. Subsequently several more BHXBs have been observed
with superluminal jets. At this stage {\em microquasars} began to be referred to as BHXBs with relativistic jets
(with bulk motion of about or larger than 90\% of the speed of light), to distinguish them from NS X-ray
binaries (NSXBs) that only have mildly relativistic jets  (with bulk motion of about or smaller than 50\% of the
speed of light) \cite{Mirabel1999}. However the discovery of relativistic jets from a NSXB Circinus X--1 made
the situation complicated: relativistic jets are no longer uniquely linked to BHs \cite{Fender2004}.

Now in retrospect, a {\em microquasar} can be literally and easily understood as the {\em micro} version of a
quasar; however a quasar may or may not be observed with collimated jets. A quasar has been already understood
as a special galaxy centered by an actively accreting supermassive BH with a mass from millions to billions of
$M_{\odot}$, and thus its total light output is dominated by the BH's accretion process, in a similar way as in
BHXBs. The production or lack of relativistic jets may have similar or even the same underlying physical
mechanisms in BHXBs and AGNs, though their surrounding environments may modify their observed morphologies
\cite{Hao2009}. It is therefore more natural to simply refer {\em microquasars} as BHXBs; in the rest of this
article, {\em microquasar} and {\em BHXB} are used interchangeably.

We therefore will focus on BHXBs and thus will not discuss binary systems producing gamma-rays and sometimes
radio jets, which are most likely high-mass NSXBs and in which jets or pulsars' winds interact with the wind of
its high-mass companion to produce the observed gamma-rays
\cite{Kniffen1997a,Paredes2000,Aharonian2005,Aharonian2005a,Mirabel2007,Cui2009,Romero2010}. Such systems show
very different observational characteristics, e.g. the long (1667 days) super-orbital modulation with phase
offset (about 280 days) between its X-ray and radio light curves found in LS I$+61^{\circ}303$ \cite{Li2012a}.

\section{Review of Reviews}

Here I attempt to review the four recent review articles \cite{Fender2012a,Remillard2006,Done2010,Done2007} I
consider most useful to readers. Additional information and updates are provided when necessary. Some overlaps
exists between these review articles, as expected and inevitable in bottom-up review articles. To avoid
repetitions as much as possible in this article, I thus put different emphasizes on different articles, with of
course my personal tastes and perceptions.

\subsection{The most recent {\it Science} Collection}

\begin{figure}
\center{
\includegraphics[angle=0,scale=0.30]{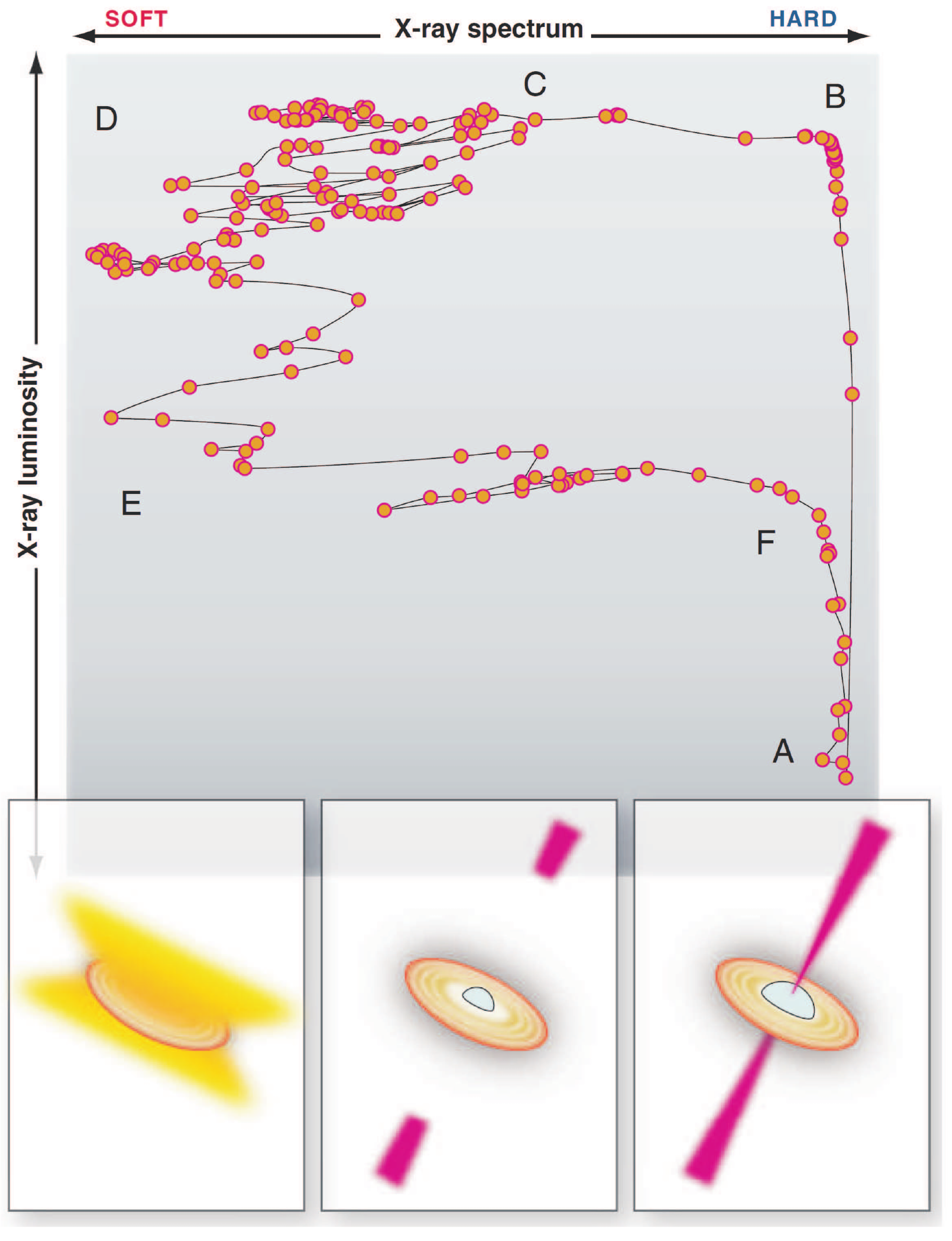}}
\caption{A typical Hardness-Intensity-Diagram (HID) of spectral evolution of a BHXB, following the
A$\rightarrow$B$\rightarrow$C$\rightarrow$D$\rightarrow$E$\rightarrow$F cycle (top). Steady or transient jets
are present during the A$\rightarrow$B or B$\rightarrow$C$\rightarrow$D stage. No jets are observed, but hot
disk winds are ubiquitous during the D$\rightarrow$E stage. (Figure~2 in Ref.\cite{Fender2012a}.)} \label{HID}
\end{figure}

\begin{figure}
\center{
\includegraphics[angle=0,scale=0.45]{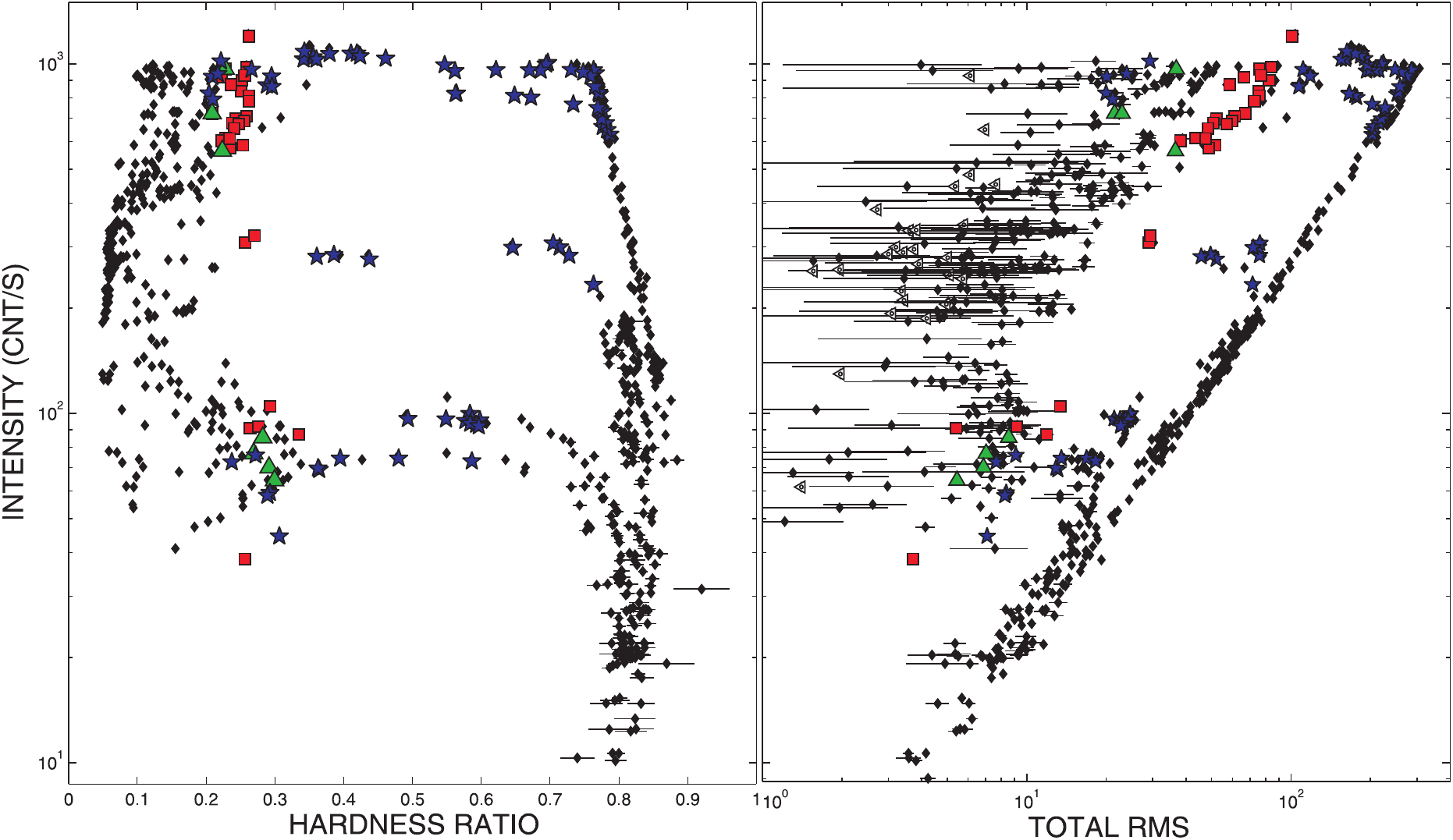}}
\caption{The Hardness-Intensity-Diagram (HID) and RMS-Intensity-Diagram (RID) of the BHXB GX~339--4. The main
difference from Figure~\ref{HID} is that there are additional horizontal tracks at intermediate intensities, a
phenomenon known as hysteresis of state transitions, which will be discussed in Section~\ref{transitions}.
(Adapted from Figure~1 in Ref.\cite{Belloni2011}.)} \label{HID_rms}
\end{figure}

\begin{figure}
\center{
\includegraphics[angle=90,scale=0.45]{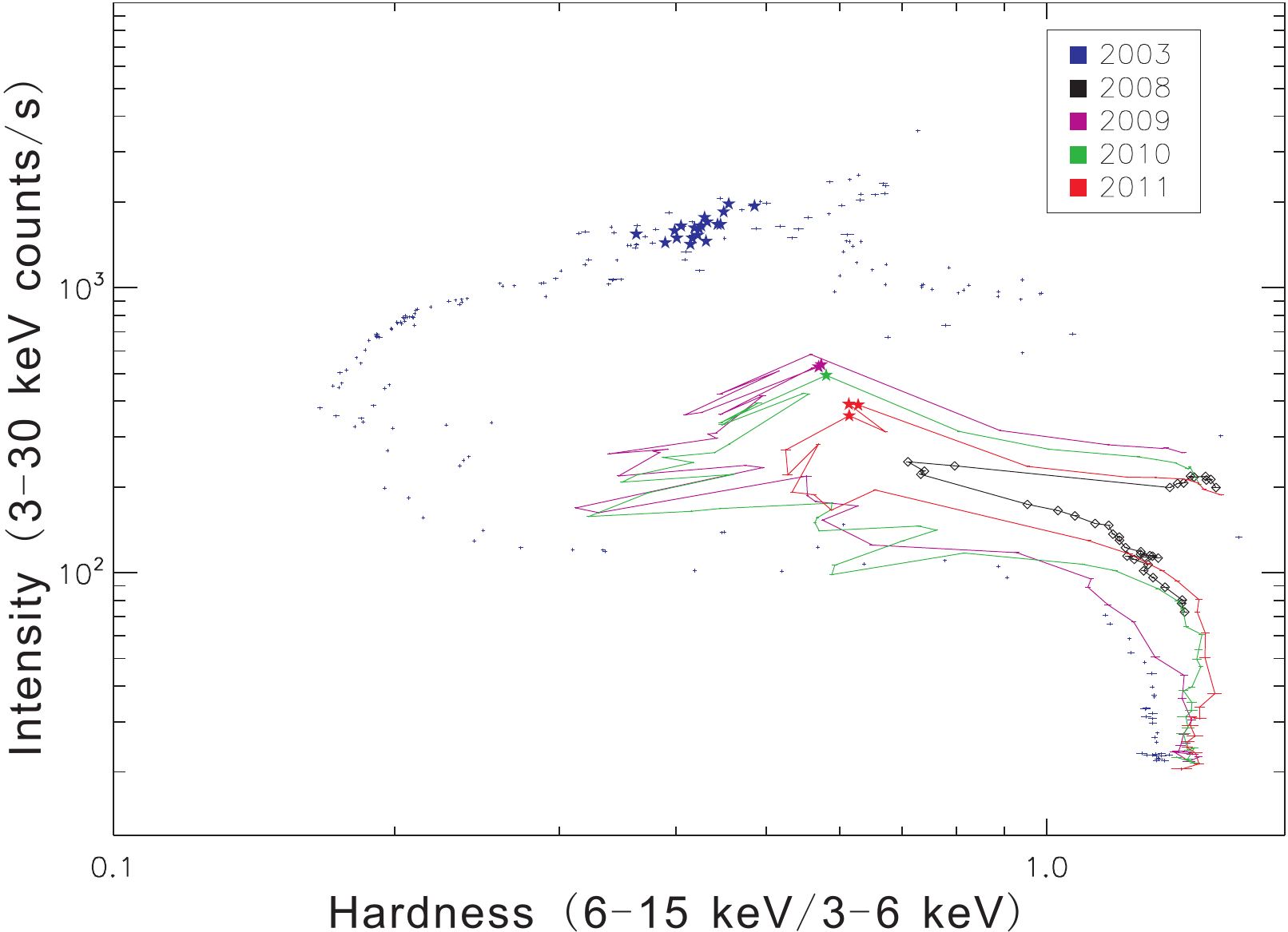}}
\caption{The Hardness-Intensity-Diagram (HID) of H1743--322. The main difference from Figure~\ref{HID_rms} is
that there is a complete track at low intensity from the 2008 outburst, which did not go into the soft state at
all. (Adapted from Figure~7 in Ref.\cite{J.N.Zhou1Q.Z.LiuY.P.ChenJ.LiJ.L.QuS.ZhangH.Q.Gao2013}.)}
\label{HID_1743}
\end{figure}

The recent collection of perspectives and reviews in the {\it Science} magazine provides excellent introductions
to and concise summaries of the current state of our understanding of BH physics and astrophysics
 \cite{Thorne2012,Witten2012,Fender2012a,Volonteri2012}. To the subjects of this article, the most relevant
article in this collection is the one by Fender and Belloni entitled ``Stellar-Mass Black Holes and
Ultraluminous X-ray Sources", which is focused on the observational characteristics of BHXBs \cite{Fender2012a}.
In particular, the article provides an excellent description of the general picture of spectral evolutions of
BHXBs with the Hardness-Intensity-Diagram (HID), which is found to be well correlated with the observed jets
from BHXBs, as shown in Figure~\ref{HID}. Actually this cycle is also well tracked by their flux variability,
represented by the measured root-mean-squares (rms) above its average flux, as shown by the
RMS-Intensity-Diagram (RID) \cite{Belloni2011} of the BHXB GX~339--4 in Figure~\ref{HID_rms}, which also shows
additional horizontal tracks at intermediate intensities. Sometimes, a full HID cycle does not go into the soft
state at all (Figure~\ref{HID_1743}), perhaps due to a failed outburst
\cite{J.N.Zhou1Q.Z.LiuY.P.ChenJ.LiJ.L.QuS.ZhangH.Q.Gao2013}.

The basic scenario is as follows. During the initial stage of an X-ray outburst of a BHXB (A$\rightarrow$B),
which is triggered by a sudden increase of accretion rate onto the BH, its spectrum is normally hard and steady
jets are always observed. After reaching its peak luminosity, its spectrum begins to soften in a chaotic way and
transient jets are normally observed (B$\rightarrow$C$\rightarrow$D). After this transition, the system calms
down with a soft spectrum and no jets are present (D$\rightarrow$E). Finally the system returns to its quiescent
state with a hard spectrum accompanied with the reappearance of jets (E$\rightarrow$F). Throughout this cycle,
the presence of hot accretion disk winds appears to be anti-correlated with its spectral hardness and jet
ejection. This empirical pattern appears to be quite universal for all BHXBs with very few exceptions, though
the underlying physics is still not well understood yet. Nevertheless putting together the above scenario is a
very significant progress in this field over the last ten years. Due to the conciseness of this article, some
other important subjects on BHXBs are not fully discussed and many original references are also missing.

Other articles in this collection \cite{Thorne2012,Witten2012,Volonteri2012} are less relevant to the subjects of this article, but are still quite
interesting to read, within the context of BH astrophysics and physics. The only other type of astrophysical BHs
known to exist in the physical universe are supermassive BHs in the center of almost each galaxy. Volonteri
 \cite{Volonteri2012} concisely summarized our current understanding on how they are formed and grow over the
cosmic time; merging of two BHs is a key process here. Thorne \cite{Thorne2012} focused on what happens when BHs
merge together to produce gravitational waves, which might be used as a new laboratory for studying
gravitational physics and a new window for exploring the universe. Witten \cite{Witten2012} then explained the
quantum properties of BHs, in particular the basic ideas behind Hawking radiation, which might not be important
at all for astrophysical BHs. Nevertheless the understanding gained through studying the quantum mechanics of
BHs actually plays very important roles in developing other theories of physics, such as that in heavy ion
collisions and high-temperature superconductors, as vividly described by Witten \cite{Witten2012}.

\subsection{The most recent ARA\&A article on BHXBs, with some updates}

\begin{figure}
\center{
\includegraphics[angle=0,scale=0.50]{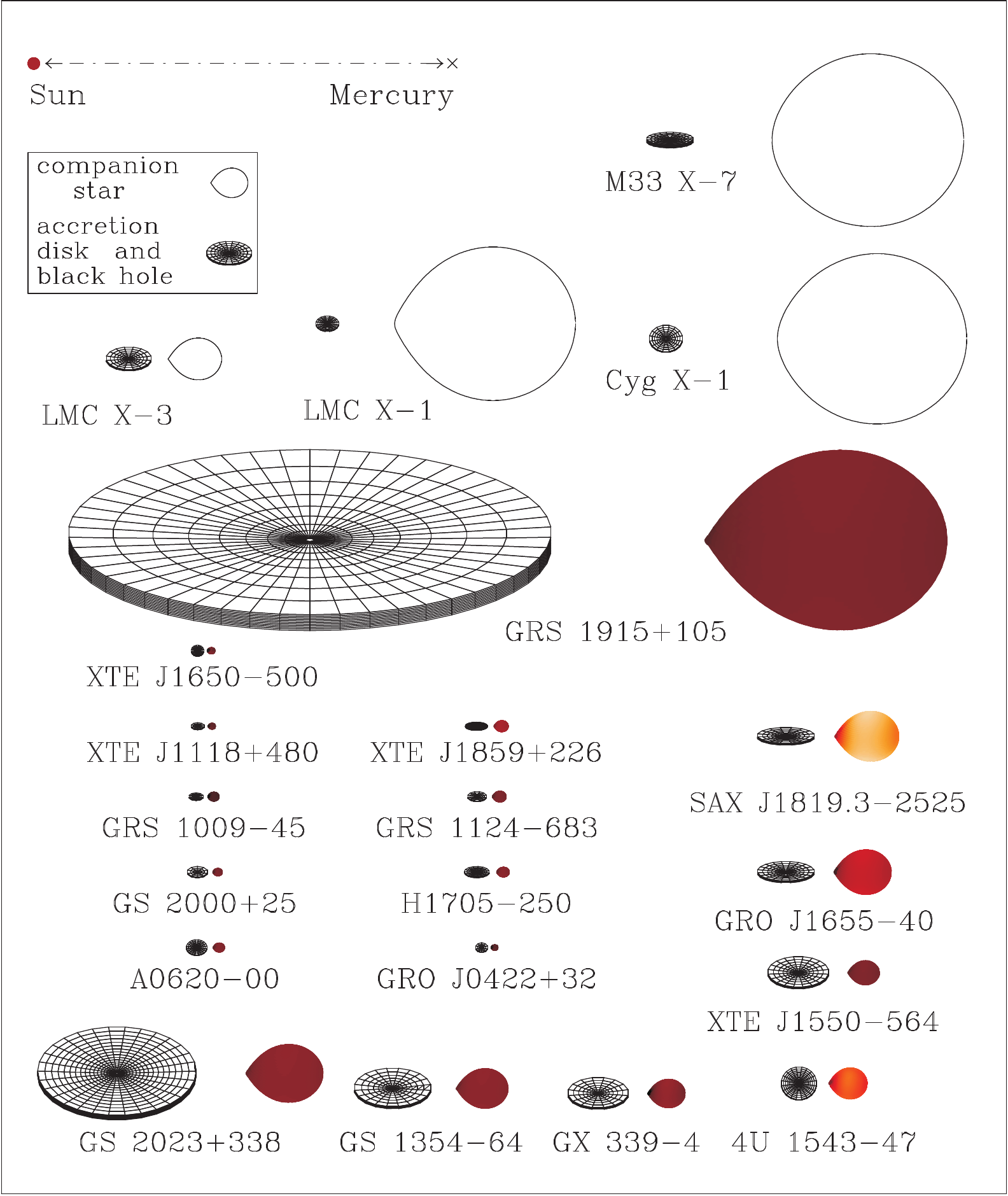}}
\caption{Schematic diagram of the dynamically confirmed BHXBs, maintained by Dr. Orosz
(http://mintaka.sdsu.edu/faculty/orosz/web/). The color scale for the 17 objects with low mass companions (i.e.
stars with masses less than about $3M_{\odot}$) represents the temperature of the star. However, the high mass
companions in Cyg X--1, LMC X--1, LMC X--3, and M33 X--7 are considerably hotter, and are thus not well
represented in color scale. (Adapted from the plot maintained by Dr. Orosz
(http://mintaka.sdsu.edu/faculty/orosz/web/ and Figure 1 in Ref.\cite{McClintock2011}) } \label{bhxbs_fig}
\end{figure}

The most recent ARA\&A article on BHXBs by Remillard and McClintock entitled ``X-Ray Properties of Black-Hole
Binaries" \cite{Remillard2006} (referred to as RM06 hereafter) provides the most complete, comprehensive and
accurate review of BHXBs, which actually covers subjects much beyond just the X-ray properties of BHXBs. The
{\it Introduction} of RM06 highlights the initial theoretical and observational studies of BHXBs, followed by
brief comments on several main review articles on BHXBs preceding this one and the basic properties of BHs
within the context of general relativity (GR). All main properties of BHXBs known at the time are summarized in
the first table and figure there. In Table~\ref{bhxbs_table}, I compile the most updated data on all BHXBs
currently known, including three BHXBs outside the Galaxy and not in the Large Magellanic Cloud (LMC); the
currently available spin measurements for these BHs are also included for completeness. In
Table~\ref{bhxbs_table},
\begin{equation}
\label{equ:mass} f(M)~\equiv~P_{\rm orb}K_{\rm C}^{3}/2\pi G~=~M_{\rm BH}\sin^3i/(1+q)^{2},
\end{equation}
where $P_{\rm orb}$ is the orbital period, $K_{\rm C}$ is the semi--amplitude of the velocity curve of the
companion star, $M_{\rm BH}$ is BH mass, $i$ is the the orbital inclination angle, and $q~\equiv~M_{\rm
C}/M_{\rm BH}$ is the mass ratio. In Figure~\ref{bhxbs_fig}, I show the updated graphical representation of most
of the BHXBs listed in Table~\ref{bhxbs_table}.

\begin{table*}%
\small \caption{Twenty four confirmed BHXBs and their BH masses and spins. Except those marked with references,
all other data are taken from Remillard \& McClintock (2006) \cite{Remillard2006} (references
therein).}\label{bhxbs_table}
\begin{tabular}{llllccc}
\toprule
Coordinate &Common$^b$&$P_{\rm orb}$ &$f(M)$ &$M_{\rm BH}$ & $a_{*}^a$ & $a_{*}^b$\\
Name&Name/Prefix &(hr)&($M_{\odot}$) &($M_{\odot}$) & ($cJ/GM^2$)& ($cJ/GM^2$)\\
\colrule
0422+32          &(GRO~J)     &5.1   &1.19$\pm$0.02   &3.7--5.0  & $-1^c$ \\
0620--003        &(A)         &7.8   &2.72$\pm$0.06   &6.35--6.85 \cite{Cantrell2010} & $0^d$  & 0.12$\pm$0.19 \cite{Gou2010}\\
1009--45         &(GRS)       &6.8   &3.17$\pm$0.12   &3.6--4.7    \\
1118+480         &(XTE J)     &4.1   &6.1$\pm$0.3     &6.5--7.2    \\
1124--684        &Nova Mus 91 &10.4  &3.01$\pm$0.15   &6.5--8.2   & -0.04\\
1354--64     &(GS)        &61.1 &5.75$\pm$0.30   & --     \\
1543--475        &(4U)        &26.8  &0.25$\pm$0.01   &8.4--10.4  &  &   0.75--0.85 \cite{Shafee2006} \\
1550--564        &(XTE~J)     &37.0  &6.86$\pm$0.71   &8.4--10.8 & &0.06--0.54 \cite{Steiner2011b}  \\
1650--500    &(XTE~J)     &7.7   &2.73$\pm$0.56   & --         \\
1655--40         &(GRO~J)     &62.9  &2.73$\pm$0.09   &6.0--6.6  & $0.93^e$ & 0.65--0.75 \cite{Shafee2006}  \\
1659--487        &GX~339--4  &42.1 &5.8$\pm$0.5    & --   \\
1705--250        &Nova Oph 77 &12.5  &4.86$\pm$0.13   &5.6--8.3    \\
1743--322        &(H)         & --   & --             & --    && 0.2$\pm$0.3 \cite{Steiner2011}  \\
1819.3--2525     &V4641 Sgr   &67.6  &3.13$\pm$0.13   &6.8--7.4    \\
1859+226         &(XTE~J)    &9.2 &7.4$\pm$1.1:$^e$  &7.6--12.0 \\
1915+105         &(GRS)     &804.0 &9.5$\pm$3.0     &10.0--18.0 &       0.998  & 0.98--1.0 \cite{McClintock2006} \\
1956+350         &Cyg~X--1    &134.4 &0.244$\pm$0.005 &13.8--15.8 \cite{Orosz2011} & $\pm 0.75^f$ & $>0.95$ \cite{Gou2011a} \\
2000+251         &(GS)        &8.3   &5.01$\pm$0.12   &7.1--7.8  &0.03  \\
2023+338         &V404 Cyg  &155.3 &6.08$\pm$0.06   &10.1--13.4  & $-1^c$ \\
\hline
0538--641        &LMC~X--3    & 40.9  &2.3$\pm$0.3     &5.9--9.2  &-0.03  & $<0.3$ \cite{Davis2006}\\
0540--697        &LMC~X--1    &93.8  &0.13$\pm0.05$   &9.4--12.5 \cite{Orosz2009}& &0.94--0.99 \cite{Gou2009,Steiner2012}\\
\hline
0020+593&     IC 10 X--1&   34.9&7.64$\pm$1.26&   $>20$ \cite{Prestwich2007,Silverman2008}\\
0055--377&     NGC 300 X--1&   32.3& 2.6$\pm$0.3&  $>10$ \cite{Crowther2010}\\
0133+305&     M33 X--7&     82.9& 0.46$\pm$0.07&  14.2--17.1 \cite{Orosz2007} & & 0.84$\pm$0.05 \cite{Liu2010}\\
\botrule
\end{tabular}
\\
{\it a}: Reported in the first paper on systematic BH spin measurements \cite{Zhang1997}; {\it b}: Reported in
the most recent literature; {\it c:} Postulated to be extreme retrograde Kerr BH, due to the lack of the thermal
disk component above 2 keV (GRO~J1719--24 also belongs to this class) \cite{Zhang1997}; {\it d}: Postulated to
be non-spinning BH, due to the observed low disk temperature; {\it e}: BH mass of 7 $M_{\odot}$ assumed
\cite{Zhang1997}; {\it f}: Based on the inner disk radius decrease by a factor of two from the hard state to the
soft state transition \cite{Zhang1997}.
\end{table*}

The X-ray properties of BHXBs are characterized by their X-ray light curves, timing and spectra. Essentially all
known BHXBs were discovered initially as bright X-ray sources, and the majority of them were detected as
transient X-ray sources with X-ray all-sky monitors. The transient properties of some of BHXBs can be
interpreted by the disk instability model (DIM) (see Ref.\cite{Lasota2001} and references therein), which
assumes a constant mass transfer rate from the mass donor to the accretion disk. However the accretion rate from
the disk to the compact object, i.e. a white dwarf (WD), a neutron star (NS), or a BH, is normally lower than
the mass supply rate in the disk, so mass is accumulated in the disk. When the accumulated mass exceeds a certain critical
value, a sudden increase of accretion rate results in a nova-like outburst. DIM is most successful in explaining
an outburst with fast rise and exponential decay (FRED). However as shown in RM06, many observed X-ray light
curves of BHXBs are far more complicated than just FRED and their recurrent time scales are also not compatible
with DIM. Perhaps disk truncation (a subject to be discussed extensively more later) and mass transfer
instability are additional ingredients \cite{Lasota2011}.

X-ray emission from a BHXB is variable at all time scales, from the free-fall or Keplerian orbital time scale of
milliseconds near the BH, to the various oscillations (some are even related to GR) in the disk with time scales
from milliseconds to minutes, to the viscous time scales of minutes, and to the various instabilities of
different time scales. Therefore timing studies of BHXBs can probe the geometry and dynamics in BHXBs. However,
lack of coherent signals, such as that observed from pulsars, makes it difficult to unambiguously identify the
underlying mechanisms from the detected X-ray variabilities. Nevertheless power density spectra (PDS) and rms
still allow us to make progress in understanding the general characteristics of a BHXB, in particular when
combined with its spectral behaviors, as already discussed briefly above.

The most successful understanding of BHXBs so far is the description of their thermal X-ray spectral component
by the classical Shakura-Sunyaev Disk (SSD) model \cite{Shakura1973}. After applying GR to SSD, the temperature
distribution in the disk can be obtained \cite{Page1974}. Applying this model to BHXBs, one can even measure the
spin of their BHs \cite{Zhang1997}, by assuming that the inner accretion disk boundary is the inner-most stable
circular orbit (ISCO) of the BH (a subject to be discussed more in Section~\ref{BH_spin}). However a power-law
(PL) component is almost ubiquitous in the spectra of BHXBs. The interplay between these two components results
in various spectral states, which are also found to be well correlated with their timing properties.

\begin{figure}
\center{
\includegraphics[angle=0,scale=0.45]{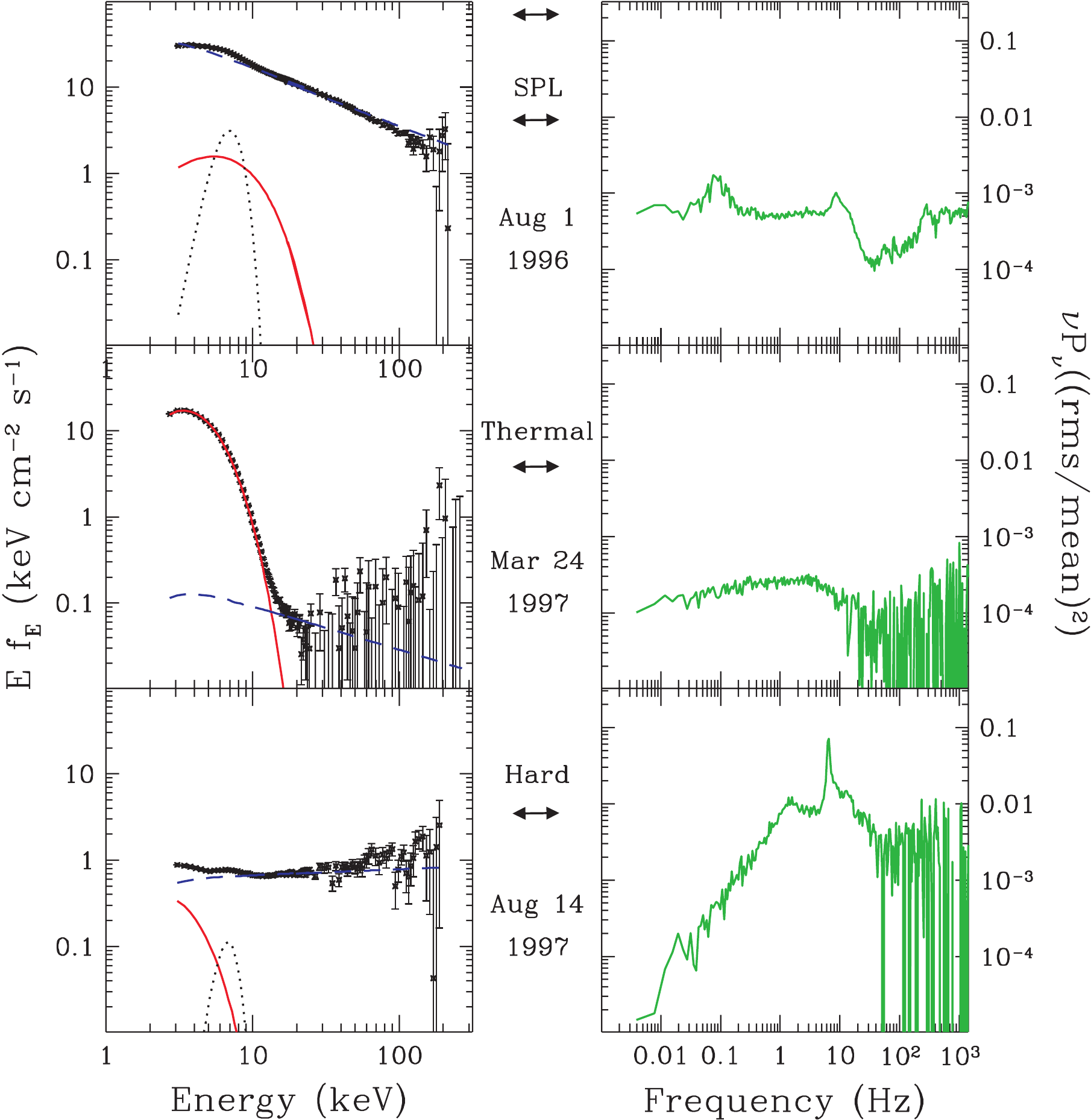}}
\caption{Characteristic spectra and PDS of the BHXB GRO~J1655--40 in three different states, namely, steep
power-law (SPL), thermal and hard states. Please refer to Figure~2 in RM06 for details. (Figure~2 in
Ref.\cite{Remillard2006}.)} \label{states}
\end{figure}

Historically these spectral states have been named in many different ways, reflecting mostly how they were
identified with the observations available at the times. In RM06, three states are defined, which show
distinctively different spectral and timing behaviors, as shown in Figure~\ref{states} based on the RXTE data on
the BHXB GRO~J1655--40. The {\it thermal state} has its X-ray spectrum dominated by the thermal disk component
and very little variability. The {\it hard state} has its X-ray spectrum dominated by the a PL component and
strong variability. The {\it steep power-law} (SPL) state is almost a combination of the above two states, but
the PL is steeper. I will keep using these definitions throughout this article for consistency. Please refer to
Table~2 in RM06 for quantitative descriptions of these three states. When a BHXB is in quiescence, i.e. not in
an outburst, its spectral shape is similar to the hard state spectrum.

Referring back to Figure~\ref{HID}, RM06 found that the HID can be reasonably well understood in terms of
transitions between the above three states; I summarize this in Table~\ref{HID_states}. After examining the
extensive RXTE data on spectral evolutions of six BHXBs, the above conclusion seems to be valid generally. Thus
a coherent picture seems to emerge: jets are produced only when the PL component in the spectrum is strong and
winds are present only when the PL component disappears; jets and winds appear to be mutually exclusive. This
suggests that transient winds should appear in the SPL state, anti-phased with transient jets, if winds quench
jets as suggested recently \cite{Neilsen2009,Ponti2012}.

The robust link between the PL component and jets provides an important clue to the jet production mechanism.
The PL component is believed to be produced in an optically thin and geometrically thick corona, in which hot
electrons up-scatter the thermal photons from the disk. As discussed in RM06, the inner region of the disk
appears to be truncated quite far away from the BH in the hard state; this should be the primary reason for the
low luminosity of this state, as evidence for the existence of BHs in BHXBs. However it is still not understood
how the corona is formed; I will discuss this later in Section~\ref{corona}.

In the final part, RM06 discussed the exciting possibilities of using BHXBs as probes of strong gravity,
including confirming these BHXBs contain true BHs, measuring the spins of BHs, relating BH spin to the Penrose
process and other phenomena, and finally carrying out tests of the Kerr Metric. They also discussed in length
how to measure the spins of BHs and commented the various methods of doing so. In Section~\ref{BH_spin} I will
re-visit the BH spin measurements in details.

\begin{table}%
\small \caption{Relations between HID, spectral states, jets and wind.}\label{HID_states}
\begin{tabular}{llll}
\toprule
HID Stage & State Transition & Jets & Winds\\
 \colrule
 A$\rightarrow$B & hard$\rightarrow$SPL & steady & no\\
B$\rightarrow$C$\rightarrow$D & SPL$\rightarrow$soft & transient & transient?\\
D$\rightarrow$E & soft decay & no & steady\\
E$\rightarrow$F &  soft$\rightarrow$hard & steady & no\\
\botrule
\end{tabular}
\end{table}

Lacking of a solid surface, a BHXB is not expected to produce and actually never observed to have coherent
pulses. However quasi-periodic oscillations (QPOs) are frequently detected in their X-ray light curves; a QPO is
defined as a `bump' feature in the PDS, if $Q=\nu/\Delta\nu>2$, where $\nu$ and $\Delta\nu$ are the peak
frequency and FWHM of the bump, respectively. The observed QPOs are further divided into low frequency (LF)
(0.1-30 Hz) and high frequency (HF) ($>30$ Hz) QPOs. Typical LFQPOs can be found as the peak around several Hz
in the top and bottom PDS of Figure~\ref{states}. In literature, LFQPOs are further divided into several
subclasses and their underlying mechanisms are far from clear at this stage, though many models have been
proposed and briefly discussed in RM06. HFQPOs have been detected from several sources at 40-450 Hz. Of
particular interests are their stable nature when detected, which may be linked to either the mass or spin or
both of the BH in a BHXB.

In some cases a 3:2 frequency ratio is found. Although the 3:2 HFQPO pairs could be interpreted in some
epicyclic resonance models \cite{Abramowicz2001,Torok2005}, there remain serious uncertainties as to whether
epicyclic resonance could overcome the severe damping forces and emit X-rays with sufficient amplitude and
coherence to produce the HFQPOs. A revised model applies epicyclic resonances to the magnetic coupling (MC) of a
BH's accretion disk to interpret the HFQPOs \cite{Huang2010}. This model naturally explains the association of
the 3:2 HFQPO pairs with the steep power-law states and finds that the severe damping can be overcome by
transferring energy and angular momentum from a spinning BH to the inner disk in the MC process.

\subsection{Two comprehensive and long articles on modeling accretion flows in BHXBs}

Page limitations to the above review articles did not allow in depth discussions on the detailed processes and
models of accretion flows in BHXBs, which are responsible for the above described energy spectra, PDS, state
transitions, jets, and wind. Here I introduce two comprehensive and long articles just on this, by Done
 \cite{Done2010} and Done, Gierli{\'n}ski \& Kubota \cite{Done2007}.

\subsubsection{A beginner's guide}

The first is intended to readers who just start research in this field \cite{Done2010}. It started with the
basic tools in plotting spectra and variability, and then described the basic ideas and methods used to infer
the inner accretion disk radius when a BHXB is in a thermal state, in order to measure the BH spin
\cite{Zhang1997}. A useful introduction is made on how to make various corrections to account for various
effects, including color correction, special and general relativistic effects, starting from the original work
\cite{Zhang1997}. Several commonly use fitting models, i.e. {\sc DISKBB}, {\sc BHSPEC} and {\sc KERRBB}, in the
XSPEC package are also briefly introduced. An example is given to demonstrate that the expected relation $L_{\rm
disk}\propto T_{\rm in}^4$ \cite{Zhang1997} agrees with the data from the BHXB GX~339--4 \cite{Kolehmainen2010},
where $L_{\rm disk}$ and $T_{\rm in}$ are the disk's total luminosity and temperature at the inner disk
boundary, respectively.

The hard state is then briefly touched upon, using the Advection Dominated Accretion Flow (ADAF) model
\cite{Narayan1995}. ADAF naturally explains the hot corona required by the observed PL component, especially at
very low accretion rate. Evaporation at low accretion rate in the inner disk region has been proposed as the
mechanism producing a geometry with a radially truncated disk and a hot inner flow; the latter might be the ADAF
\cite{Liu1999}. This also means that when a significant PL component is present in the spectrum, BH spin
measurement cannot be done with the inferred inner disk boundary (see, however, counter evidence discussed in
Sections~\ref{BH_spin} and \ref{corona}). This geometry is considered a paradigm that can account for many of
the observed diverse phenomena from spectral evolution to timing properties.

A brief, yet interesting discussion is given on scaling up the above models to Active Galactic Nuclei (AGNs),
which host actively accreting supermassive BHs in the centers of galaxies. Applying the insights learnt from
BHXBs on their spectral evolution, changing disk-temperature with accretion rate and BH mass, and disk-jet
connections, one might be able to understand many phenomena beyond the simple AGN unification scheme, in which
the different observational appearance is all attributed to an viewing angle difference.

\begin{figure}
\center{
\includegraphics[angle=0,scale=0.40]{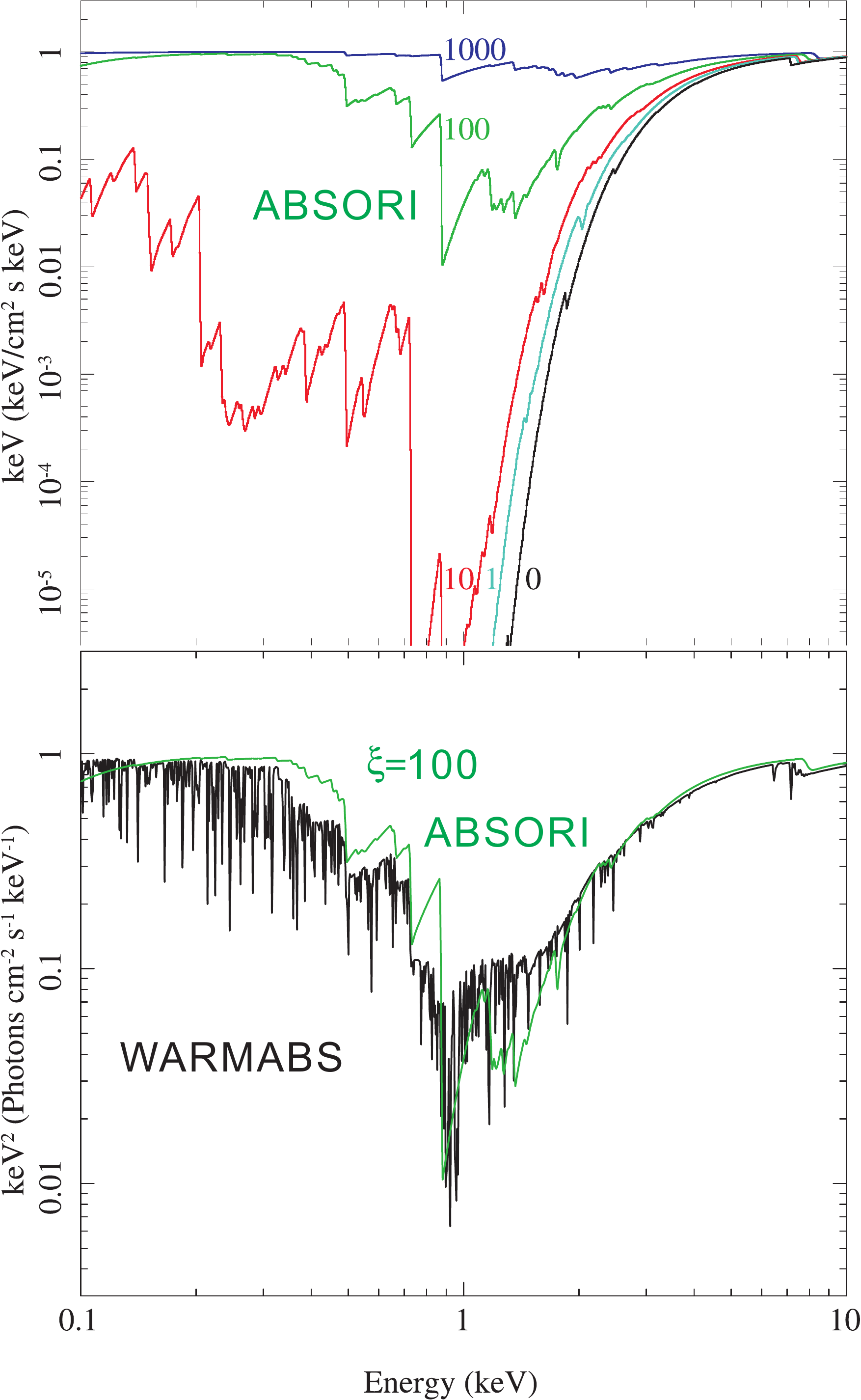}}
\caption{Calculated absorption structures of a column density $N_{\rm H}=10^{23}~{\rm cm}^{-2}$ for different
ionization parameter $\xi=L/nr^2$, where $L$ is the illuminating luminosity, $n$ is the density of the medium,
and $r$ is the distance from the radiation source. {\it Top panel}: with the ABSORI model that does not include
absorption lines. {\it Bottom panel}: for $\xi=100$ with both ABSORI and WARMABS, the latter is based on the
XSTAR photo-ionization package thus absorption lines are taken into account properly.(Adapted from Figure~17 in
Ref.\cite{Done2010}.)} \label{absorb}
\end{figure}

The continuum emissions of both the disk thermal and PL components are modified by absorptions and added by
additional spectral features along the line of sight (LOS). Absorptions in neutral media produce various
photoelectric absorption edges, but in ionized media result in both absorption edges and lines. In addition to
recovering the original X-ray emission of a BHXB, modeling the absorption features is important in learning the
physical properties of the absorption media, such as column density, ionization and velocity along LOS. Winds
from BHXBs discussed above are always detected this way. Several XSPEC fitting models for various kinds of
absorption edges and lines are also introduced here, i.e. {\sc TBABS}, {\sc ZTHABS}, {\sc TBVARABS}, {\sc
ZVPHABS}, {\sc TBNEW}, {\sc ABSORI} and {\sc WARMABS}. Figure~\ref{absorb} shows the calculated absorption
structures with {\sc ABSORI} and {\sc WARMABS}; note that {\sc ABSORI} does not include absorption lines, but
{\sc WARMABS} does.

X-rays interacting with the surrounding medium can ionize it and heat the free electrons up by Compton
scattering; hot electrons can also loose energy by interacting with lower energy photons. A Compton temperature
of the plasma is reached when the above two processes reaches an equilibrium. The temperature is determined by
only the spectral shape of the continuum, so the heated plasma can escape as winds from the disk if the velocity
of the ions exceed that of the escape velocity at that radius, which is usually very far away from the inner
disk boundary. As the continuum luminosity approaches the Eddington limit, the radiation pressure reduces the
escape velocity substantially in the inner disk region such that continuum driven winds can be launched almost
everywhere in the disk of a BHXB, forming a radiation driven wind, which is also called thermal wind;
alternatively winds may also be driven magnetically in the inner disk region, but this is much less understood
yet. In contrast, in an AGN the peak continuum emission is in the UV band, which has a much larger opacity than
Compton scattering in neutral or weakly ionized medium during both photoelectric and line absorptions. This
means the effective Eddington luminosity is reduced by large factors and UV line driven wind is easily produced
at high velocity. This explains the relatively lower velocity (hundreds km/s) and highly ionized winds from
BHXBs, but much higher velocity (thousands km/s to 0.2 $c$) and weakly ionized winds from AGNs.

\begin{figure}
\center{
\includegraphics[angle=0,scale=0.45]{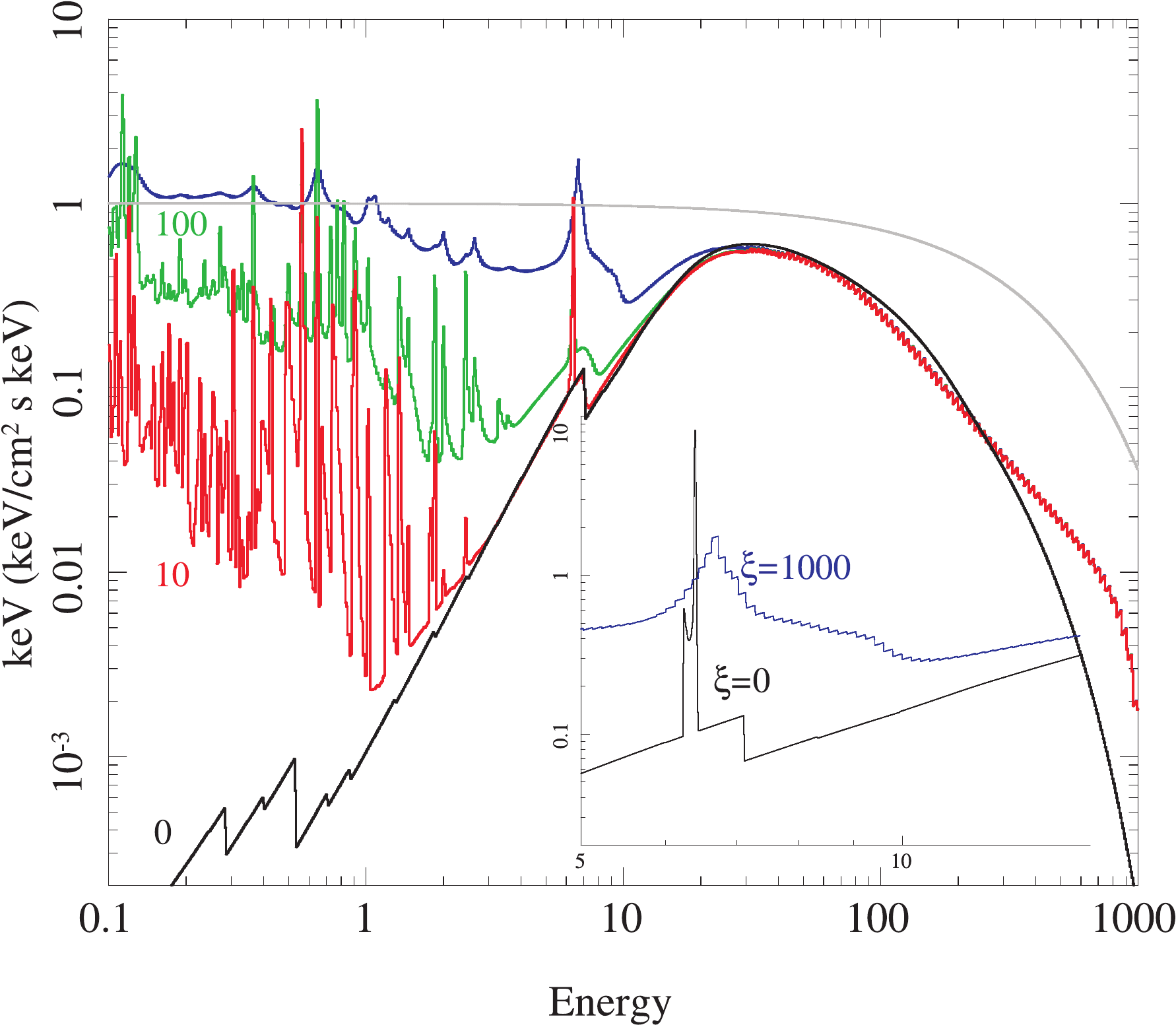}}
\caption{Ionized reflections from a constant density slab, calculated with the XSPEC {\sc ATABLE} model, {\sc
REFLIONX.MOD}, which includes the self-consistent line and recombination continuum emission, for different
values of $\xi$. The inset shows a detailed view of the iron line region. For neutral material, around 1/3 of
the line photons are scattered in the cool upper layers of the disk before escaping, forming a Compton
down-scattered shoulder to the line. For highly ionized reflection, the upper layers of the disk are heated to
the Compton temperature so that the spectral features are broadened. (Adapted from Figure~22 in
Ref.\cite{Done2010}.)} \label{reflion}
\end{figure}

\begin{figure}
\center{
\includegraphics[angle=0,scale=0.45]{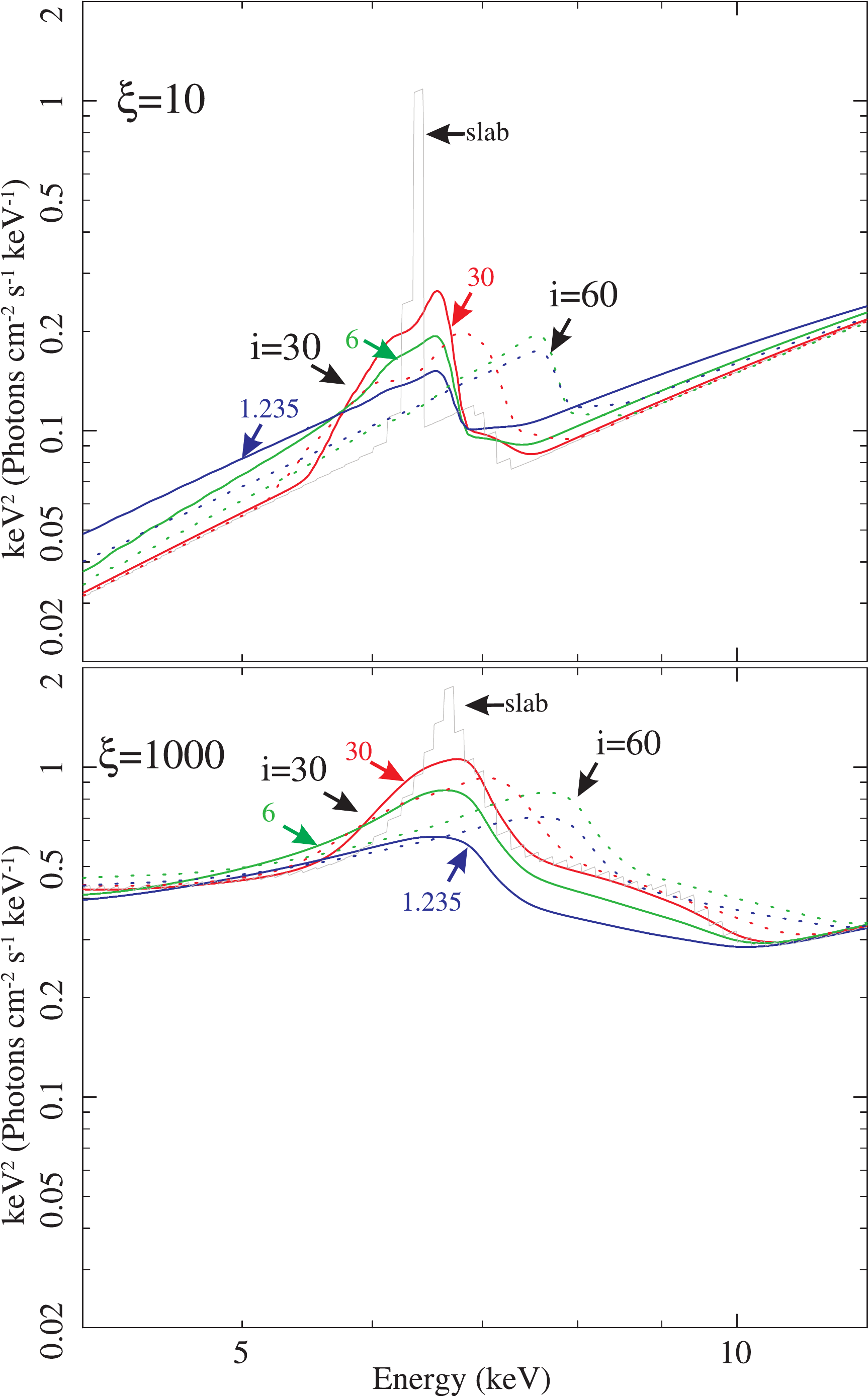}}
\caption{Relativistic smearing of the iron line region. The sharp peak shown in solid grey lines are the
original features. Other colored solid lines have different $r_{\rm in}$ marked in units of $r_{\rm g}=GM/c^2$
for $i=30^\circ$ (solid lines); similarly those for $i=60^\circ$ are shown in dotted lines. (Adapted from
Figure~25 in Ref.\cite{Done2010}.)} \label{iron}
\end{figure}

\begin{figure}
\center{
\includegraphics[angle=0,scale=0.85]{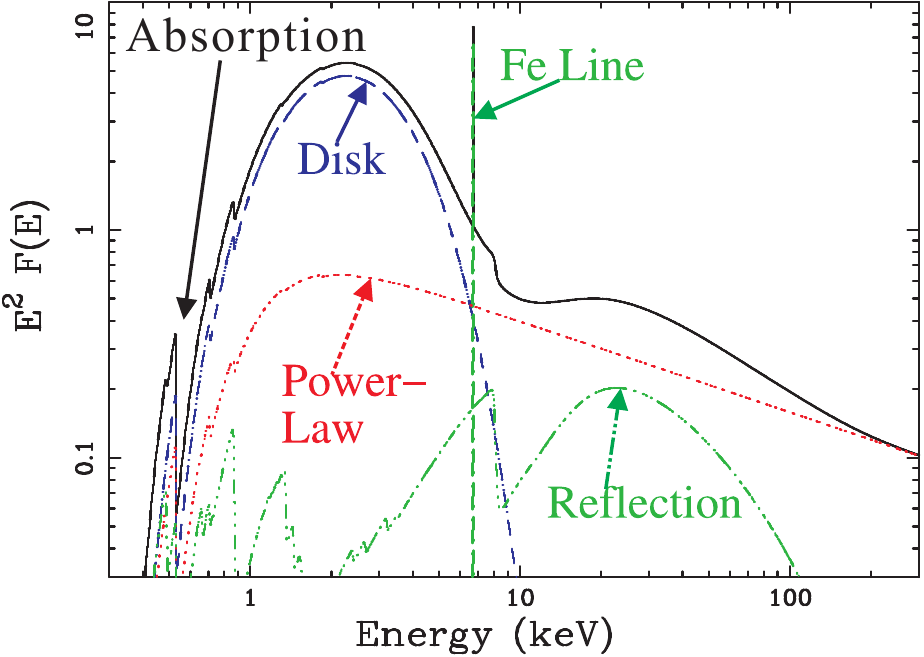}}
\caption{Illustration of all components of the broad band energy spectrum of a BHXBs; interstellar absorption is
normally prominent at low energies. (Adapted from Figure~3b in Ref.\cite{Miller2007}.)} \label{broad_band}
\end{figure}

Material illuminated by X-rays produce both fluorescence lines and reflection features, which depends on both
the continuum spectral shape and ionization state of the material, as shown in Figure~\ref{reflion} calculated
with the XSPEC {\sc ATABLE} model REFLIONX.MOD, which includes the self-consistent line and recombination
continuum emission. Note that for highly ionized reflection, the Compton heated upper layers of the disk
broadens the spectral features; these effects are not included in the simpler XSPEC model {\sc PEXRIV}.
Replacing the stationary slab by a disk around a BH, both the special and general relativistic effects will
further smear (broaden) the spectral features, as shown in Figure~\ref{iron} for the iron line region with
different values of $\xi$, inner disk radius $r_{\rm in}$, and viewing angle $i$; these are calculated with the
{\sc REFLIONX.MOD} and then convolved with {\sc KDBLUR} in the XSPEC package. Other similar XSPEC fitting models
are {\sc DISKLINE}, {\sc LAOR}, and {\sc KY}. The inner disk radius can in principle be determined by modeling
the observed broad iron line features, that in turn can be used to measure BH spins, as will be discussed
briefly in Section~\ref{BH_spin}. Figure~\ref{broad_band} shows all components of a broad band spectrum of a
BHXBs, including interstellar absorption.

We thus have two ways discussed so far to determine $r_{\rm in}$ of BHXBs by either fitting its thermal
continuum or the fluorescent iron line feature. However results are not always consistent, obtained with
different methods, even using the same data by different authors. The author believes that $r_{\rm in}$
increases at low accrete rate, to make room for the corona occupying this space. Nevertheless a fairly good
summary on the current conflicts and confrontations is made on this issue, which I will discuss further in
Section~\ref{BH_spin}.

\subsubsection{An expert's handbook}

The second one can be considered as a handbook on accretion flows on BHXBs, and really lives up to its subtitle
``Everything you always wanted to know about accretion but were afraid to ask" \cite{Done2007} (DGK07
hereafter). Besides its much longer length of 66 pages, the main difference from the above review articles is
that it is focused on confrontations between theories and observations and intends to depict a coherence picture
of the accretion physics in BHXBs. In the following I will summarize briefly the main points and conclusions
reached in DGK07; those I have reviewed above and will discuss more later will be skipped for brevity.

\begin{figure}
\center{
\includegraphics[angle=0,scale=0.6]{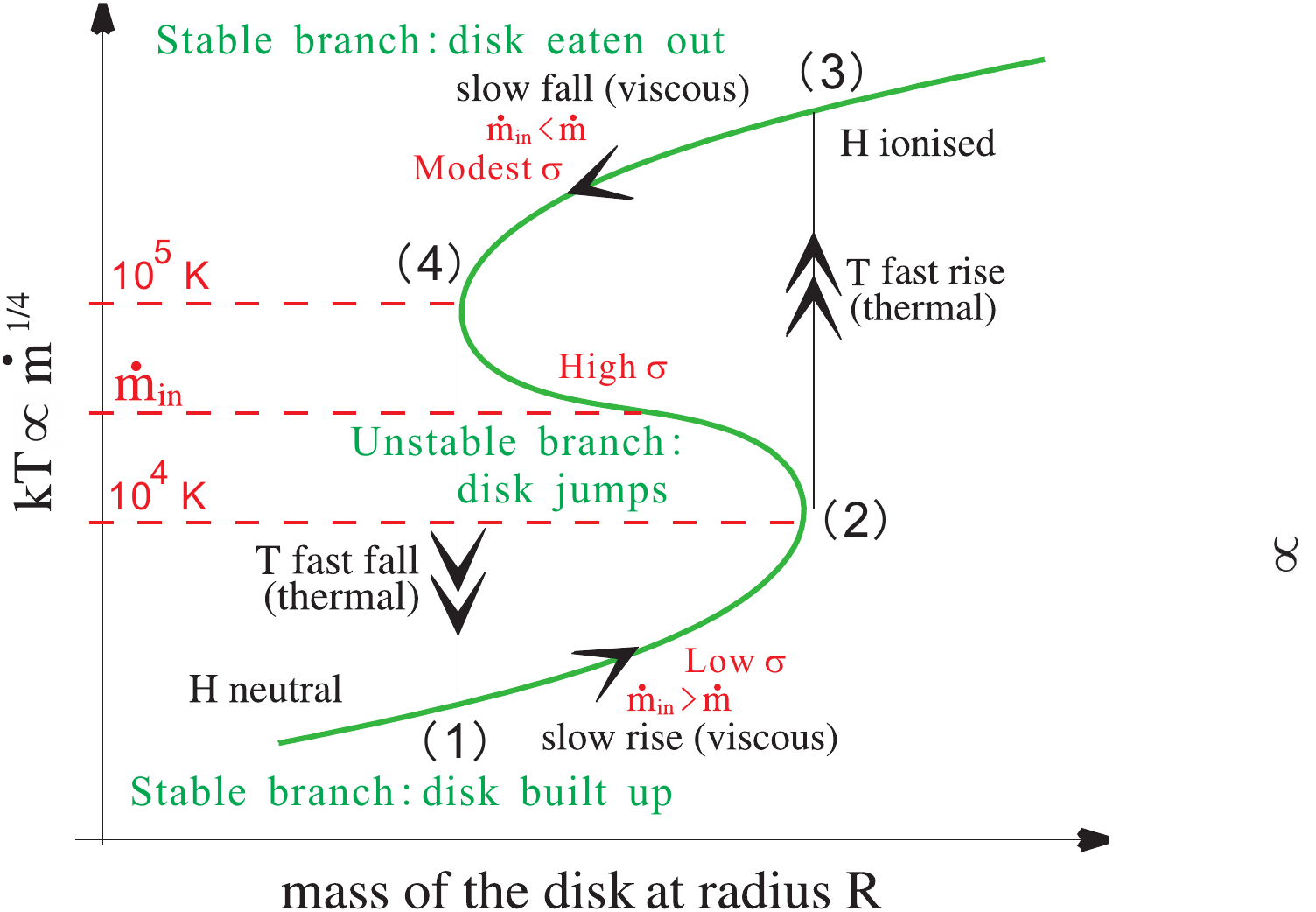}}
\caption{``S"-curve due to the Hydrogen-ionization instability. At a radius $R$, the SSD solution gives
$kT\propto \dot{m}^{1/4}$, where $T$ is the local disk temperature and $\dot{m}$ is the local mass accretion
rate. A positive slope means cooling can balance heating, so $T$ changes slowly and the disk is stable. In the
unstable middle branch, thermally emitted photons are absorbed to ionize the hydrogen atoms and are thus trapped
in the disk, causing a large opacity ($\sigma$). If the mass supply rate $\dot{m}_{\rm in}$ happens in the
middle range, then the disk experiences a limit-cycle instability from (1) to (4): (1) $T< 10^4$~K, neutral
hydrogen has low $\sigma$, $\dot{m}_{\rm in}>\dot{m}$, so the disk is built up and $T$ increases slowly; (2)
$T\sim 10^4$~K, hydrogen atoms begin to be ionized and so has high $\sigma$, $T$ increases rapidly until
hydrogen is fully ionized; (3) $T> 10^5$~K, ionized hydrogen has modest $\sigma$, $\dot{m}_{\rm in}<\dot{m}$, so
the disk is eaten out and $T$ decreases slowly; (4) $T\sim 10^5$~K, protons and electrons begin to recombine,
$T$ decreases rapidly until reaching the lower stable branch again. (Adapted from Figure~1 in
Ref.\cite{Done2007}.)} \label{s_curve}
\end{figure}

The underlying physics of DIM for triggering the outbursts in BHXBs discussed above is the hydrogen-ionization
instability, which produces the so-called ``S"-curve, as shown in Figure~\ref{s_curve}; irradiation by the inner
hot disk to keep the outer disk hot is required to produce the slow flux decays, e.g. the exponential decays,
frequently observed in them. The outer disk radius is obviously another key parameter, which is determined by
the tidal instability in a binary system. This mechanism can explain why BHXBs with high mass companions are all
persistently bright, since their outer disks are always in the upper branch due to the combination of their
higher average mass transfer rate and inner disk irradiation. Similarly it also explains the differences and
similarities between the light curve properties of neutron star X-ray binaries (NSXBs) and BHXBs, since a NS has
a lower mass. It should be noted that the additional surface emission from the NS may also help to maintain the
outer disk hot and stay in the upper stable branch \cite{King1997}.

The SSD prescription assumes that the stress is proportional to pressure. Because the gas pressure is $P_{\rm
gas}\propto T$ but radiation pressure is $P_{\rm rad}\propto T^4$, so a small temperature increase causes a
large pressure increase when $P_{\rm rad}\geq P_{\rm gas}$, and thus large stress increase, which in turn heats
the disk even more. The opacity cannot decrease effectively to cool it down, so the disk becomes unstable when
$L\geq 0.06 L_{\rm Edd}$, where $L_{\rm Edd}$ is the Eddington luminosity of a BHXB. However in most BHXBs their
disk emissions appear to be stable up to around $0.5 L_{\rm Edd}$. One way out is to assume that the stress is
proportional to $\sqrt{P_{\rm rad}P_{\rm gas}}\propto T^{5/2}$, so the stress increases slower in the radiation
pressure dominated regime. Beyond this the disk should become unstable, as evidenced by the sometimes
``heart-beat" bursts of the super-Eddington BHXB GRS~1915+105; other BHXBs with $L_{\rm max} > L_{\rm Edd}$
(e.g. V404 Cyg and V4641) were not observed with such instabilities, perhaps due to the combination of
insufficient observational coverage and sensitivity. However similar ``heart-beat" bursts were also observed
recently from IGR~J17091--3624, which is likely substantially sub-Eddington unless it is located much beyond 20
kpc and/or its BH mass is quite small \cite{Altamirano2011}.

Super-Eddington accretion flow (SEAF) can become stable again, if the radiation instability is overcome by an
optically thick ADAF, i.e. the slim disk model, in which the trapped photons in the flow is advected inwards,
thus balancing the heating generated by viscosity. Strong radiation driven winds can be easily produced; this
can happen in BHXBs and NSXBs (e.g. Z-sources). Evidence exists that truncated inner disk (TID) is quite common
in SEAF and outflow even dominates over inflow in SEAF \cite{Weng2011}.

The observed PL component in BHXBs cannot be explained by the SSD-like models and thus requires a hot accretion
flow (HAF). At low accretion luminosity (e.g. the quiescent state), the HAF may be the advection
dominated accretion flow (ADAF), the convection dominated accretion flow (CDAF), or the advection dominated
accretion inflow/outflow solution (ADIOS). At higher luminosity (e.g. the hard state), the original hot and
optically thin disk solution (i.e. the SLE solution) is unstable, because the electron heating efficiency by the
Coulomb coupling between protons and electrons is too low. The luminous HAF (LHAF), however, has the advection
as a heating source to electrons, so the heating efficiency increases and thus electrons can cool the flow more
effectively. Outflows can be produced in ADAF/ADIOS; collimated jets can also be produced if magnetic fields are
involved, so an accretion flow may even be jet dominated (i.e. JDAF).

\begin{figure*}
\center{
\includegraphics[angle=0,scale=0.6]{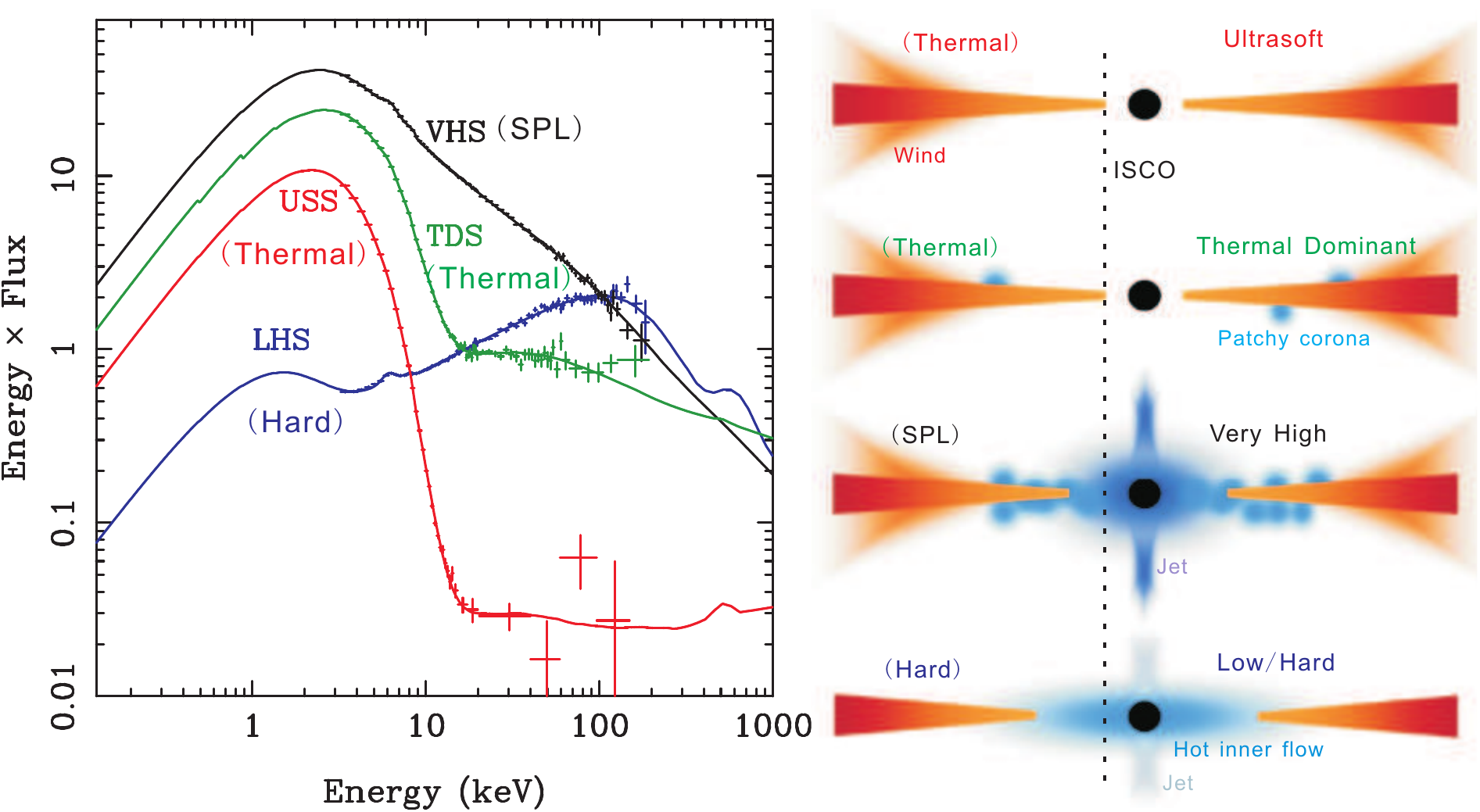}}
\caption{Left panel: a selection of spectra in different states taken from the 2005 outburst of GRO~J1655--40;
state names in parentheses are defined in RM06. Right panel: proposed accretion flow structures in these
different spectra; note the SSD is truncated away from ISCO in the bottom two states. (Adapted from Figure~9 in
Ref.\cite{Done2007}.)} \label{4states}
\end{figure*}

The interplays between the SSD and HAF may be responsible for the observed different states in BHXBs discussed
above. If the PL component is produced by Compton up-scattering, then the combination of the optical depth
$\tau$ and ${\cal L}_{\rm h}/{\cal L}_{\rm s}$, where ${\cal L}_{\rm h}$ is the heating power in electrons and
${\cal L}_{\rm s}$ is the cooling power in seed photons, can describe the observed variety of spectra. For
example, the hard state has ${\cal L}_{\rm h}/{\cal L}_{\rm s}\gg 1$, but the thermal and SPL state have ${\cal
L}_{\rm h}/{\cal L}_{\rm s}\leq 1$. The location of $r_{\rm in}$ is proposed to be closely related to ${\cal
L}_{\rm h}/{\cal L}_{\rm s}$, as illustrated in Figure~\ref{4states}. As discussed above, the steeper PL in the
thermal and SPL states is mostly non-thermal in nature, so non-thermal Comptonization is required. Detailed
comparisons with data suggest thermal Comptonization also cannot be ignored even in the thermal and SPL states.

In Figure~\ref{4states}, the hot inner flow (HIF) and patchy corona are responsible for thermal and
non-thermal Comptonization, respectively. When a source transits from hard to SPL state, $r_{\rm in}$ decreases,
HIF is reduced but the patchy corona becomes dominant. The transition from SPL to soft state is then marked by
the disappearance of HIF and significant reduction of the patchy corona. The above scenario obviously depends on
two fundamental assumptions: (1) slab-HIF produces the non-thermal PL component; (2) slab-HIF is mostly located
between the TID and the BH. Both assumptions are examined exhaustively based on the existing observations and
their spectral modeling. Slab-HIF is found to be consistent with essentially all data. An alternative to
slab-HIF is that the PL is produced from the jet base and beamed away from the disk; however observations
suggest the PL emission is quite isotropic, thus in conflict with this. TID is found also consistent with data
when the PL component becomes important, since the observed $L_{\rm disk}\sim T_{\rm in}^4$ deviates
significantly from a linear relation, suggesting $r_{\rm in}$ is larger when the PL component becomes important.
However some studies showed that $r_{\rm in}$ is unchanged in the initial hard state, if the Compton
up-scattering process is treated properly in Monte-Carlo simulations to recover the lost disk photons
\cite{Yao2005}; this issue will be further discussed in Section~\ref{BH_spin}.

The above developed HIF/TID model based on the spectral evolutions of BHXBs can also be applied to explain the
majority of time variability PDS of BHXBs consistently. The TID acts as a low pass filter as it cannot response
effectively to variations in the HIF, so $r_{\rm in}$ controls the low frequency break of the PDS: $\nu_{\rm
LFB}\sim 0.2 (r/6)^{-3/2} (m/10)^{-1}$~Hz, where $r=r_{\rm in}/r_{\rm g}$ and $m=M_{\rm BH}/M_\odot$; this
predicts that $\nu_{\rm LFB}$ changes from 0.03 to 0.2~Hz as observed during transitions from the hard to SPL
and soft states if $r$ decreases from 20 to 6. The HIF/TID model can also explain the observed LFQPO variations
with luminosity, if the LFQPOs are some kinds of characteristic frequencies related to $r_{\rm in}$. To some
extend, the shape of the PDS can also be explained by this model. However the observed variability rms $\sim$
flux linear correlation and log-normal distribution of fluctuations may need additional ingredient, such as the
proposed propagation fluctuation model. However I point out that such correlations and distributions are also
observed in the light curves of gamma-ray bursts and solar flares \cite{Zhang2007b,Wang2008}, which can be
produced by the generic self-organized criticality mechanism \cite{Zhang2007b}. Similar rms $\sim$ flux
correlation has also been found for blazars, in agreement with the minijets-in-a-jet statistical model
\cite{Biteau2012}.

The spectral and timing properties of weakly magnetized NSXBs are known to have many similarities and
differences from BHXBs; this is particularly true for the atoll sources that have similar ranges of $L/L_{\rm
Edd}$ to BHXBs. The essential distinction is that a NS has a solid surface, but a BH does not. Observationally
the continuum spectra of atoll sources can be modeled as composed of SSD, PL emission, and blackbody emission
from the NS surface (or the boundary layer between the TID and NS surface); the former two components are quite
similar to BHXBs. The spectral evolutions of these NSXBs are thus driven similarly by the combinations of $\tau$
and ${\cal L}_{\rm h}/{\cal L}_{\rm s}$. However the blackbody emission is an additional source of ${\cal
L}_{\rm s}$, so the PL is not as hard as that in BHXBs and $r_{\rm in}$ variations are less effective in
changing ${\cal L}_{\rm h}/{\cal L}_{\rm s}$; the latter means it is more difficult to find evidence of TID from
modeling only the spectral evolutions in NSXBs. On the other hand, the TID/HIF model in NSXBs can produce timing
behaviors in the same way as in BHXBs discussed above, consistent with observations; additional timing
behaviors, such as coherent X-ray pulsations and kilo-Hz QPOs observed in these NSXBs, are caused by the rapid
spins of the hard surfaces of the NSs.

\section{Further developments on BH spin measurements}\label{BH_spin}

A BH predicted in GR can only possess three parameters, namely, mass, spin and electric charge, known as the
so-called BH no hair theorem. Even if a BH was born with net electric charge, its electric charge can be rapidly
neutralized by attracting the opposite charge around it in any astrophysical setting, because the strength of
electromagnetic interaction is many orders of magnitude stronger than that of gravitational interaction.
Therefore an astrophysical BH may only have two measurable properties, namely, mass and spin, making BHs the
simplest macroscopic objects in the universe. Practically, only Newtonian gravity is needed in measuring the BH
mass in a binary system. However, GR is needed in measuring the BH spin.

\begin{figure}
\center{
\includegraphics[angle=0,scale=0.4]{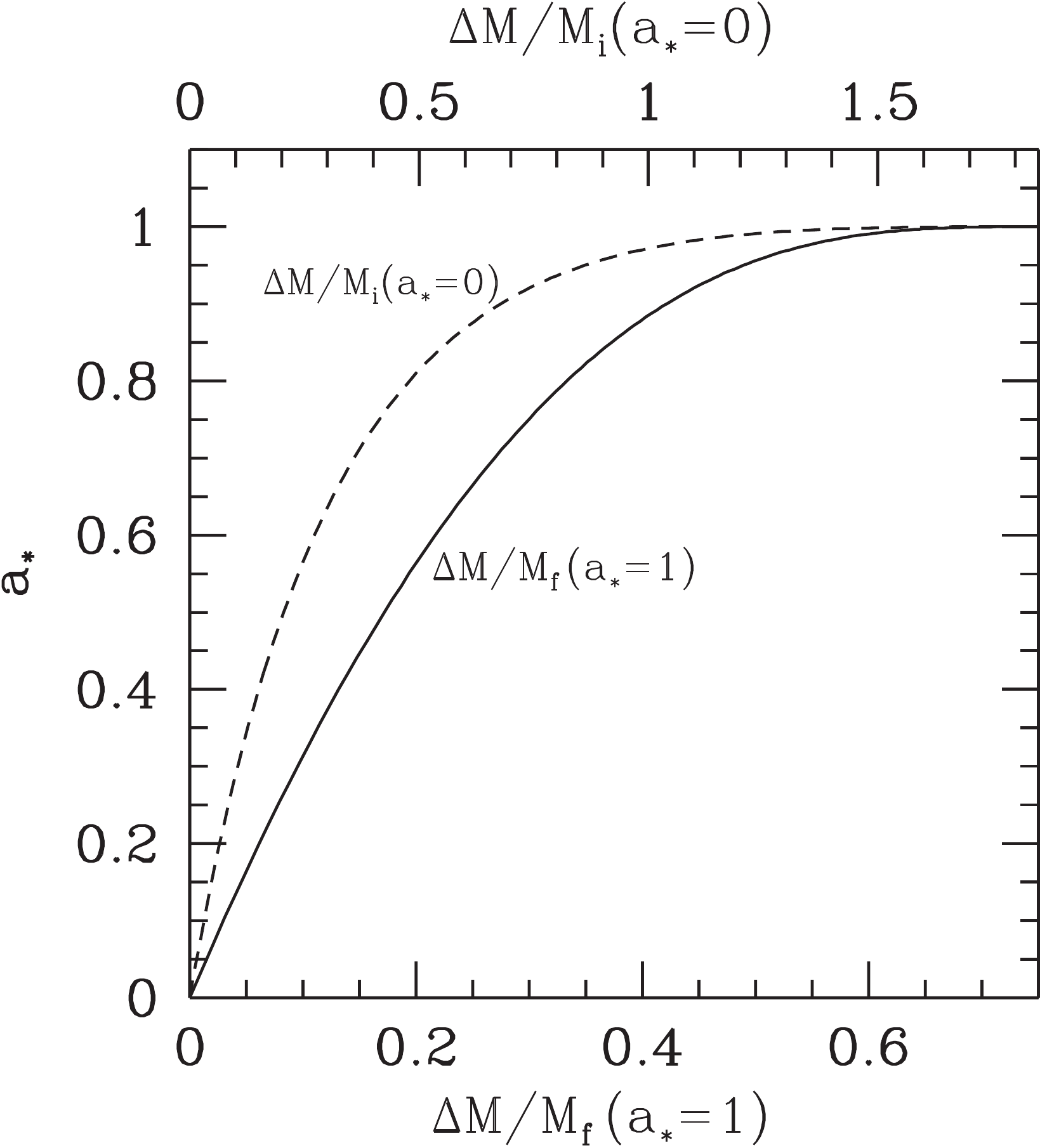}}
\caption{BH spin $a_*$ vs.\ accreted rest-mass $\Delta M$, in units the final mass $M_{\rm f} (a_*=1)$ (solid
line, bottom axis), and in units of the initial mass $M_{\rm i}(a_*=0)$ (dashed, top axis). (Adapted slightly
from Figure~3 Ref.\cite{King1999}.} \label{delta_M_f}
\end{figure}

The mass and spin of a BH has different astrophysical meanings. Its mass can be used to address the question of
``{\it How much} matter (and energy) has plunged into the BH?". However its spin can be used to address the
question of ``{\it How did} the matter (and energy) plunge into the BH?". This is because matter and energy
plunged into a BH can carry angular momentum, which is a vector with respect to the spin axis of the BH. In
order to increase the total gravitating mass-energy from $M_{\rm i}$ with zero spin to $M_{\rm f}$, the added
rest-mass must be \cite{Bardeen1970,Thorne1974}
\begin{equation}
 \Delta M
= 3M_{\rm i}[\sin^{-1}(M_{\rm f}/3M_{\rm i}) - \sin^{-1}(1/3)], \label{delta_M}
\end{equation}
and its final spin becomes
\begin{equation}
 a_* \equiv cJ/GM_{\rm f}^2 = \biggl({2\over 3}\biggr)^{1/2}{M_i\over M_{\rm f}}
  \biggl[4 - \biggl({18M_{\rm i}^2\over M_{\rm f}^2} - 2\biggr)^{1/2}\biggr],
\label{a*}
\end{equation}
where $J$ is the BH's angular momentum. Clearly we have $a_*=1$ when $M_{\rm f}/M_{\rm i} = 6^{1/2}$; further
accretion simply maintains this state \cite{Thorne1974}. Therefore the required additional rest-mass to spin a
BH from zero to maximum spin is $\Delta M \simeq 1.85 M_{\rm i} = 0.75 M_{\rm f}$; this is a lower limit to the
accreted mass \cite{King1999}. Figure~\ref{delta_M_f} shows $a_*$ as a function of $\Delta M$. Ignoring Hawking
radiation of a macroscopic BH, the only way to extract the energy of a BH and thus reducing its gravitating mass
is by extracting its spin energy. Recently, evidence of BH spin energy extraction to power relativistic jets has
been found, from the observed correlation between the maximum radio luminosity and its BH spin of a microquasar
\cite{Narayan2012,Steiner2013}; however the average jet power is not correlated with BH spin \cite{Fender2010}.
This indicates that jets may be produced by both Blandford-Payne (BP) \cite{Blandford1982} and Blandford-Znajek
(BZ) \cite{Blandford1977} mechanisms; but the BZ mechanism is more powerful and responsible for producing the
peak radio luminosity. This provides another possible way to estimate a BH's spin \cite{Steiner2013}, similar
to a recent proposal of using the peak luminosity of the disk emission to estimate a BH's spin \cite{Xue2011}.
However there is so far no independent demonstration of validness of either of the above two new methods, which
are thus not discussed further in this section.

\begin{figure}
\center{
\includegraphics[angle=0,scale=0.35]{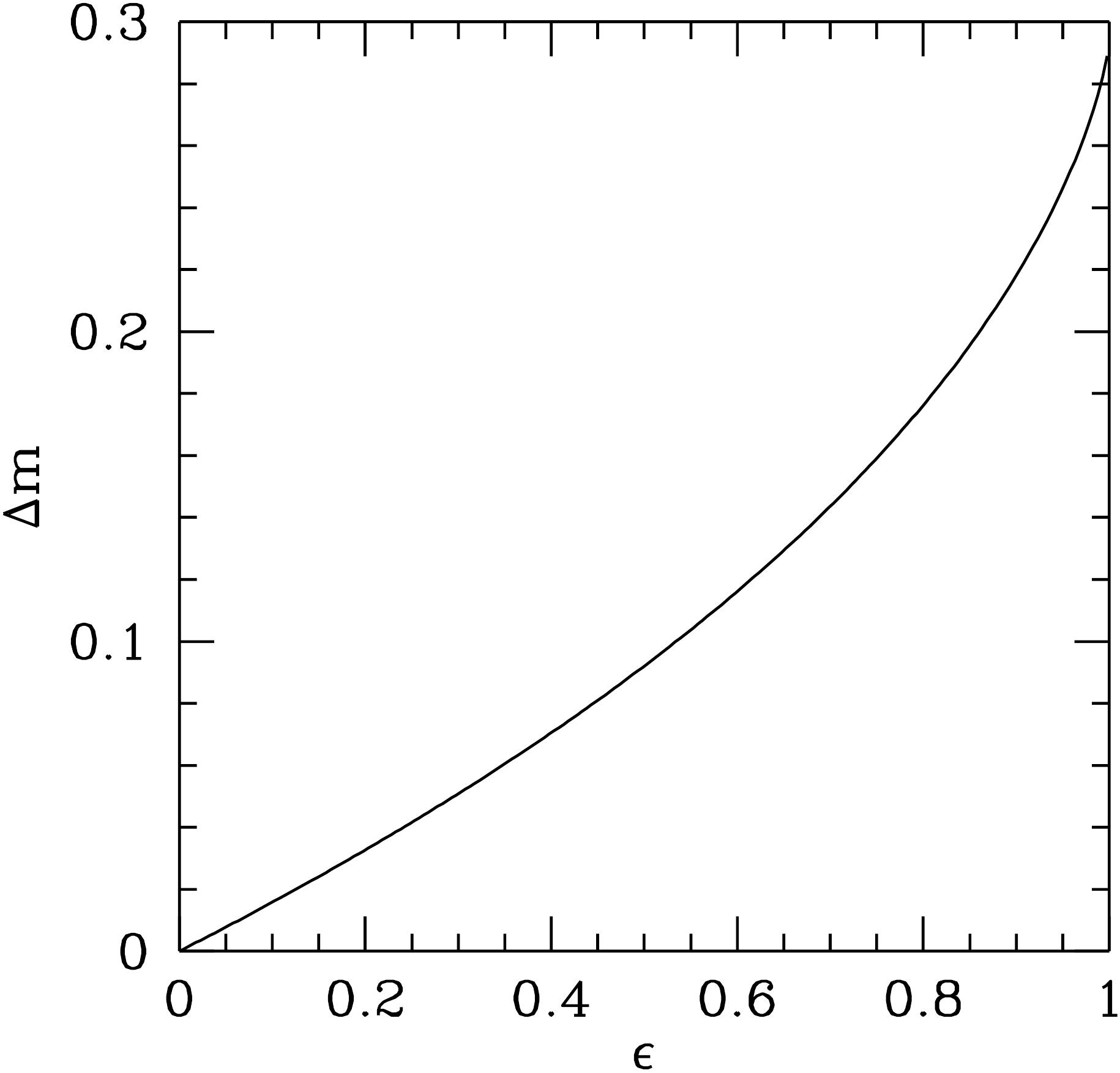}}
\caption{Extracted fractional energy $\Delta m=(M_{\rm max} - M_{\rm f})/M_{\rm max}$ as a function of
efficiency $\epsilon$ of extraction of a BH's rotational energy. (Figure~4 in Ref.\cite{King1999}.}
\label{extraction}
\end{figure}

Figure~\ref{extraction} shows the mass reduction, $\Delta m=(M_{\rm max} - M_{\rm f})/M_{\rm max}$ ($M_{\rm
max}$ and $M_{\rm f}$ are the BH's maximum mass and final mass, respectively), as a function of extracting
efficiency $\epsilon$ of an extreme Kerr BH. For a $10M_{\odot}$ BH with $\epsilon=1$ ($\Delta m\sim 0.3$), the
total extracted energy is $\Delta m M_{\rm f} c^2 \simeq 10^{54}$ erg and its total gravitating mass is reduced
to $M_{\rm max}/\sqrt{2}$ \cite{King1999}; this energy could be sufficient to power gamma-ray bursts (GRBs). It
is thus plausible that supercritical accretion onto a newly born BH may spin it up and extract its spin energy
to power ultra-relativistic jets; multiple spin-up and spin-down cycles may also happen during one GRB, if the
collapsing material is clumpy.

\begin{figure}
\center{
\includegraphics[angle=0,scale=0.4]{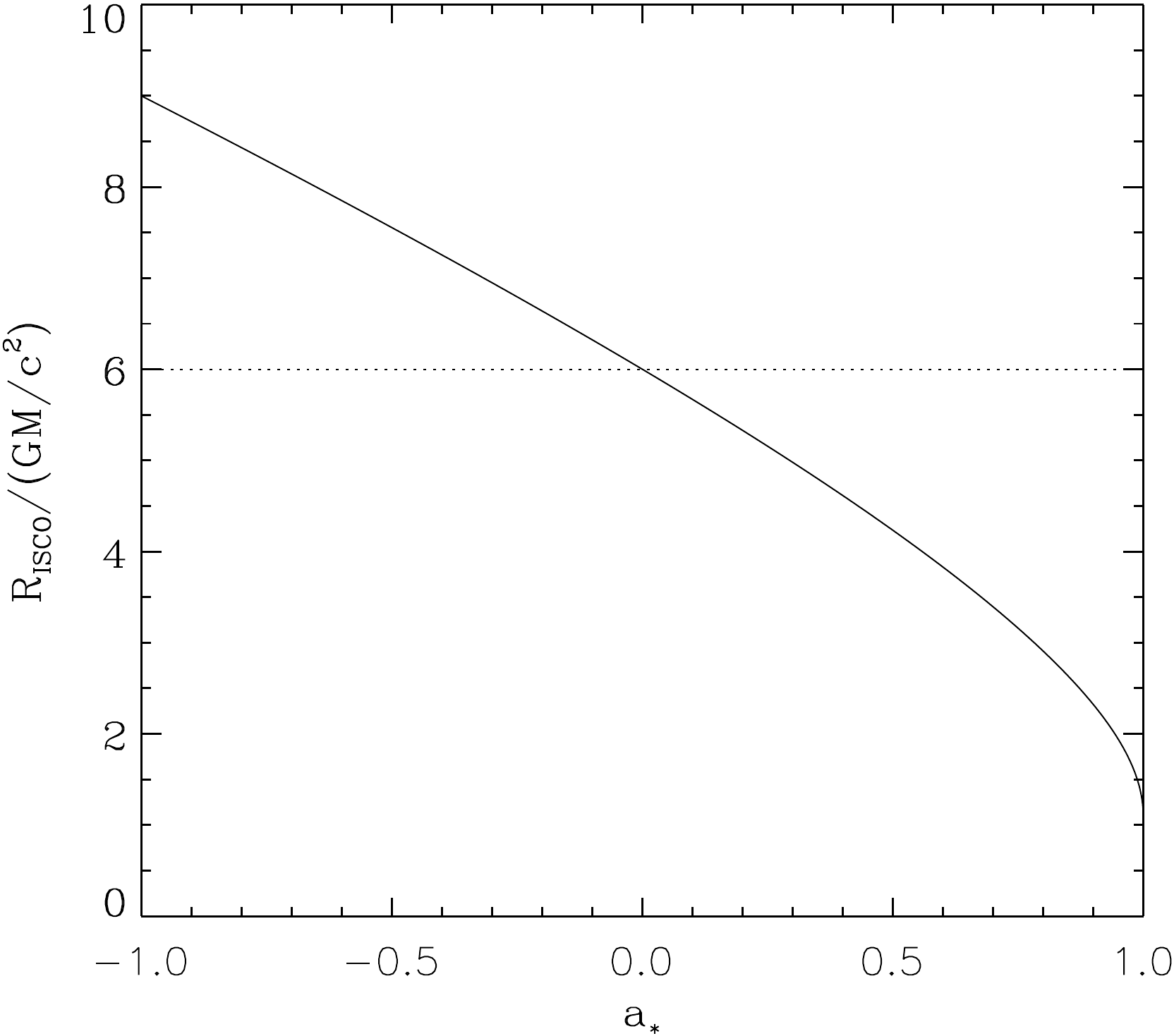}}
\caption{The radius of the innermost stable circular orbit ($R_{\rm ISCO}$) of a BH as a function of the spin
parameter ($a_*$) of the BH, i.e., the dimensionless angular momentum; a negative value of $a_*$ represents the
case that the angular momentum of the disk is opposite to that of the BH, i.e., the disk is in a retrograde
mode. Therefore $a_*$ can be measured by determining the inner accretion disk radius, if the inner boundary of
the disk is the ISCO of the BH.} \label{r_isco}
\end{figure}

\begin{figure}
\center{
\includegraphics[angle=0,scale=0.4]{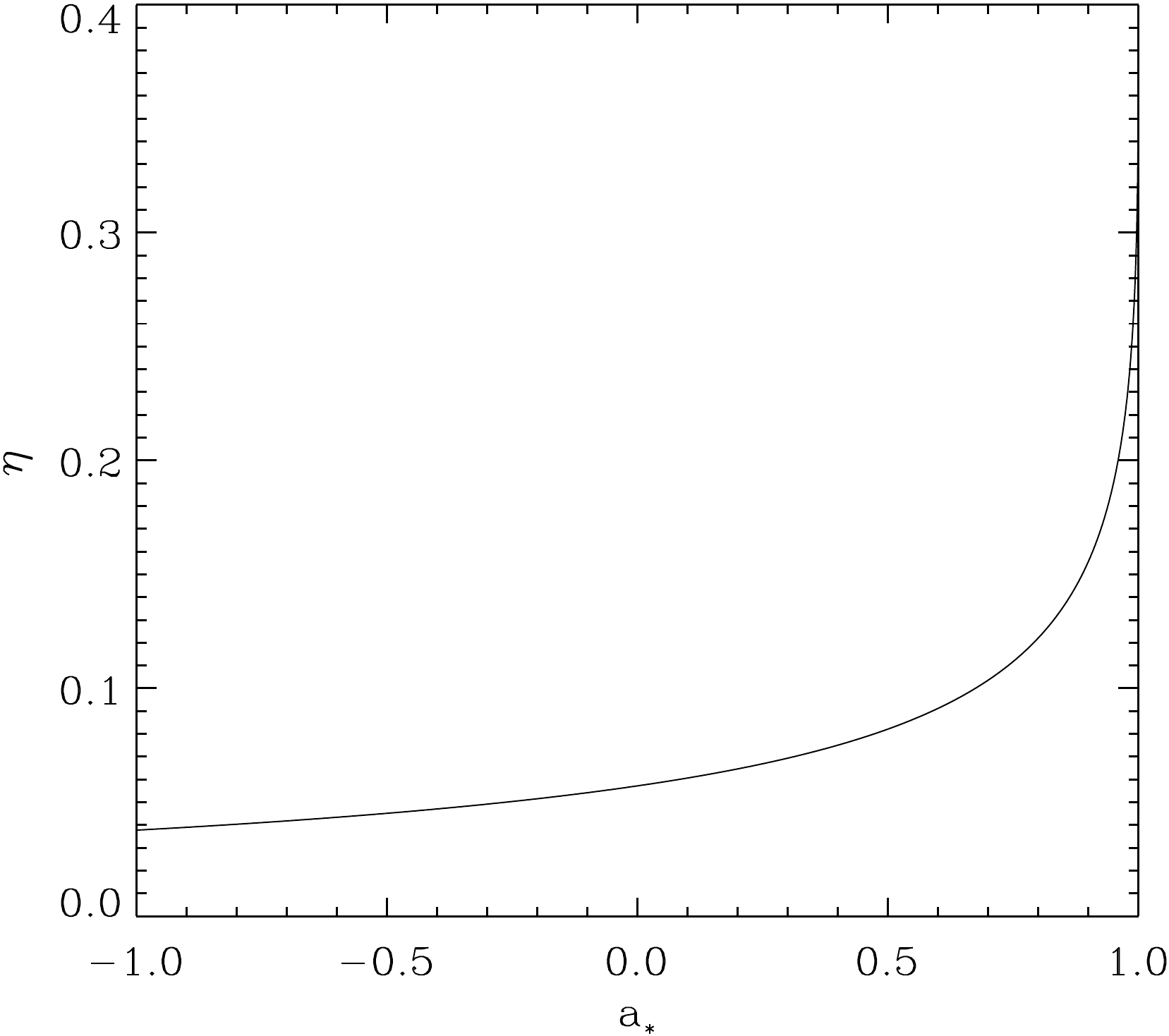}}
\caption{Radiative efficiency $\eta$ as a function of BH spin parameter $a_*$.} \label{eta}
\end{figure}

\begin{figure}
\center{
\includegraphics[angle=0,scale=0.4]{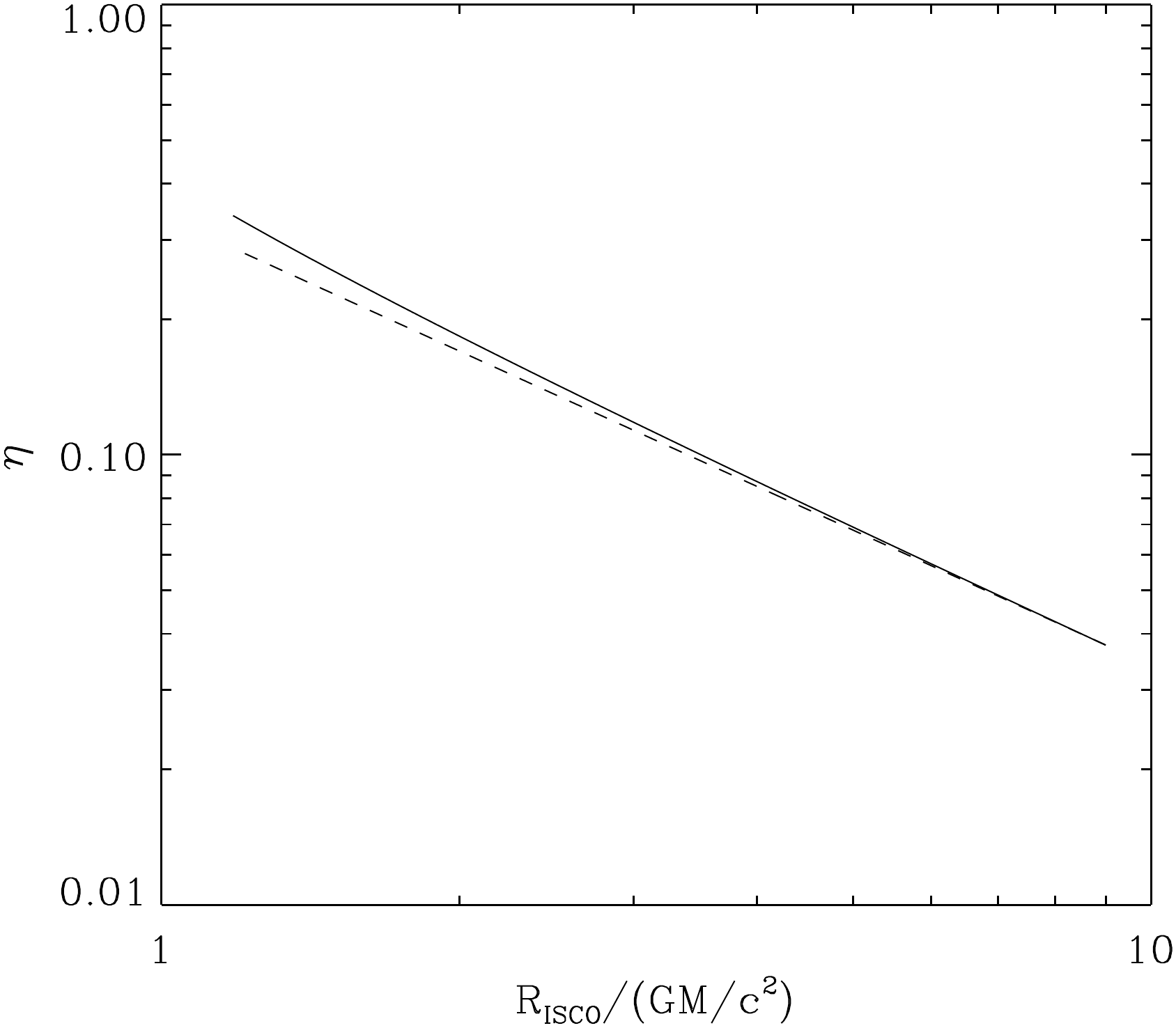}}
\caption{The solid line shows numerical curve of the radiative efficiency $\eta$ as a function of $R_{\rm
ISCO}$. The dashed line shows $\eta\sim 1/R_{\rm ISCO}$, which approximates the numerical curve.} \label{eta_r}
\end{figure}

As shown in Figure~\ref{r_isco}, the radius of the ISCO of the BH, $R_{\rm ISCO}$, is a monotonic function of
the BH spin \cite{Bardeen1972}, beyond which radius a test particle will plunge into the BH under any
perturbation; however, in Newtonian gravity a stable circular orbit can be found at any radius. It is thus
reasonable to assume that the accretion disk around a BH terminates at this radius, i.e., $r_{\rm in}=R_{\rm
ISCO}$. Therefore, $a_*$ can be inferred if one can measure $R_{\rm ISCO}$ in units of its gravitational radius
$r_{\rm g}=GM/c^2$. Currently three methods have been proposed to measure the BH spin in BHXBs, and all these
methods rely essentially on measuring $R_{\rm ISCO}$. In case the radiative efficiency ($\eta\equiv
L/\dot{M}c^2$) can be measured, $a_*$ can also be determined this way, as shown in Figure~\ref{eta}. Actually
$\eta$ is a very simple function of $R_{\rm ISCO}$, i.e, $\eta\sim 1/R_{\rm ISCO}$, as shown in
Figure~\ref{eta_r}.

\subsection{The first black hole spin measurement with X-ray spectral continuum fitting}

In 1997, I and my colleagues proposed a method of measuring the BH spin in BHXBs. It started when we tried to
measure the mass of the BH in GRO J1655--40 \cite{Zhanga} by measuring $r_{\rm in}$ from its X-ray continuum
spectral fitting and assuming the BH is not spinning; we found the mass of the BH is around $4M_{\odot}$
\cite{Zhang1997a}. Coincidentally just when this paper was going to press, Orosz and Bailyn \cite{Orosz1997}
announced their accurate measurement of the BH mass of GRO J1655-40, which is significantly larger than what we
found. A co-author of our paper, Rashid Sunyaev urged me to resolve this apparent discrepancy. I immediately
realized that a spinning BH of $7M_{\odot}$ in GRO J1655-40 would be consistent with the inferred with X-ray
data, and added a note in proof in the paper suggesting this possibility \cite{Zhang1997a}. This turns out to be
the first BH spin measurement on record. I then invited two of my close friends and collaborators, Wei Cui and
Wan Chen, to join me to apply this method systematically to other BHXBs; a new main conclusion in this work was
that the first microquasar in the Milky Way, GRS 1915+105, also contains a spinning BH \cite{Zhang1997}.

At the time {\sc DISKBB} was the only available fitting model in the XSPEC package for determining $r_{\rm in}$
from an observed X-ray continuum spectra, if the disk continuum is described by the SSD model: $L_{\rm
disk}=4\pi\sigma r_{\rm in}^2 T_{\rm in}^2$, where $T_{\rm in}$ in {\sc DISKBB} is the disk temperature at
$r_{\rm in}$ (i.e. the peak disk temperature in SSD) and $L_{\rm disk}$ can be calculated from the disk flux
(after the correction to absorption), the distance to the source and disk inclination (an issue to be discussed
later). To determine the physical inner disk radius from the {\sc DISKBB} parameter $r_{\rm in}$, several
effects must be considered: (1) electron scattering in the disk modifies the observed X-ray spectrum; (2) the
temperature distribution in the disk is not accurately described by the Newtonian gravity as assumed in SSD; (3)
the observed temperature distribution is different from the locally emitted one; (4) the observed flux is
different from the locally emitted one. The latter three are all due to GR effects \cite{Page1974}. For each of
the above effects, we introduced a correction factor, using the best available knowledge at the time. Since
then, several improvements have been made to correct for these effects and this continuum fitting (CF) method is
now quite mature in making accurate BH spin measurements, given sufficiently high quality X-ray continuum
spectral measurements and accurate system parameters of the observed BHXB.

\subsection{Further developments and applications of the continuum fitting method}
This CF method of measuring BH spin has since been applied widely to essentially every BHXB with a well measured
X-ray continuum spectrum showing a prominent thermal accretion disk component. In particular, this method has
been improved and incorporated into the widely used X-ray spectral fitting package XSPEC, e.g., {\sc KERRBB}
\cite{Li2005}, {\sc BHSPEC} \cite{Davis2006a}, and {\sc KERRBB2} \cite{McClintock2006,Gou2011a}. Both the {\sc
KERRBB} and {\sc BHSPEC} are relativistic models, but they have their own drawbacks and advantages (See
Ref.\cite{McClintock2006} for a detailed comparison). {\sc KERRBB} includes all the relativistic effects, but it
requires to fix the spectral hardening factor. In contrast, {\sc BHSPEC} could calculate the spectral hardening
factor on its own; however, it does not include the returning radiation effect, which turns out to be an
important factor in BH spin determination in BHXBs. {\sc KERRBB2} combines both models by generating the
spectral hardening factor table from {\sc BHSPEC} and using the table as the input for {\sc KERRBB}. The
research group led by Ramesh Narayan of Harvard University, Jeffrey McClintock at Smithsonian Astronomical
Observatory (SAO), and Ronald Remillard of Massachusetts Institute of Technology (MIT) \cite{McClintock2009} has
since applied this method and contributed to most of the BH spin measurements available in the community, as
shown in Table~\ref{bhxbs_table}.

The CF method relies on two fundamental assumptions: (1) The measured $r_{\rm in}$ is uniquely related to
$R_{\rm ISCO}$ of the BH. (2) There is no or negligible X-ray radiation from the plunging matter onto the BH
beyond $R_{\rm ISCO}$. The latter has been studied with numerical simulations that include the full physics of
the magnetized flow, which predict that a small fraction of the disk¡¯s total luminosity emanates from the
plunging region \cite{Penna2010}. However, in the context of BH spin estimation, it has been found that the
neglected inner light in the CF method only has a modest effect, i.e., this bias is less than typical
observational systematic errors \cite{Kulkarni2011,Zhu2012}.

\begin{figure}
\center{
\includegraphics[angle=0,scale=0.4]{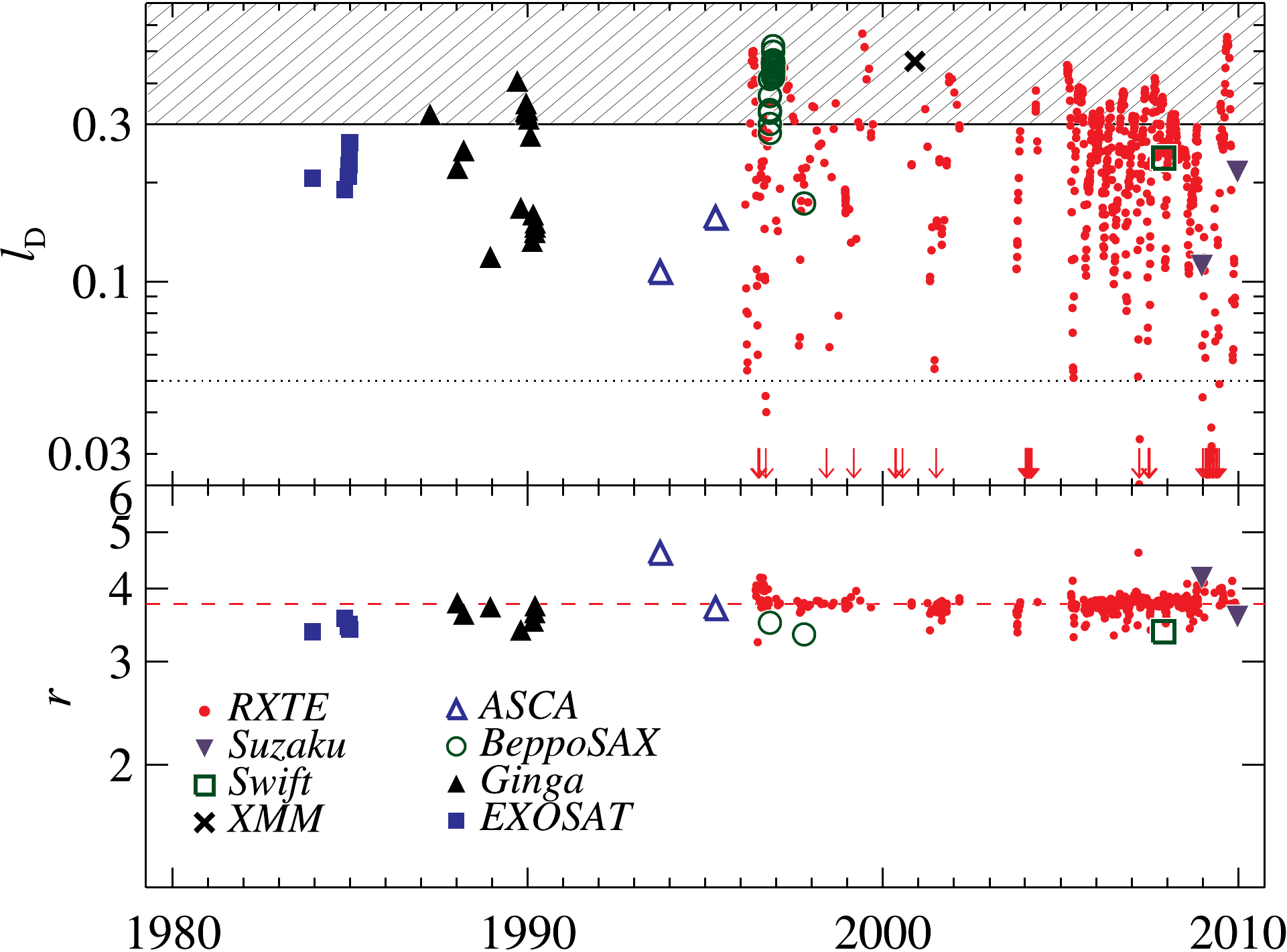}}
\caption{The measured $r=r_{\rm in}/r_{\rm g}$ and disk luminosity $l_{\rm D}=L_{\rm disk}/L_{\rm Edd}$
(assuming $M_{\rm BH}=10M_{\odot}$) as functions of observation time, for the BHXB LMC X--3. (Slightly adapted
from Figure~1 in Ref.\cite{Steiner2010})} \label{lmcx3_1}
\end{figure}

The first assumption above requires that the measured $r_{\rm in}$ remains stable as a BHXB changes its spectral
state and luminosity. However it was noticed that $r_{\rm in}$ measured is usually much smaller, sometimes even
smaller than $R_{\rm ISCO}$ of a extreme Kerr BH in a prograde orbit, when the X-ray spectrum contains a
significant hard PL component, which is believed to be produced by inverse Compton scattering of the thermal
disk photons in a hot corona. We realized that the inferred smaller $r_{\rm in}$ could be due to the lost
thermal disk photons in the scattering process. We then investigated this problem and confirmed that the
inferred $r_{\rm in}$ can be made consistent with $r_{\rm in}$ inferred from the thermal disk component
dominated spectrum, if the scattered photons are recovered properly by doing detailed radiative transfer in the
corona \cite{Yao2005}; the same conclusion was also reached by the Harvard/SAO/MIT group independently without
knowing our much earlier results \cite{Steiner2009}. Therefore the method of BH spin measurement by X-ray
continuum fitting can also be applied to some SPL state with strong PL component. The stable nature of the
measured $r_{\rm in}$ is proven with the textbook case of LMC X--3, when its X-ray luminosity varied over more
than one order of magnitude observed in nearly two decades with many different X-ray instruments, as shown in
Figure~\ref{lmcx3_1} \cite{Steiner2010}.

\begin{figure}
\center{
\includegraphics[angle=0,scale=0.5]{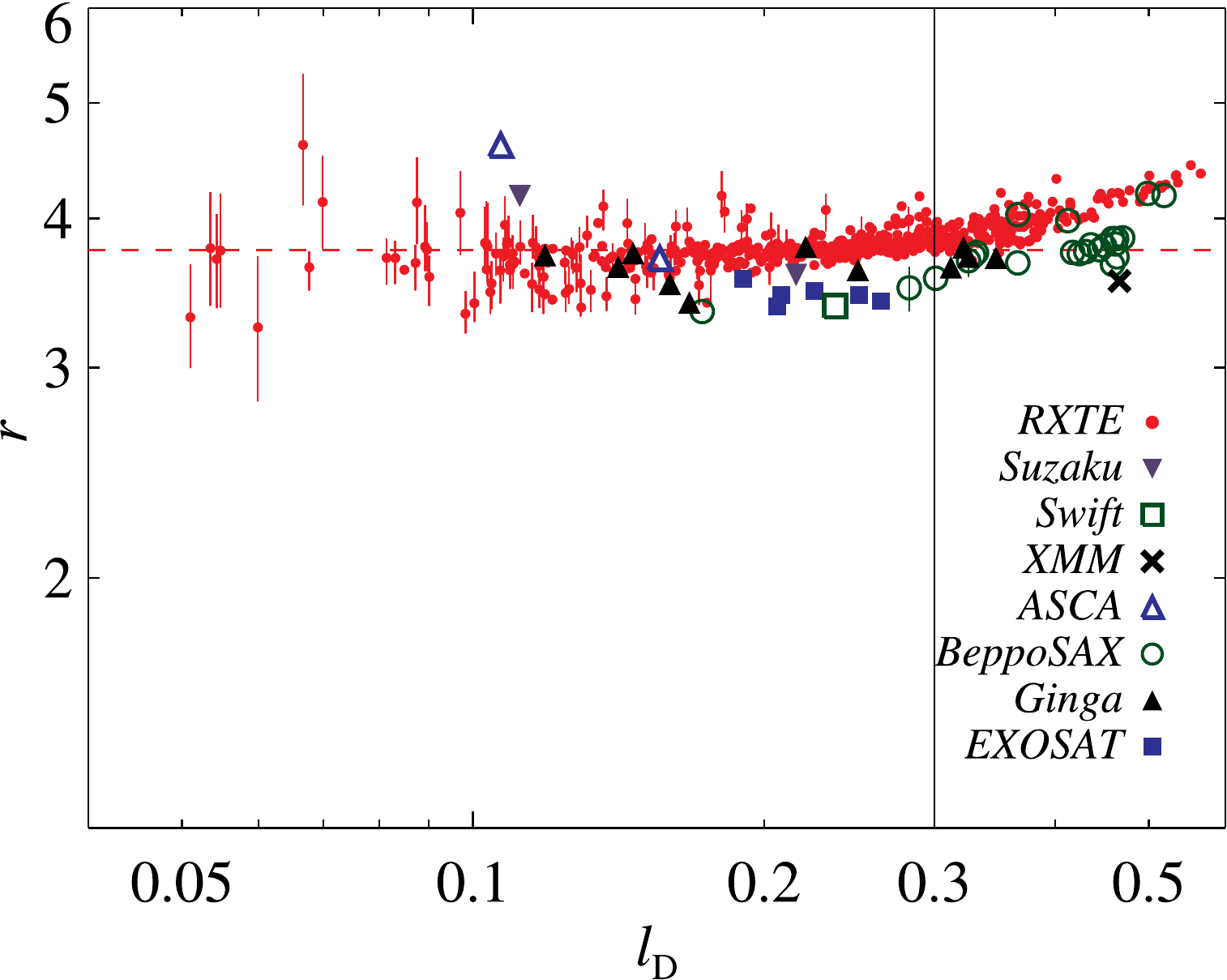}}
\caption{The measured $r=r_{\rm in}/r_{\rm g}$ as a function of disk luminosity $l_{\rm D}=L_{\rm disk}/L_{\rm
Edd}$ (assuming $M_{\rm BH}=10M_{\odot}$), for the BHXB LMC X--3. (Slightly adapted from Figure~2 in
Ref.\cite{Steiner2010})} \label{lmcx3_2}
\end{figure}

\begin{figure}
\center{
\includegraphics[angle=0,scale=0.45]{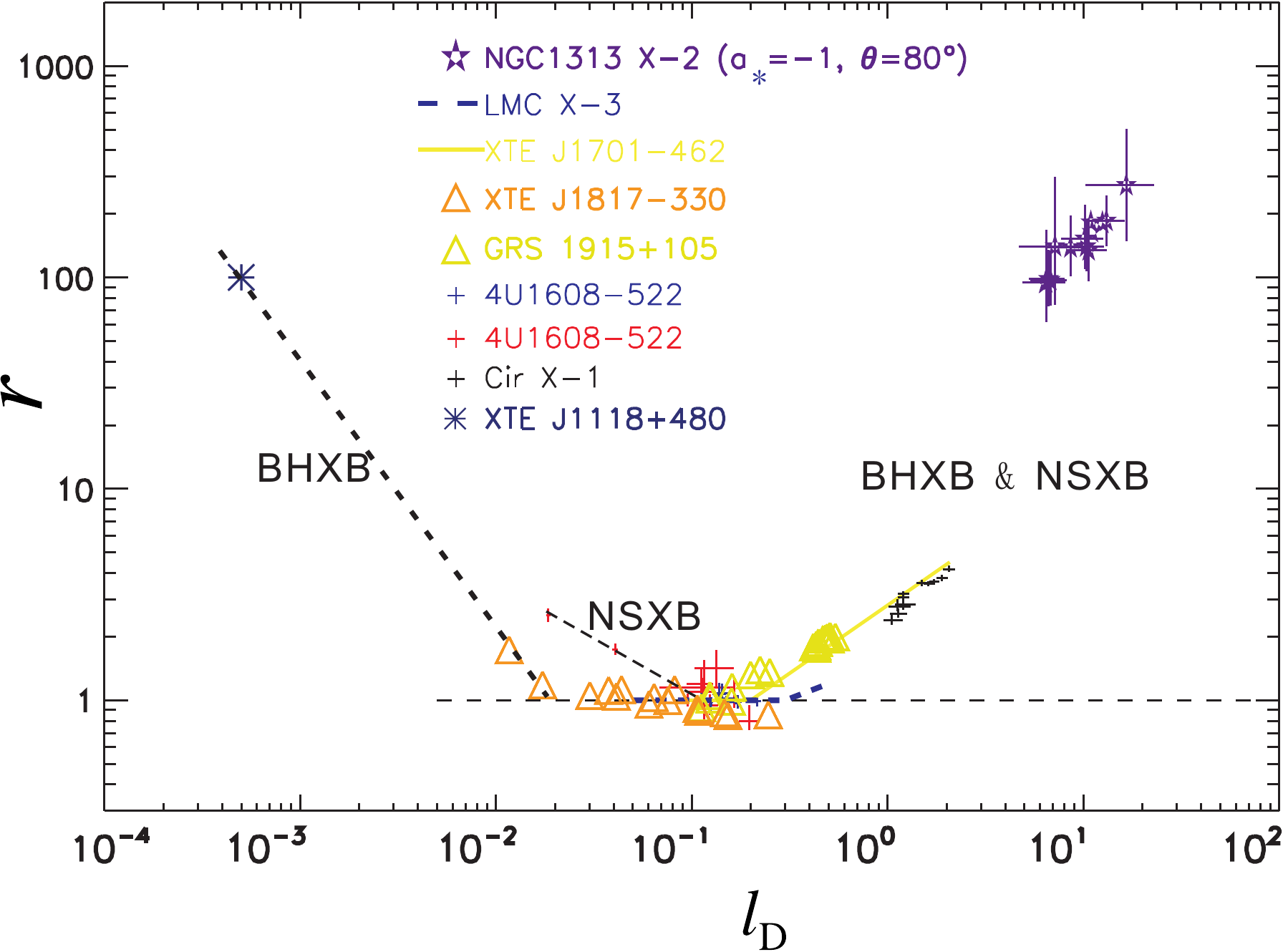}}
\caption{The measured $r=r_{\rm in}/r_{\rm g}$ as a function of disk luminosity $l_{\rm D}=L_{\rm disk}/L_{\rm
Edd}$, for several NSXBs and BHXBs, and the super-Eddington accreting ultra-luminous X-ray source NGC1313~X--2.
All data points are taken from Ref.\cite{Weng2011}, except for that on XTE~J1118+480 \cite{Yuan2005} and
NGC1313~X--2 \cite{Weng2012}.} \label{ngc1313}
\end{figure}

However, when ordered by the observed disk luminosity $l_{\rm D}=L_{\rm disk}/L_{\rm Edd}$, the measured $r_{\rm
in}$ shows a clear increasing trend when $l_{\rm D}> 0.3$, as shown in Figure~\ref{lmcx3_2} \cite{Steiner2010}.
The similar trend has also been found in another BHXB GRS~1915+105  \cite{McClintock2006} and two NSXBs
\cite{Lin2009,Weng2011}. It was found that $r=r_{\rm in}/r_{\rm g}$ indeed increases physically when $l_{\rm D}>
0.3$, by comparing the evolution of $r$ as a function of $l_{\rm D}$ over a large range for several BHXBs and
NSXBs. Using the blackbody surface emissions of the NSs in these NSXBs is critical in evaluating any possible
disk thickening due to high luminosity that would block at least part of the NS surface emission, as well as
determining the actual NS mass accretion rate, which turns out to be much less than the disk mass accretion
rate; this suggests that the increased radiation pressure is responsible for the increase of $r$ and significant
outflow when $l_{\rm D}> 0.3$ \cite{Weng2011}. The same trend is much more pronounced in the super-Eddington
accreting ultra-luminous X-ray source NGC1313 X--2, as shown in Figure~\ref{ngc1313} together with the data from
other BHXBs and NSXBs \cite{Weng2012}. However the exact value of $r$ obtained this way should be taken with
caution, since the non-negligible energy advection at high accretion rate can modify the disk structure in
non-trivial ways, thus making the SSD prescription inaccurate in this case \cite{Gu2012}.

\begin{figure}
\center{
\includegraphics[angle=0,scale=0.45]{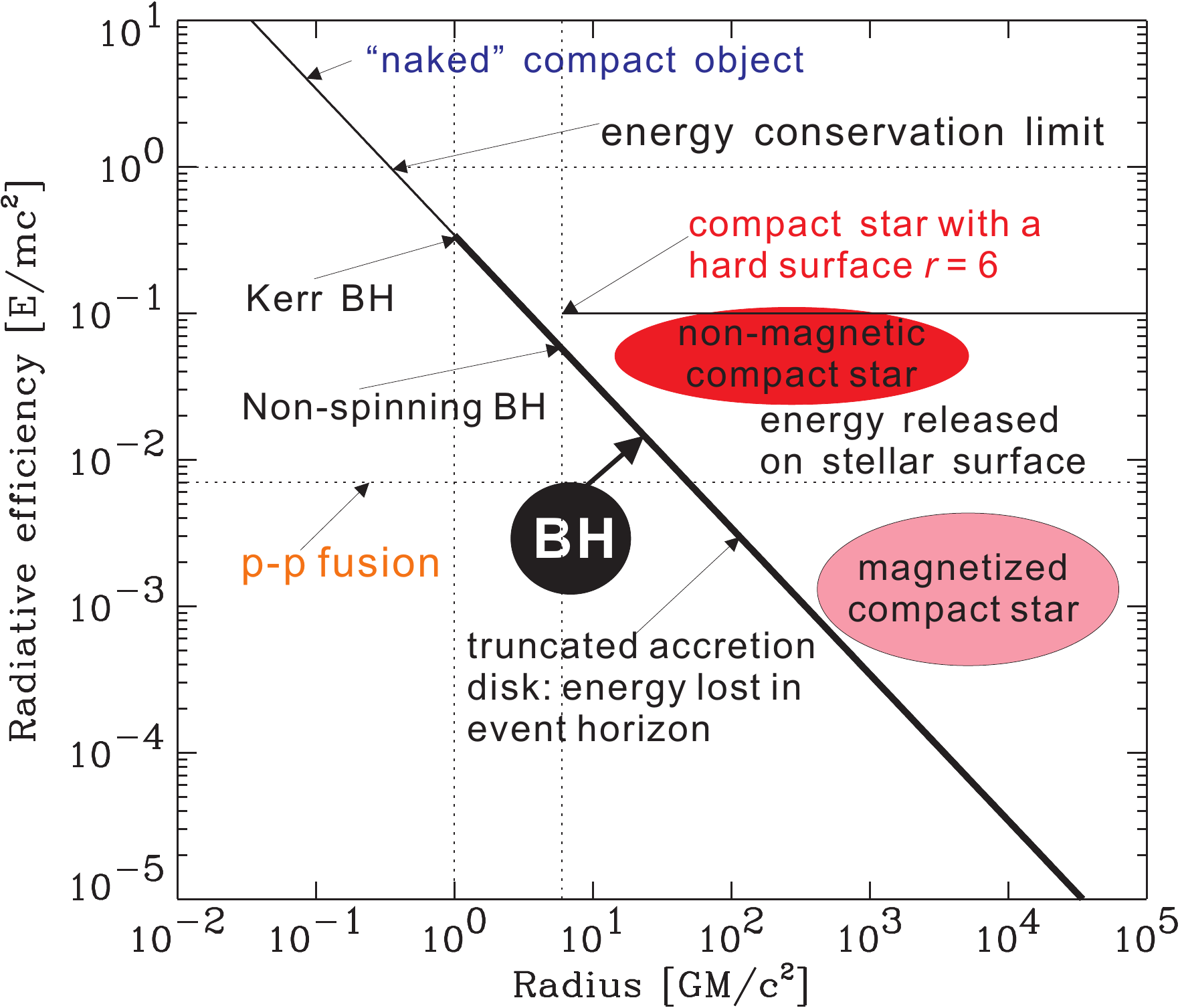}}
\caption{The diagonal line shows a $1/r$ scaling, calibrated to take the value of 0.057 when r = 6. The thick
black line is for a BH accreting systems. The range of  $r = 1-9$ corresponds to the ISCO of a BH with different
$a_*$; the radiative efficiency ranges between a few to several tens of percents, far exceeding the p-p fusion
radiative efficiency taking place in the Sun. The case for $r > 9$ corresponds to TID, whose radiative
efficiency can be extremely low, because energy is lost into the event horizon of the BH. The thin solid black
horizontal line is for the 10\% efficiency when matter hits the surface of a neutron star where all
gravitational energy is released as radiation. The thin solid black diagonal line above the point marked for
``Kerr BH" is for a speculated ``naked" compact object \cite{Zhang2011b}, whose surface radius is extremely
small and thus the radiative efficiency can be extremely high. (Figure 10.7 in Ref.\cite{Zhang2011b}}
\label{efficiency}
\end{figure}

Figure~\ref{ngc1313} also shows that as $l_{\rm D}$ decreases, $r$ again starts to increases. However $r$
increases at higher $l_{\rm D}$ and with a different slope for a NSXB than for a BHXB, which can be naturally
explained as due to the ``propeller" effect of the interaction between the NS's magnetosphere and its accretion
disk \cite{Chen2006a} and the ``no-hair" of the BH \cite{Weng2012}. Figure~\ref{efficiency} shows the radiative
efficiencies of various systems; BHXBs may have either higher or lower efficiencies than NSXBs, because a BH has
neither solid surface nor magnetic field \cite{Zhang2011b}.

It is interesting to compare the BH spin results in our first paper \cite{Zhang1997} and the most recent
literature for the same BHXBs as listed in Table~\ref{bhxbs_table}: (1) For GRS~1915+105, A0620--003 and
LMC~X--3, both results are fully consistent; (2) For GRO~J1655--40, the original result points to an extreme
Kerr BH ($a_* \sim 0.93$) \cite{Zhang1997}, somewhat different from the most recent result of mildly spinning BH
($a_* =0.65-0.75$) \cite{Shafee2006}. However, the original result was obtained using the BH mass of
$7M_{\odot}$, about 10\% larger than the currently best estimate that was used to obtain an updated BH spin in
the most recent literature \cite{Shafee2006}. From Figure~\ref{r_isco}, it can be seen that $a_*$ would be
decreased from 0.93 to ~0.87, if the BH mass is decreased by about 10\%; actually $a_* =0.85$ is also allowed in
the new estimate \cite{Shafee2006}; (3) For Cygnus~X--1, our original conclusion that $a_* =0.75$ in the
high/soft state and $a_* =-0.75$ in the low/hard state was based on the assumption that $r_{\rm in}$ decreased
by a factor of two when the source made a transition from its normal low/hard state to the unusual high/soft
state \cite{Zhang1997b}. However the more realistic constraint is that $r_{\rm in}$ changed by more than a
factor of 1.8-3.2 during the state transition \cite{Zhang1997b}, which implies that $a_*$ switched from
$|a_*|>0.85$, fully compatible with the latest result of $a_*>0.95$ in the high/soft state \cite{Gou2011a}.

The above would suggest that a very low temperature disk component exists in its hard state spectrum, if
Cygnus~X--1 indeed harbors a Kerr BH and the accretion disk switches from a retrograde mode to a prograde mode
when it makes the hard-to-soft state transition. For a supermassive BH in the center of a galaxy, it is well
understood that its mass is mostly gained through accretion in its AGN phase \cite{Yu2002}. Therefore random
accretion (between prograde and retrograde modes) tends to make its final BH spin close to zero, regardless its
initial BH spin. For Cygnus~X-1, its current BH mass of around 10-20 $M_{\odot}$ cannot be gained through
post-formation accretion, since its observed average mass accretion rate $\dot{M}\sim 2\times 10^{-9} M_{\odot}$
/yr and the age of its companion is much less that 10$^8$ yr; actually the age of the companion is estimated to
range from 4.6 and 7.8 million years \cite{Wong2012}. This means that its post-formation BH mass growth is much
less than a fraction of its current mass and thus its current BH spin must be quite close to that at birth. Even
if accreting at Eddington rate, to grow its BH spin from 0 and the final mass to be the current observed value,
the timescale is around $3.1\times10^7$ yr and the accreted mass is roughly 7.3 $M_{\odot}$ \cite{Gou2011a}. If
its accretions alternates between prograde and retrograde modes, then its current BH spin should be even closer
to its initial spin than that in low mass BHXBs that only stay in one accretion mode due to roche-lobe
overflows. Therefore the current high BH spin in Cyg~X--1 must be natal; this conclusion is also true for the
other highly spinning BHs in other BHXBs \cite{McClintock2011}.

In summary, BH spin in BHXBs can now be measured reliably with the CF method, when the luminosity of a BHXB is
between $\sim 0.02$ and $\sim 0.3$ in Eddington unit and their system parameters are well-known {\it a prior}.
The observed thermal disk spectrum can be modeled directly to obtain the BH spin with the available {\sc
KERRBB2} model in XSPEC when the X-ray spectrum in dominated by this component, i.e., the source is in the
thermal state. When a significant power-law component is present, the inverse Compton scattering process must be
taken into account to recover the disk photons scattered into this PL component, with for example the {\sc
SIMPL/SIMPLR} model \cite{Steiner2009a,Gou2011a} in XSPEC. Recently another way to measure BH spin using the
outburst properties of BHXBs has been proposed \cite{Xue2011}, which is mentioned briefly in Section~\ref{disk};
however the effectiveness of this method needs to be tested.

\subsection{Uncertainties of the continuum fitting method}

In spite of the tremendous progress made so far on BH spin measurements in BHXBs, most of these BH spin
measurements suffer from considerable uncertainties. Actually the major source of these uncertainties comes
primarily from the uncertainties in their BH masses, accretion disk inclination angles, and distances. The
accurate BH mass measurement is required because $R_{\rm ISCO}$ must be in units of $r_{\rm g}=GM/c^2$ in
Figure~\ref{r_isco}. The disk inclination angle and distance are also required because the total luminosity of
the disk emission $L_{\rm disk}$ is needed, in order to estimate the absolute value of $a_*$.

So far, all BH masses in X-ray binaries (XRBs) have been estimated using the Kepler's 3rd law of stellar motion,
expressed in the so-called the mass function given in Equation~(\ref{equ:mass}). Since the only direct
observables are $P_{\rm orb}$ and $K_{\rm C}$, both $M_{\rm C}$ and $i$ have to be determined indirectly in
order to obtain the BH mass estimate reliably. The companion's mass $M_{\rm C}$ can be determined relatively
reliably by the observed spectral type of the companion star and $i$ can be estimated by modeling the observed
ellipsoidal modulation of the companion's optical or infrared light curve. The observed ellipsoidal modulation
is a consequence of exposing different parts of the pear-shaped companion star to the observer at different
orbital phases (see Figs.~{\ref{BHXB} and {\ref{bhxbs_fig}); the pear-shape is caused by the tidal force of the compact
star, which also heats the side of the companion star facing it. For details of BH mass estimates using this
method, please refer to \cite{Remillard2006}.

However model dependence and other uncertainties (such as accretion disk contamination) cannot be circumvented
completely and thus systematic error may exist in determining their system parameters. For example, three
optical states, namely ``passive", ``loop" and ``active" states, have been identified in the normally called
``quiescent" state of A0620--003 when its X-ray luminosity is very low; only during the passive state its
optical light curve modulation is purely ellipsoidal, i.e., accretion disk contamination is completely
negligible, as shown in Figure~\ref{optical_states} \cite{Cantrell2008}. This means that considerable systematic
errors in determining its inclination angle may occur unless only the ``passive" state data are used.
Unfortunately previous observations of BHXBs used to determine their inclinations did not always occur during
the passive state, thus systematic errors may be common in previous results \cite{Kreidberg2012a}. Even for
GRO~J1655--40, of which all previous observations were made during its passive state \cite{Kreidberg2012a}, its
BH mass measured with different observations, or the same data analyzed by different groups are not exactly the
same, and even not completely consistent between them, as shown in Figure~\ref{groj1655}, which show a scatter
of about 20-30\% to the estimated BH mass, much larger than its statistical error of a few percents.

\begin{figure}
\center{
\includegraphics[angle=0,scale=0.4]{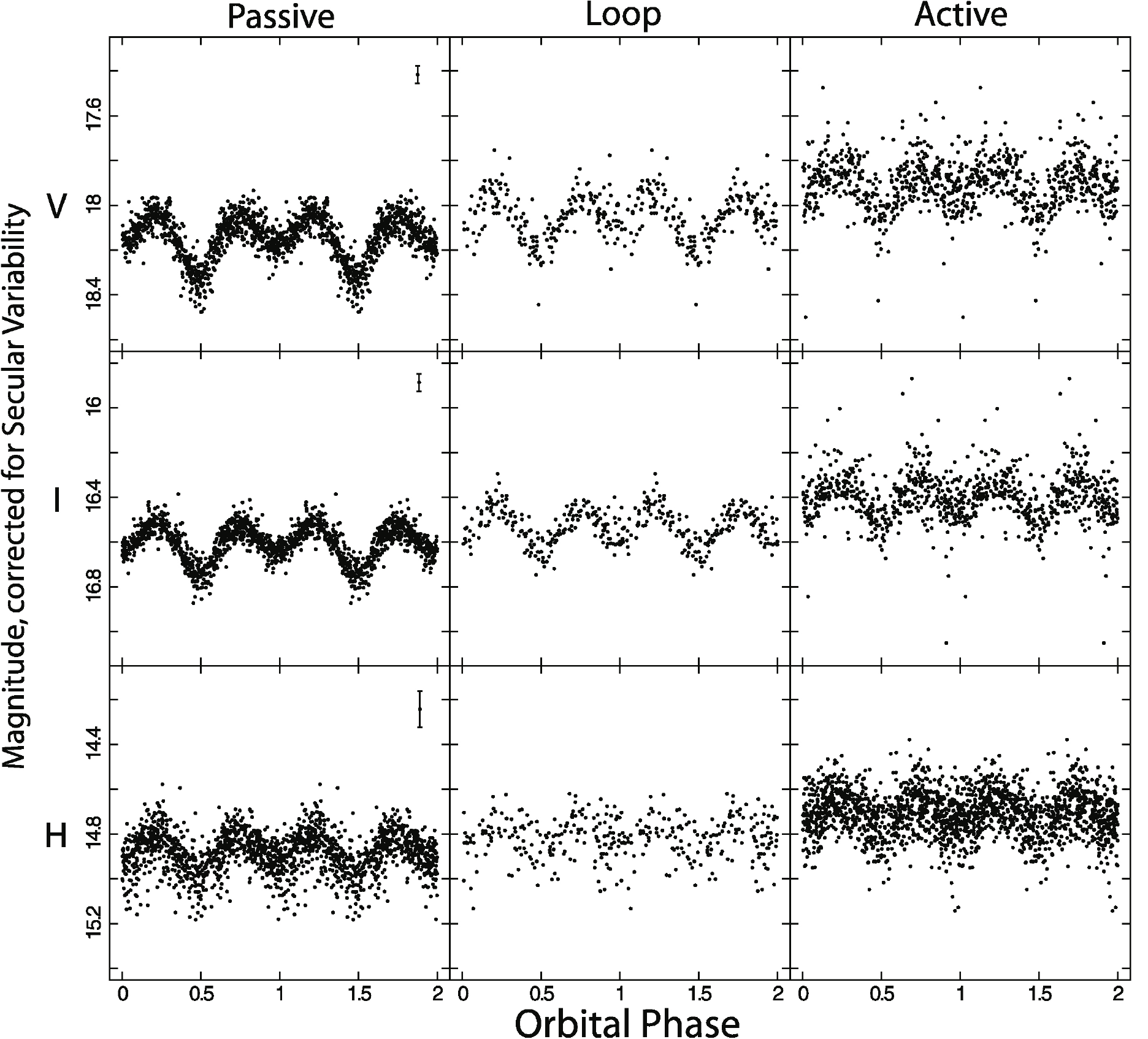}}
\caption{Phased VIH ellipsoidal light curves corresponding to the three states, namely ``passive", ``loop" and
``active" states; the data in each state and each wave band are plotted twice for clarity. Typical differential
photometric error bars are shown in the upper right corner of the passive-state light curve for each band. To
produce purer ellipsoidal light curves, variability on timescales greater than 10 days has been removed from
loop- and active-state data. (Figure~2 in Ref.\cite{Cantrell2008}.)} \label{optical_states}
\end{figure}

\begin{figure}
\center{
\includegraphics[angle=0,scale=0.4]{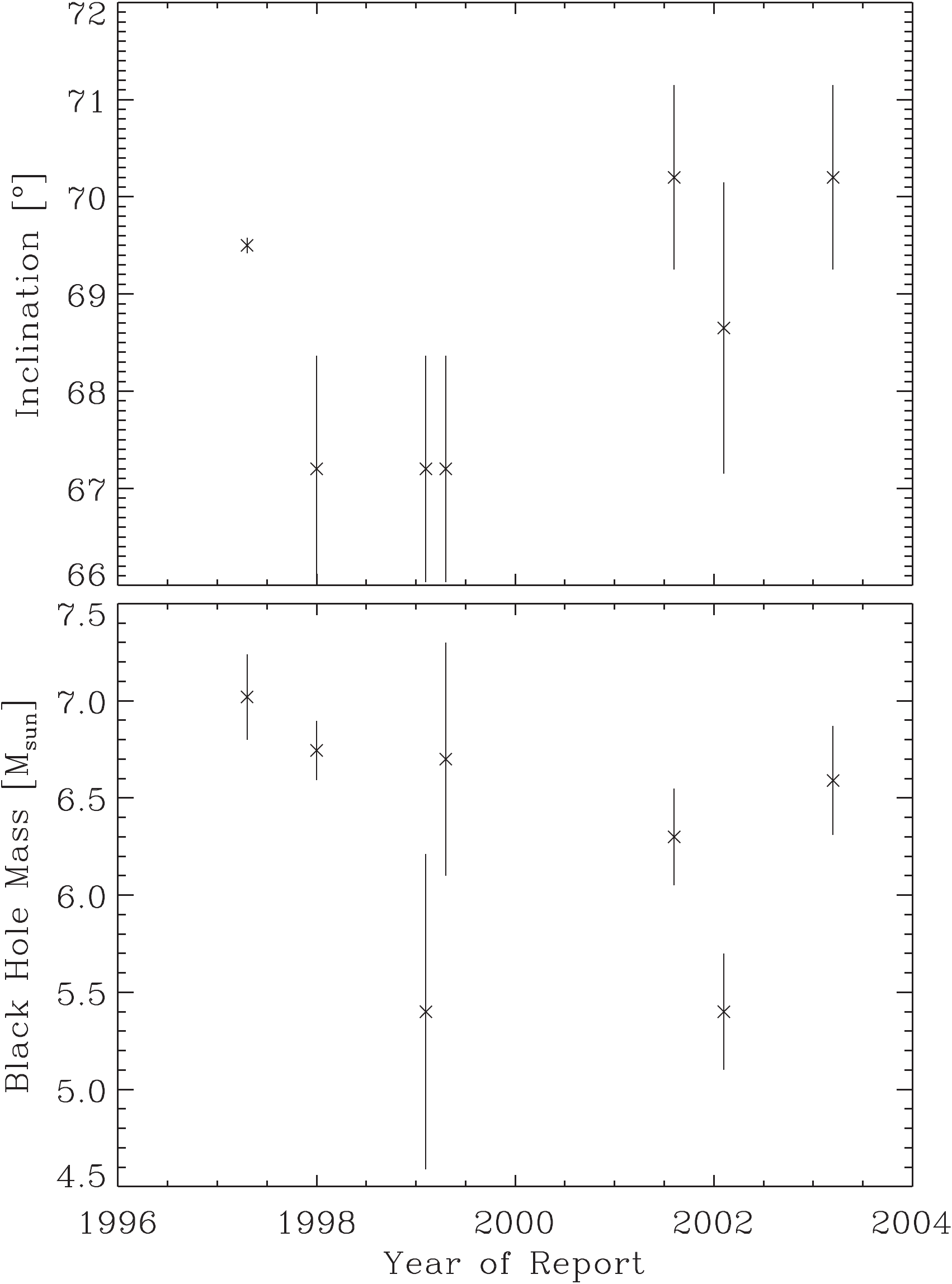}}
\caption{The BH mass and system inclination of GRO~J1655-40 reported at different times, ordered by their
publication dates \cite{Orosz1997,Hooft1998,Phillips1999,Shahbaz1999,Greene2001,Beer2002,Shahbaz2003}. Note that
the third and fourth reports \cite{Phillips1999,Shahbaz1999} adopted the inclination in the second report
\cite{Hooft1998}, and the last (seventh) report \cite{Shahbaz2003} adopted the inclination in the fifth report
\cite{Greene2001}.} \label{groj1655}
\end{figure}

It should be pointed out that the inclination $i$ in Equation~(\ref{equ:mass}) refers to that of the orbital
plane of the binary system. Because the accreting material initially carries the angular momentum inherited from
the companion star, the formed disk should be co-planar with the orbital plane of the binary system. However,
the BH spin in a BHXB cannot be changed significantly by accretion \cite{King1999} (and see discussion above on
Cygnus X--1). Therefore its spin axis may not coincide with the normal direction of the orbital plane of the
binary system. In this case the Bardeen-Petterson effect \cite{Bardeen1975}, due to frame-dragging by the
spinning BH, can rapidly align the normal axis of the inner disk region with the spin axis of the BH, making the
inner disk and binary system mis-aligned. Circumstantial evidence already exists for mis-alignment between the
two axes, because the orbital inclination of GRO~J1655--40 shown in the upper panel of Figure~\ref{groj1655} is
significantly different from $\sim 80^\circ$ inferred from its relativistic jets \cite{Hjellming1995}, if the
jets are powered by extracting the spin energy of the BH via the Blandford-Znajek mechanism
\cite{Blandford1977}. Nevertheless, the orbital inclination is normally used in place of the disk inclination,
which is needed in BH spin measurement using the CF method, in order to infer the total disk luminosity and
calculate GR correction factors. However, the inner disk inclination is currently not available essentially for
any of the known BHXBs. For example, a mis-alignment of $10^{\circ}$ from $i=70^{\circ}$ can cause nearly 50\%
error to the total disk luminosity, which will translate into nearly 30\% error in $r_{\rm in}$. A Schwarzschild
BH may be estimated to have $a_*$ falling any where between $[-0.5,0.5]$, if $r_{\rm in}$ is uncertain within
about 30\%, according to Fig.(\ref{r_isco}).

Accurate determinations of distances of astrophysical objects in the Milky Way are difficult, e.g., for BHXBs
that are not standard candles. Normally some absorption features in their spectra, in conjunction with their
positions in the galactic coordinates, are used to infer their distances. For example, the distance of
GRO~J1655--40 is commonly taken as $3.2\pm 0.2$ kpc, based primarily on observed absorption lines and somewhat
on the dynamics of the observed jets. Critical examinations of all available data related to its distance,
however, are in favor of a much closer distance of less than 2 kpc and more likely just 1 kpc
\cite{Foellmi2006,Foellmi2009}. Similar conclusion is also reached to the distance of A0620--003, revising its
distance from the commonly accepted $1.2\pm 0.4$ kpc to $\sim 0.4$ kpc, making its possibly the closest BHXB
known so far \cite{Foellmi2009}; however, a distance of $1.06\pm 0.12$~pc was preferred in a more recent study
\cite{Cantrell2010}. Similarly the currently adopted distances of many other BHXBs may also have considerable
systematic errors. If true, this would change significantly the current estimates on their masses and spins.

Therefore future improvements of the continuum fitting method depend upon the improved measurements on their BH
masses, accretion disk inclination, and distances.

\subsection{Future improvements of the continuum fitting method}

The mass ratio $q$ can be determined directly according to the law of momentum conservation, i.e.,
\begin{equation}
\label{equ:ratio} M_{\rm C}/M_{\rm BH}=K_{\rm BH}/K_{\rm C},
\end{equation}
if the semi--amplitude of the velocity curve of the BH $K_{\rm BH}$ can be observed directly, as illustrated in
Figure~\ref{BHXB}. The orbital inclination $i$ can be then calculated using Equation~(\ref{equ:mass}), avoiding
any systematics related to the ellipsoidal light curve modeling.

Since a BH is not directly observable, we can only hope to observe any emission or absorption line feature
co-moving with it. The accretion disk certainly moves with the accreting BH. However any line feature of the
inner accretion disk suffers from the broadening of disk's orbital motion and distortions by relativistic
effects around the BH, thus making it practically impossible, or difficult at least, for detecting the binary
orbital motion of the BH. Orbital motion of double-peaked disk emission lines were observed for NSXB Sco X--1
\cite{Steeghs2002a}, the BHXBs A0620--003 \cite{Haswell1990,Orosz1994}) and GRS 1124--68 \cite{Orosz1994}.
Unfortunately a significant phase offset of velocity modulation was found from that expected based on the
observed orbital motion of the companion, though the velocity semi-amplitude is consistent with the expected
mass ratio \cite{Orosz1994}. Soria et al. (1998) observed the orbital motion of the double-peaked disk emission
line He~{\scriptsize II}~$\lambda$4686 from GRO~J1655--40, and found its velocity modulation phase and
semi-amplitude in agreement with the kinematic and dynamical parameters of the system \cite{Soria1998}. However
one major problem in accurately measuring the orbital motion of the primary from the observed double-peaked
emission lines is how to determine reliably the line center, because the lines are typically asymmetric and also
variable.

We have recently proposed to observe the Doppler shifts of the absorption lines of the accretion disk winds
co-rotating with the BH around its companion star \cite{Zhang2012}, since in many XRBs accretion disk winds are
ubiquitous and appear to be rather stable when observed (e.g., in Ref.\cite{Miller2006}). We verified this
method using {\it Chandra} and {\it HST} high resolution spectroscopic observations of GRO~J1655--40 (shown in
Figure~\ref{bh_mass_wind}) and LMC~X--3. Unfortunately the currently available data only covered small portions
of their orbital phases and thus do not allow better constraints to their system parameters. Future more
observations of these two sources and other sources with detectable absorption lines from their accretion disk
winds will allow reliable and precise measurements of the BH masses and orbital inclination angles in accreting
BHXBs.

\begin{figure}
\center{
\includegraphics[angle=0,scale=0.4]{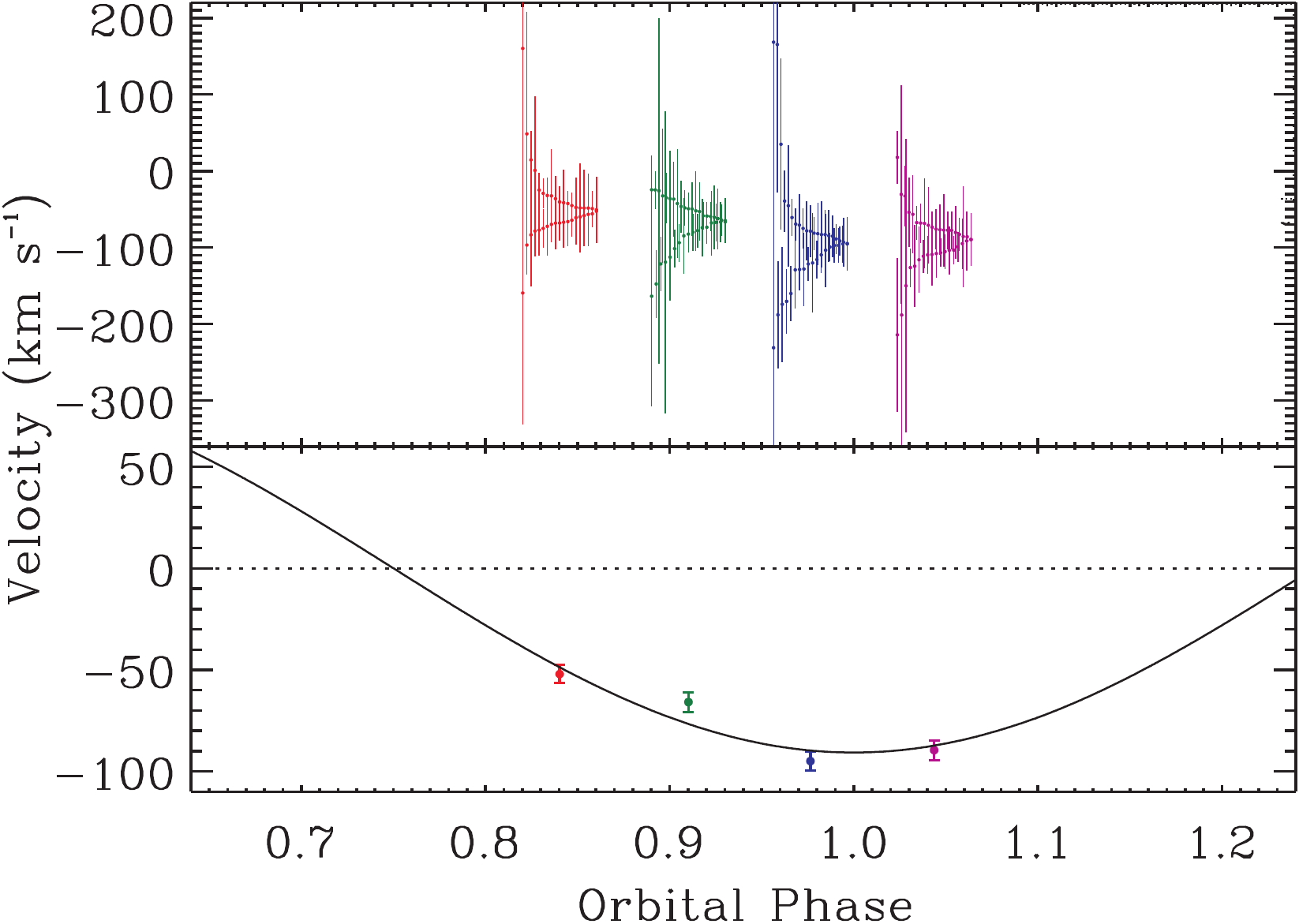}}
\caption{Velocity curve of the 39 observed X-ray absorption lines with {\it Chandra} from GRO~J1655--40, after
subtracting the line of sight intrinsic velocity of each line at each orbital phase. The upper panel marks the
velocity of each line with its 1-$\sigma$ error bar, slightly shifted horizontally for visual clarity. The
bottom panel shows the weighted average velocity of all lines in the upper panel at each phase; the solid curve
is the fitted velocity curve with its orbital period and phase fixed at the values observed previously.
(Slightly adapted from Figures 2 in Ref.\cite{Zhang2012})} \label{bh_mass_wind}
\end{figure}

Since the accretion disk has very high density and is ionized near the BH, where the majority of the observed
disk emission is produced due to the very deep gravitational potential near the BH, scattering of the primary
disk emission is inevitable. The scattered light is polarized and its polarization fraction and position angle
depend on the viewing direction (inclination ), scattering optical depth and the radius where the scattering
occurs \cite{CONNORS1977,Connors1980,Li2009,Schnittman2009}. Ignoring many details, it can be shown that the
polarization fraction, $P(i)$, of the observed disk photons (initial disk emission plus the scattered emission)
is given by
\begin{equation}
\frac{1}{P(i)}=1+A\frac{\cos i}{1-\cos^2i},
\end{equation}
where $A$ is a constant depending upon the scattering optical depth; detailed calculations made by Chandrasekhar
 \cite{Chandrasekhar1960} gives $P(75^{\circ})=0.04$. Note that here the disk photons are from the
Raleigh-Jeans part of the multi-color blackbody spectrum with a characteristic shape of $f(\nu)\propto
\nu^{1/3}$, i.e., no GR effect is included. We can therefore find the disk inclination angle by measuring the
polarization fraction of this part of the disk emission, as shown in Figure~\ref{polar}.  At energies above the
Raleigh-Jeans part of the multi-color blackbody spectrum, the polarization is strongly effected by both the
inclination and BH spin, as shown in Figure~\ref{polar_spec}. The continuum spectra are clearly degenerated for
the different combinations of inclination and BH spin, but the polarization fraction and angle as functions of
energy can clearly distinguish between them \cite{Li2009}. Therefore X-ray spectra-polarimetry observations of
BHXBs will certainly make important progresses in measuring BH spin.

\begin{figure}
\center{
\includegraphics[angle=0,scale=0.4]{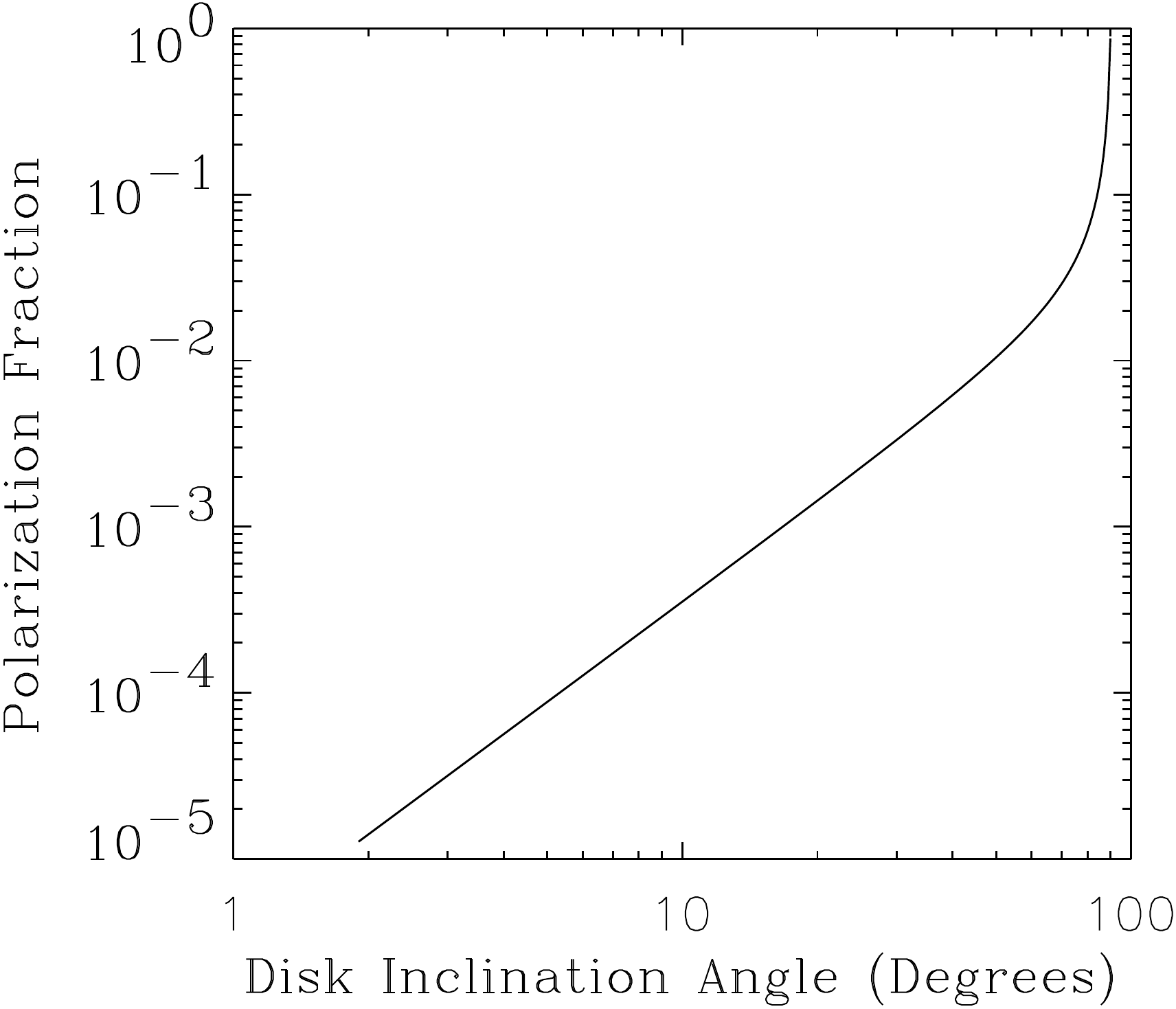}}
\caption{Polarization fraction of observed accretion disk emission as a function of its inclination.}
\label{polar}
\end{figure}

\begin{figure}
\center{
\includegraphics[angle=0,scale=0.8]{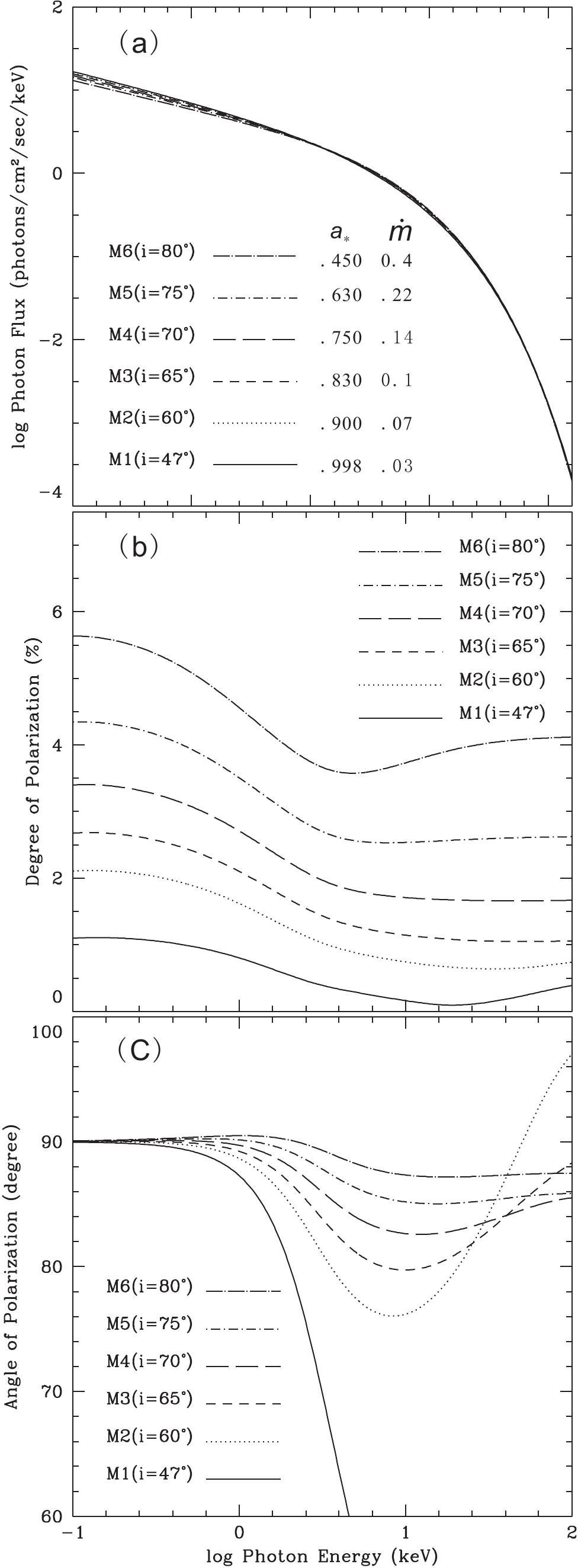}}
\caption{Disk continuum flux (a), polarization fraction (b) and polarization angle (c) as functions of photon
energy for different model parameters. $i$, inclination angle of the inner disk in degrees; $\dot{m}$, mass
accretion rate in units of Eddington rate by assuming a 10\% conversion efficiency from rest mass to radiation
in all cases. Other parameters: mass of the BH is $10 M_\odot$; distance to the BH is $10$~kpc; spectral
hardening factor is $1.6$. (Adapted from Figures 4-6 and Table 1 in Ref.\cite{Li2009})} \label{polar_spec}
\end{figure}

Besides using polarization measurements to obtain inner disk inclination (and BH spin), the broad iron K-alpha
measurements can also be used to do so \cite{Miller2008,Miller2009,Reis2009}, because iron K-alpha emission is
believed to come from the fluorescent emission of the disk as a result of illumination by a hard X-ray source
above the disk. Naturally this emission is sensitive to the disk inclination (and BH spin). However, compared to
the polarization measurement method, this method is less straight forward and may suffer from systematic
uncertainties in deriving the disk inclination (and BH spin), because complicated modeling of the hard X-ray
component and line emissivity from the disk is required. Therefore it is essential to use all methods discussed
above to measure both inclination angles (orbital plane and inner disk) and BH spin. Studying the relationship
between the results obtained with different methods is also important in its own rights, in order to understand
accretion disk physics, its interaction with the BH in its center, production of relativistic jets, the origin
of the BH spin, and ultimately the formation mechanisms of BHs and BHXB systems.

The recent dispute on the distances of GRO~J1655-40 and A0620--003 exemplifies the difficulty of determining the
distances of BHXBs using mostly absorption lines \cite{Foellmi2006,Foellmi2009}. We have recently suggested a
method of using the delay time between the X-ray fluxes of an XRB and its X-ray scattering halo by interstellar
dust to infer its distance \cite{Ling2009,Ling2009a}. However this method may suffer from our incomplete
knowledge of the distribution of interstellar medium. Ideally precise astrometry can determine their distances
model-independently, by measuring their trigonometric parallaxes. Recently the distance to Cygnus~X--1 was
determined reliably and accurately this way ($1.86^{+0.12}_{-0.11}$~kpc; \cite{Reid2011}); which is key to the
consequent measurement of its BH mass and spin \cite{Orosz2011,Gou2011a}. Currently it remains challenging to
measure the trigonometric parallaxes of objects at distances beyond several kpc where most BHXBs are located.
Future high precision astrometry missions are expected to improve the distance estimates to these BHXBs
significantly.

Therefore we expect that the improved measurements discussed above on their BH masses, inner disk inclination,
and distances will allow future improvements of BH spin measurement with the CF method.

\subsection{Possible application of the continuum fitting method to AGNs}

More recently, the CF method is also applied to constrain the BH spin in an active galactic nucleus (AGN), which
is powered by matter accretion onto the central supermassive BH in its center \cite{Czerny2011}. However the BH
spin inferred this way for an supermassive BH is quite uncertain, because: (1) The peak temperature of the
accretion disk inversely increases with the mass of a BH. Therefore, for a supermassive, its temperature would
be in the ultraviolet energy range which will be strongly absorbed and hard to observe; (2) The system
parameters of a supermassive BH (e.g., the mass of the BH and the inclination angle of the accretion disk) have
larger uncertainties; (3) The uncertain mechanism for some components (e.g., the soft X-ray excess) in AGN
spectra also increases the difficulty; (4) Some emission and absorption lines may distort the continuum spectrum
substantially; and (5) In some cases the contribution of its host galaxy to the observed total continuum
spectrum cannot be removed satisfactorily.

The polarized continuum of an AGN should be a pure accretion disk continuum, at least in the optical to near
infrared band \cite{Kishimoto2008}. One possible way to measure the BH spin in an AGN with the CF method is to
combine the observed polarized optical to near infrared continuum spectrum with the observed total UV continuum
spectrum to get a broad band continuum spectrum of an AGN \cite{Hu2012}.  In principle the broadened Balmer edge
features and the total UV spectrum can be used to constrain the disk inclination angle and fraction of host
galaxy contamination, respectively \cite{Hu2012}. However the quality of the currently available data is still
insufficient to allow accurate determination of BH spin in AGNs with this method.

Nevertheless the principal method of measuring BH spin in AGNs should be using the reflected broad iron line and
continuum components \cite{Miller2008,Miller2009,Reis2009}. The main reason is that $i$ can also be determined
simultaneously and $r_{\rm in}$ obtained this way is already in units of $r_{\rm g}$, thus avoiding naturally
the uncertainties caused by the BH mass and inclination in the CF method. However cross calibration can be done
between the CF and reflection fitting methods if both can be applied to the same supermassive BH.

\section{Further developments on hot accretion flows}\label{hot_flow}

In this section I briefly summarize some further developments on hot accretion flows, which are believed to be
responsible for the PL component of the spectra in BHXBs. The multi-waveband spectra of the hard state of the
BHXB XTE J1118+480 was modeled with the TID with hot accretion flow (ADAF) geometry (panel (a) in
Figure~\ref{xte1118}) and TID (with ADAF) plus jet model (panel (b) in Figure~\ref{xte1118}), with $\dot m_{\rm
D}=0.05$ and $\dot m_{\rm jet}=5\times 10^{-3}\dot m_{\rm D}$ \cite{Yuan2005}. The steep UV spectrum provides
clear evidence for a large truncation radius for the SSD ($r=600$), and the radio to infrared spectrum dominates
the jet emission, which also contribute to the hard PL component \cite{Yuan2005}. Such low accretion rate gives
very low SSD luminosity $l_{\rm D}\sim 5\times 10^{-4}$ (since the disk radiation efficiency $\eta \sim 1/r$;
see Figure~\ref{eta_r}), which is far below the turn-over luminosity of $l_{\rm D}\sim 10^{-2}$ shown in
Figure~\ref{ngc1313}. Therefore the inferred truncation radius of XTE J1118+480 agrees with the extrapolation of
data points of XTE J1817--330 down to very low disk luminosity. Large truncation radii are also reported from
several other sources in the hard state (e.g., in Ref.\cite{Zhang2010a}). However, the exact values of these
truncation radii may have large uncertainties, since no direct detection of the inner disk peak emission was
available, unlike the strong case of XTE J1118+480 \cite{Yuan2005}. For this reason I did not include these
reported values in Figure~\ref{ngc1313}.

\begin{figure}
\center{
\includegraphics[angle=0,scale=0.4]{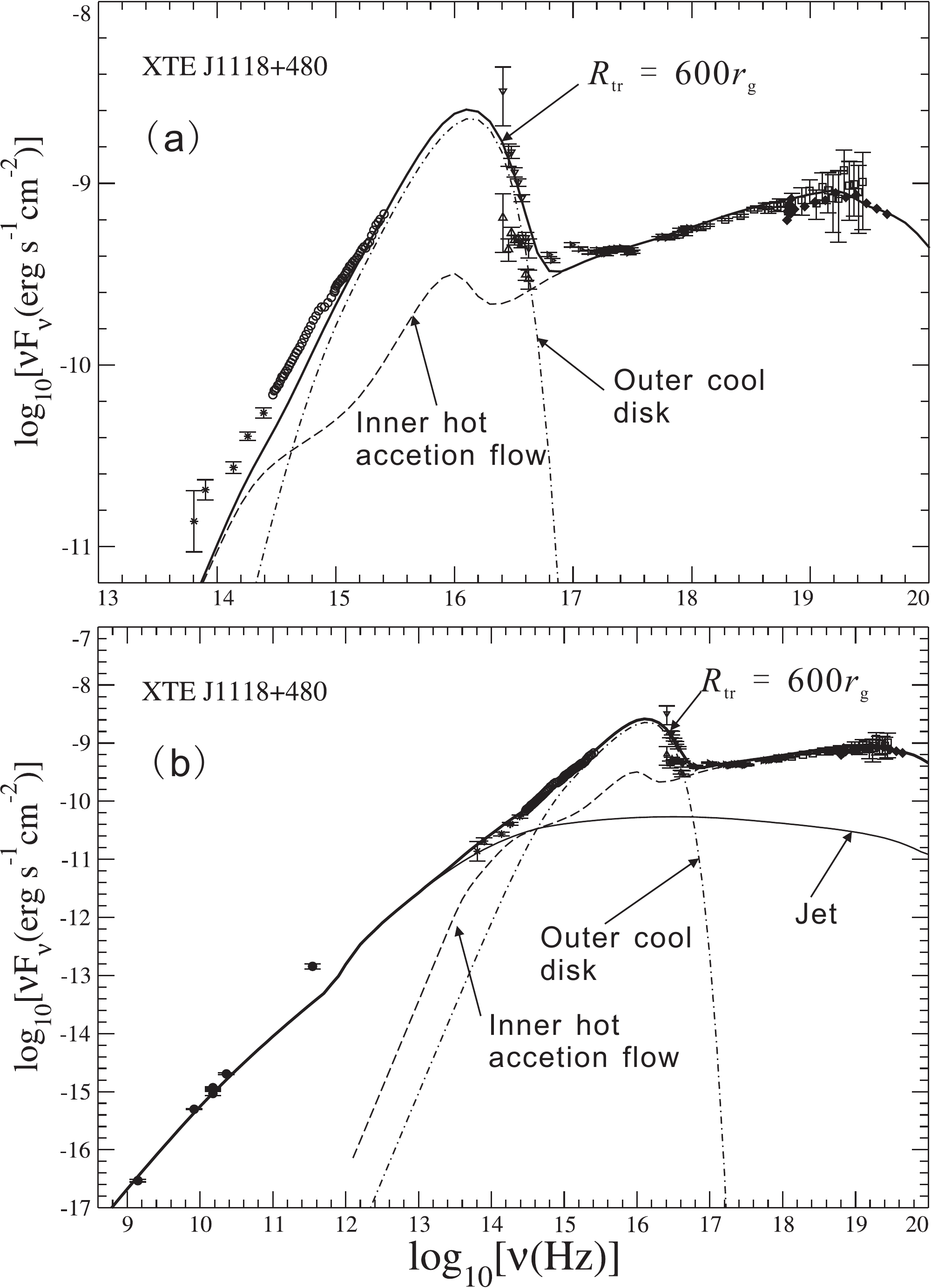}}
\caption{Spectral modeling results for XTE J1118+480. (a) Outer cool disk plus inner hot accretion flow model.
(b) An additional jet component is included. $\dot m_{\rm D}=0.05$ and $\dot m_{\rm jet}=5\times 10^{-3}\dot
m_{\rm D}$. (Adapted from Figures 1-2 in Ref.\cite{Yuan2005})} \label{xte1118}
\end{figure}

Panel (b) in Figure~\ref{xte1118} also shows that the radiation from both the hot accretion flow and the jet
contribute to the X-ray emission. However the former is roughly proportional to $\dot{m}^2$, whereas the latter
to $\dot{m}$. Thus with the decrease of $\dot{m}$, the contribution from the jet becomes more and more
important, thus the X-ray radiation will be dominated by the jet (when $l \lesssim 10^{-5}-10^{-6}$), as shown
in Figure~\ref{disk_jet} \cite{Yuan2005}. The observational data of very low-luminosity AGNs clearly show a
correlation between radio and X-ray with a correlation index of $\sim 1.2$ \cite{Yuan2009}, in excellent
agreement with the prediction shown in Figure~\ref{disk_jet} \cite{Yuan2005}.

\begin{figure}
\center{
\includegraphics[angle=0,scale=0.35]{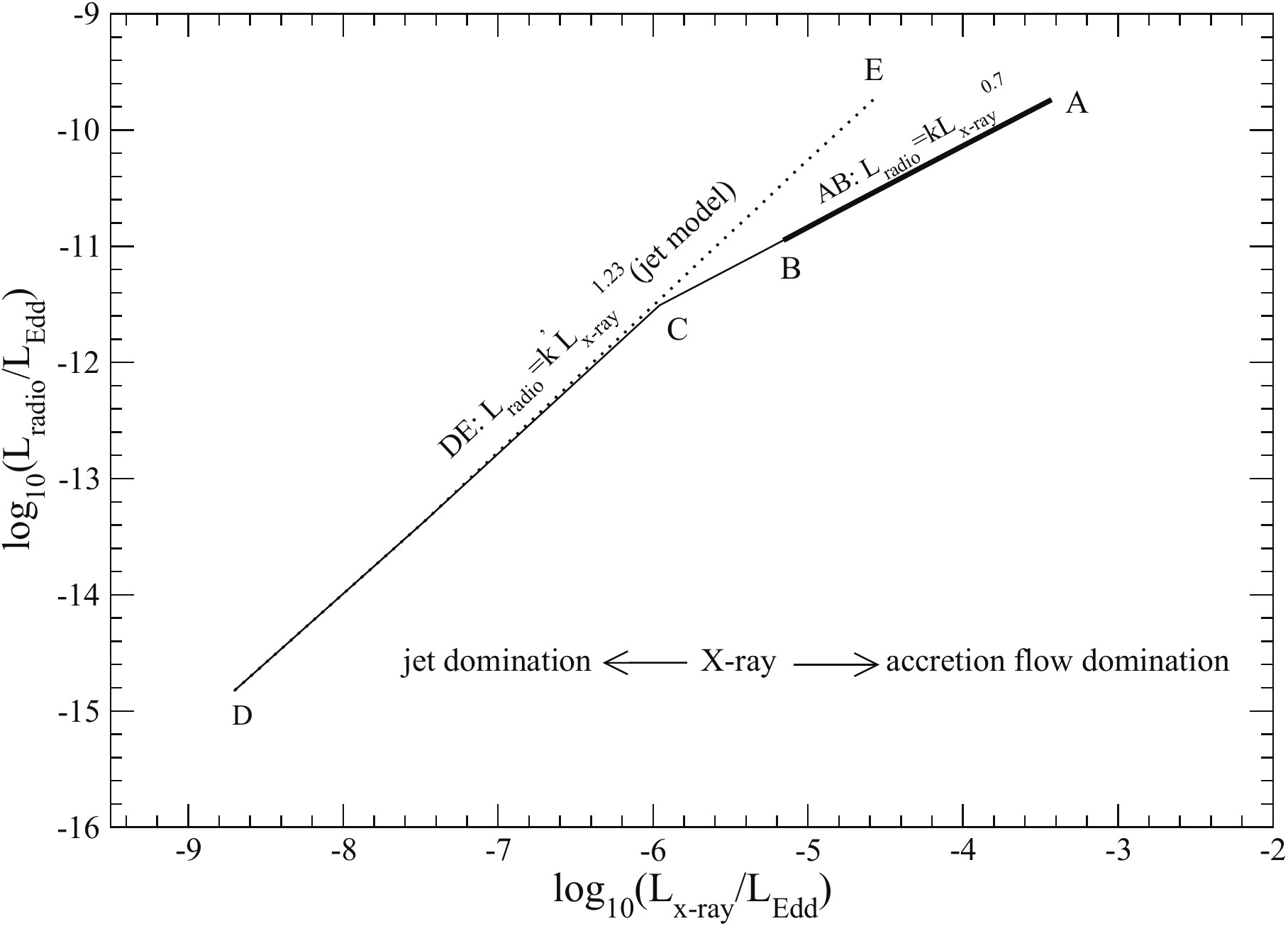}}
\caption{Radio (8.6 GHz)¨CX-ray (2¨C11 keV) correlation for BHXBs. The observed correlation is shown by the
segment AB. Segments BCD show the predicted correlation at lower luminosities, which approaches that of a
pure-jet model, as shown by the segment DE. Note that below the point C ($\sim 10^{-6}l_{\rm D}$), the X-ray
emission is dominated by the jet and the correlation steepens.  (Figures 1 in Ref.\cite{Yuan2005a})}
\label{disk_jet}
\end{figure}

It is well known that the highest luminosity an ADAF can produce is only about $3\%L_{\rm Edd}$. However the
observed highest luminosity a hard state can reach can be $10\%L_{\rm Edd}$ or even higher, which can be
described by LHAF shown in Figure~\ref{lhaf}, where several previously known solutions of accretion flows are
all unified in a single scheme \cite{Yuan2003}. The hard state spectrum of XTE~J1550--564 with $L\sim 6\% L_{\rm
Edd}$ is explained by LHAF very well, including the X-ray spectral slope and the value of the cutoff energy
\cite{Yuan2007}.

\begin{figure}
\center{
\includegraphics[angle=0,scale=0.4]{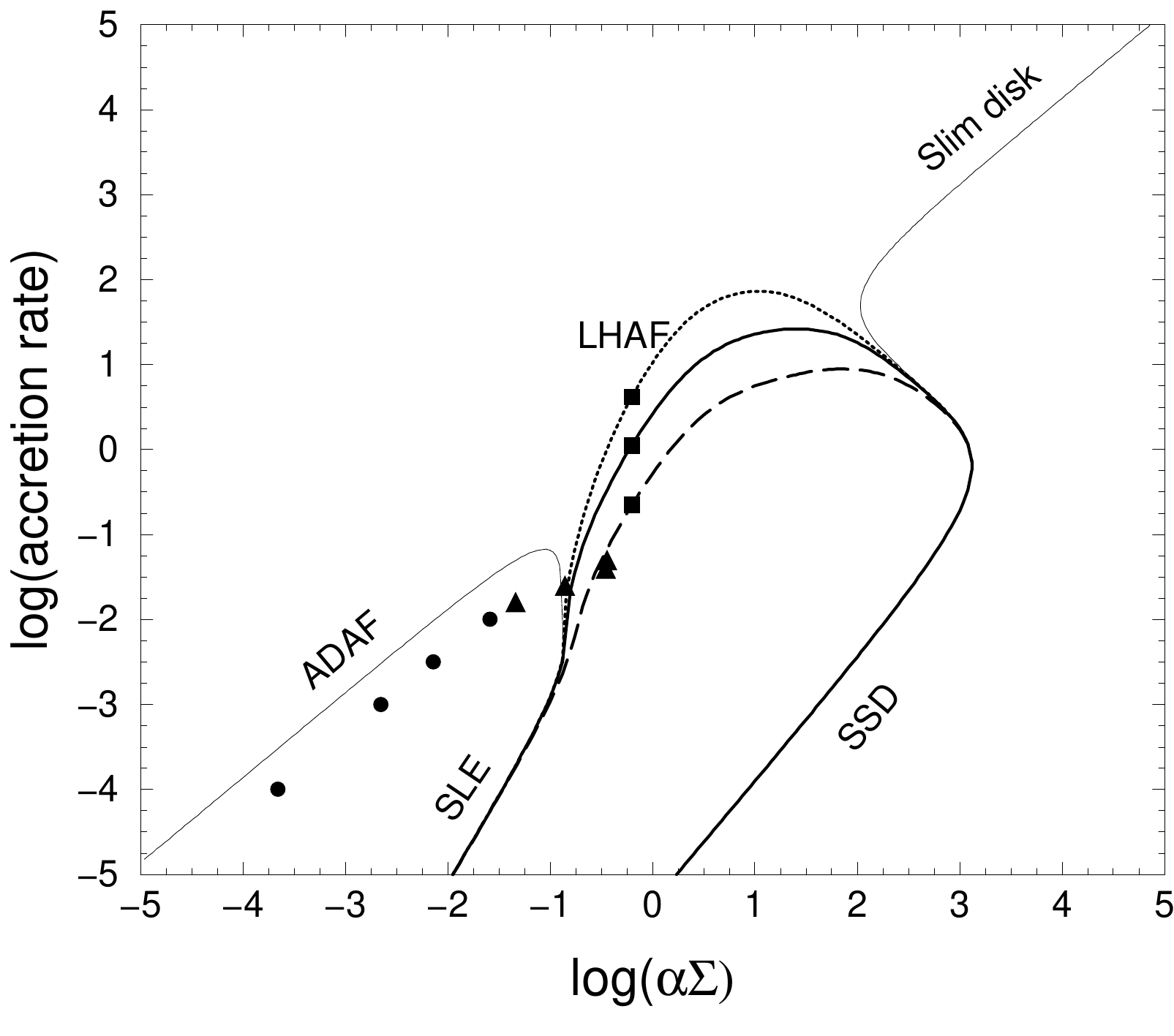}}
\caption{The thermal equilibrium curve of various accretion solutions. The accretion rate is in units of
$\dot{M}_{\rm Edd}\equiv 10L_{\rm Edd}/c^2$ and the units of $\Sigma$ is g~cm$^{-2}$. The parameters are $M_{\rm
BH}=10M_{\odot}, \alpha=0.1$, and $r=10$. (Figures 1 in Ref.\cite{Yuan2003})} \label{lhaf}
\end{figure}

\section{Further developments on corona formation}\label{corona}

While the formation of the SSD in a BHXB is reasonably well understood, the formation of the corona, which is
the hot accretion flow discussed in Section~\ref{hot_flow}, remains less understood. Over the last more than 10
years, Liu and her collaborators have developed a model to explain the formation and evolution of the corona in
a BHXB or AGN, in which the mass accretion rate $\dot m$ in units of the Eddington ratio drives the variations
of the complex accretion flows by the interaction between the cold SSD and the hot corona
\cite{Liu1999,Meyer2000,Meyer2000a,Liu2002,Liu2007,Liu2011}. Specifically, the coupling between the hot corona
and the cold disk leads to mass exchange between them.  The gas in the thin disk is heated up and evaporates
into the corona as a consequence of thermal conduction from the hot corona, or the corona gas condenses into the
disk as a result of overcooling by, for example, external inverse Compton scattering. If $\dot m$ is low,
evaporation occurs and can completely remove the thin disk, leaving only the hot corona in the inner region and
a truncated thin disk in outer region; this provides a mechanism for ADAF at low $\dot m$. If $\dot m$ is high,
the gas in the corona partially condenses to the disk due to strong Compton cooling, resulting in disk dominant
accretion. The model naturally explains the different structures of accretion flow in different spectral states
as shown in Figure~\ref{fig-bifang}
\cite{Liu1999,Meyer2000,Liu2002,Liu2006,Liu2007,Meyer2007,Taam2008,Qiao2009,Liu2011,Qiao2012}. The hysteresis
observed in spectral state transitions can also be explained by different irradiations from different evolution
history under the same scenario \cite{Liu2005,Meyer-Hofmeister2005,Meyer-Hofmeister2009}.

Figure~\ref{fig-bifang} is significantly different from the illustration of accretion flow structures in
different spectral state in Figure~\ref{4states} on three aspects: (1) At intermediate $\dot m$, the SSD is
broken by ADAF into two parts, an outer disk and an inner disk; (2) The inner disk boundary here is always
located very close to the BH, except in the very low $\dot m$ hard or quiescent state, which is very different
from the TID scenario depicted in Figure~\ref{4states}; (3) The corona here covers essentially the whole
accretion disk, especially the inner disk region, whereas in Figure~\ref{4states} the corona is mostly located
inward from the inner disk boundary. The observed soft X-ray component in the low/hard state can be explained by
the existence of the inner disk \cite{Liu2006,Taam2008,Liu2011,Qiao2012}. As I have discussed above, the inner
disk boundary radius inferred with the CF method in the presence of a strong PL component is actually consistent
with that in the soft state when the PL component is weak, after taking into account the Compton scattering in
the corona \cite{Yao2005,Steiner2009}. Actually the essential assumption behind the broad iron line/reflection
fitting method of determining BH spin is that the inner disk boundary is at the ISCO when the PL component is
strong. All these tend to support the existence of the inner disk at intermediate $\dot m$. However it remains
to be demonstrated that if the whole SSD is indeed broken into the two parts at intermediate $\dot m$, as
illustrated in Figure~\ref{fig-bifang}.

\begin{figure}
\center{
\includegraphics[angle=0,scale=0.4]{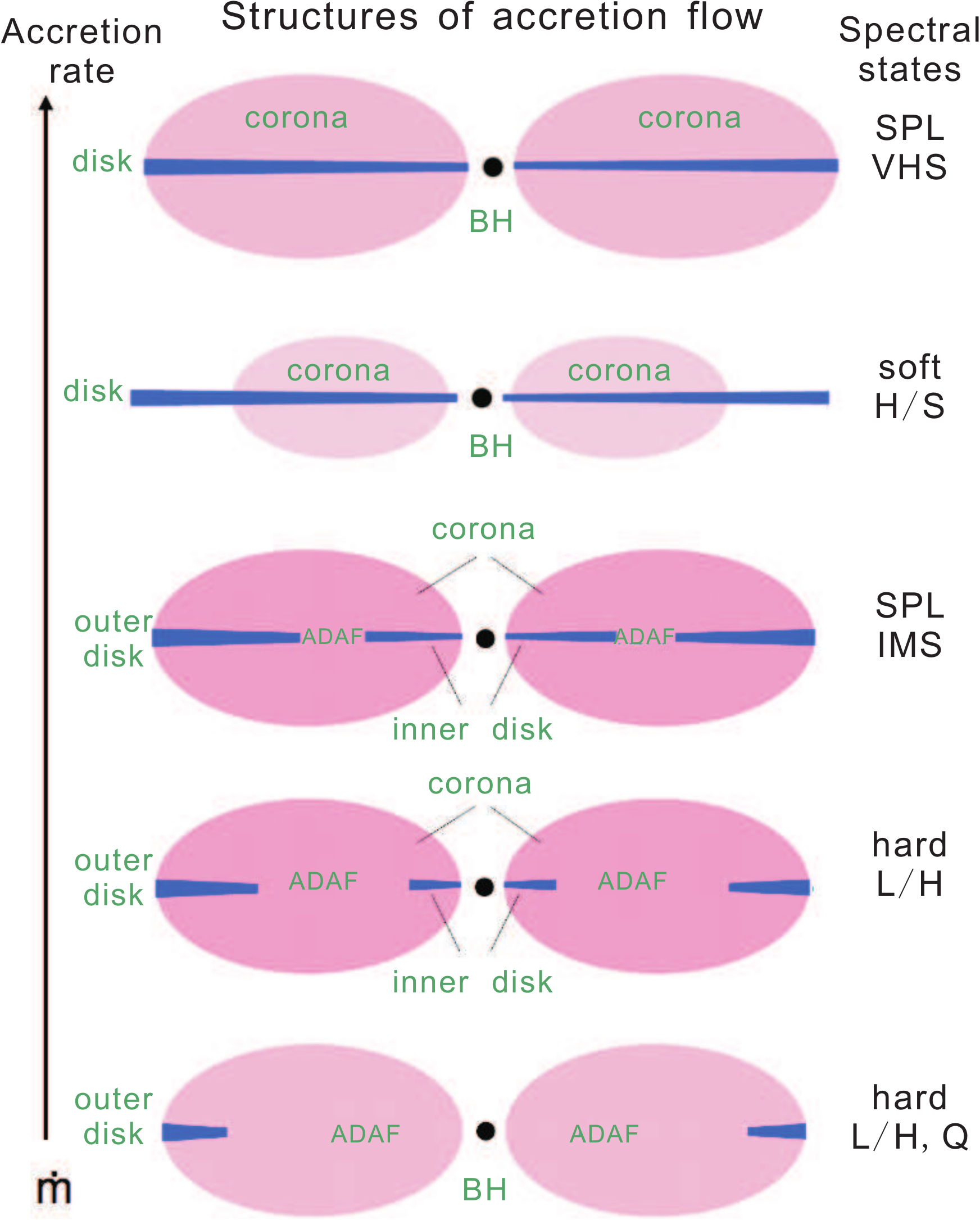}} \caption{\label{fig-bifang} A schematic description of the accretion flow structures in different
spectral states as consequences of disk-corona interaction, which is primarily driven by mass accretion rate
$dot m$. SPL: steep power-law state; VHS: very high state; soft: soft state; H/S: high/soft state; hard: hard
state; L/H: low/hard state; Q: quiescent state. (Adapted from a similar figure provided by Prof. Bifang Liu.)}
\end{figure}

Recently it has been suggested that the ADAF and corona shown in Figure~\ref{fig-bifang} may have clumpy
structures, as shown in Figure~\ref{clumpy} \cite{Wang2012}. The ``clumpy" model (Figure~\ref{clumpy}) has
mainly two different consequences from the ``uniform" model (Figure~\ref{fig-bifang}): (1) The inner disk is
transient in the ``clumpy" model; (2) The ``clumpy" model can explain the variabilities observed in X-ray
binaries (such as the state transitions discussed in Section~\ref{transitions}) and radio-loud AGNs (such as BL
Lac objects).

\begin{figure}
\center{
\includegraphics[angle=0,scale=0.45]{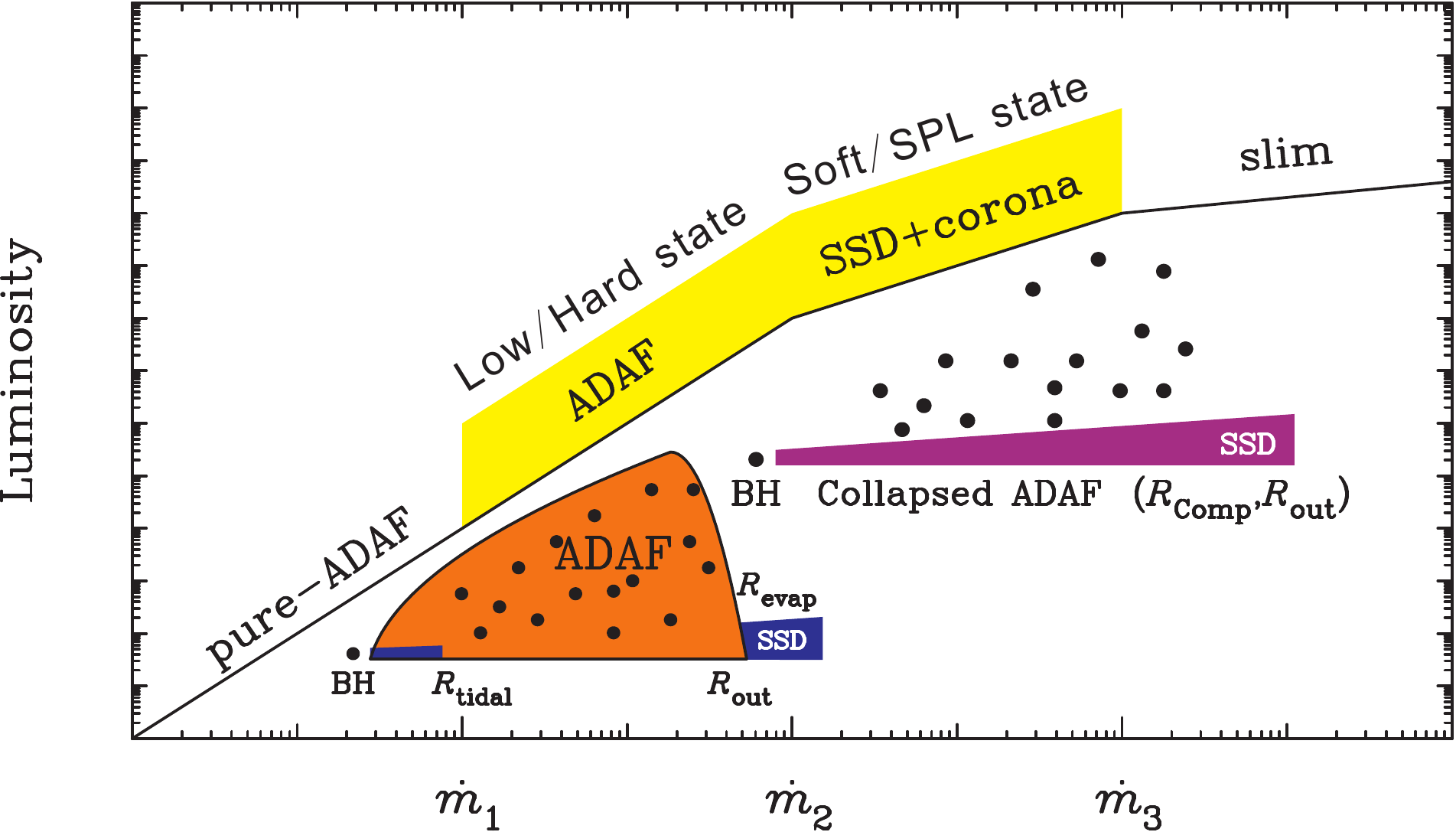}} \caption{\label{clumpy} Solid lines: accretion disk models at different accretion rates.
Clumpy structures in either ADAF or corona (shown as insets) may be developed at intermediate accretion rates,
corresponding to different spectral states.  (Adapted from Figures 1 and 2 in Ref.\cite{Wang2012})}
\end{figure}

Determining the structure of the corona in an BHXB observationally remains difficult; the fundamental issue on
whether the corona covers mostly the accretion disk or the central compact object still remains unclear so far.
Since the observed similar spectral and state transitions between some NS and BH XRBs are quite similar
\cite{Zhang1996}, it is reasonable to assume that they have similar coronae. Recently we have used type I X-ray
bursts from low-mass NSXBs to show that X-ray bursts experience negligible Comptonization and that the corona
cools rapidly during the rising phase of X-ray bursts and is then heated up rapidly during the rising phase of
X-ray bursts, as shown in the upper panel of Figure~\ref{bursts} for IGR~J1747--721 \cite{Chen2012}. These
results suggest that the corona cannot cover the central compact object completely (lower panel of
Figure~\ref{bursts}) and that the destruction and formation time scales of the corona are as short as seconds;
such short time scales are quite difficult to understand in the above discussed evaporation model, in which the
time scales are related to the viscous time scales of the accretion disk. However, this short time scale is
consistent with a corona produced by magnetic reconnections in the accretion disk, in a similar way to the solar
corona heating \cite{Zhang2000}; this conclusion was based on the inferred accretion flow structure of a BHXB
shown in panel (A) of Figure~\ref{three_layer}, in comparison with the atmospheric structure of the Sun shown in
panel (B) of Figure~\ref{three_layer}.

\begin{figure}
\center{
\includegraphics[angle=0,scale=0.35]{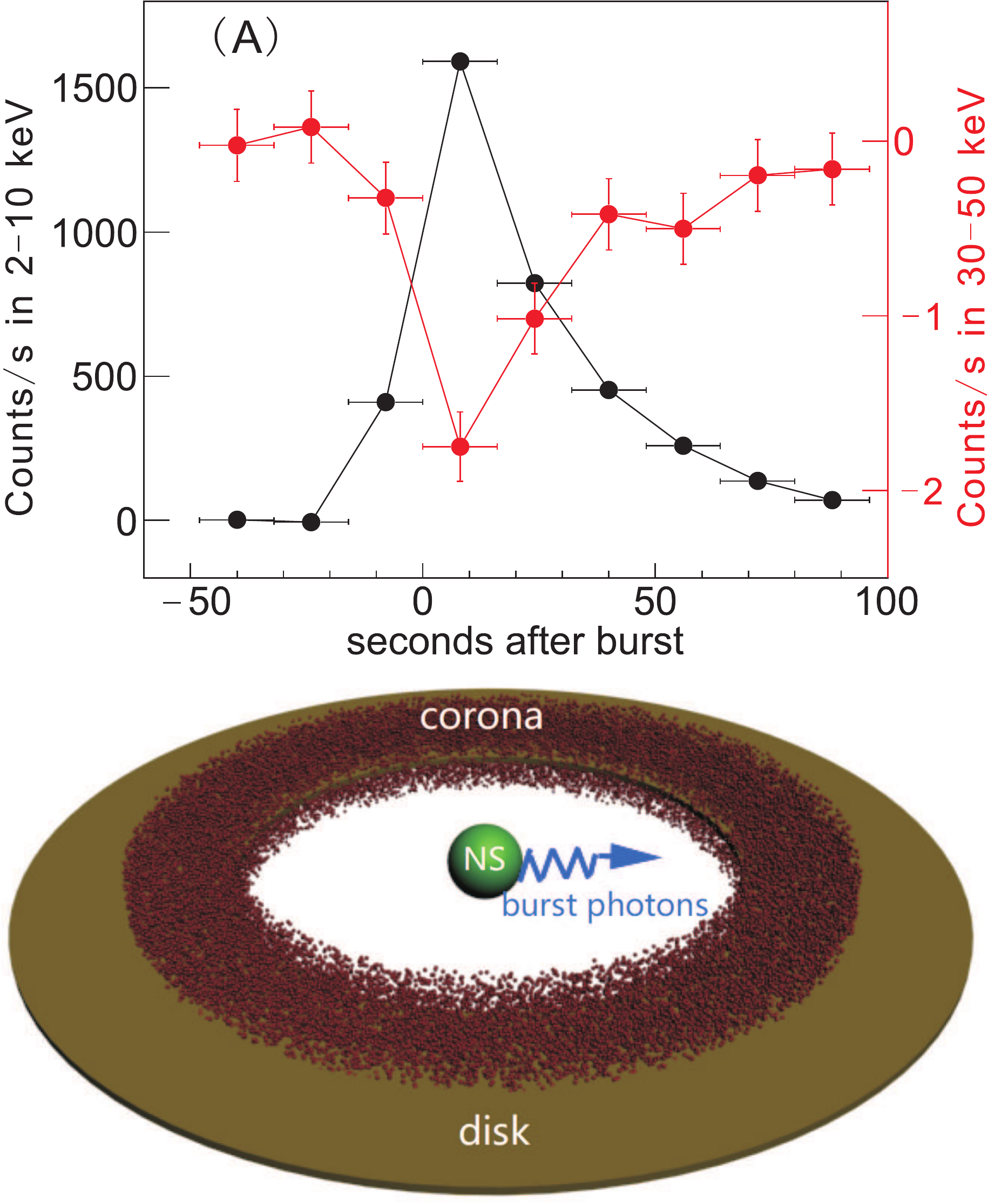}} \caption{\label{bursts} Upper panel: Anti-correlation between the observed
type I X-ray bursts from surface of the NS in IGR~J1747--721 (Adapted from Figures~3 and 2 in
Ref.\cite{Chen2012}).}
\end{figure}

\begin{figure}
\center{
\includegraphics[angle=0,scale=0.3]{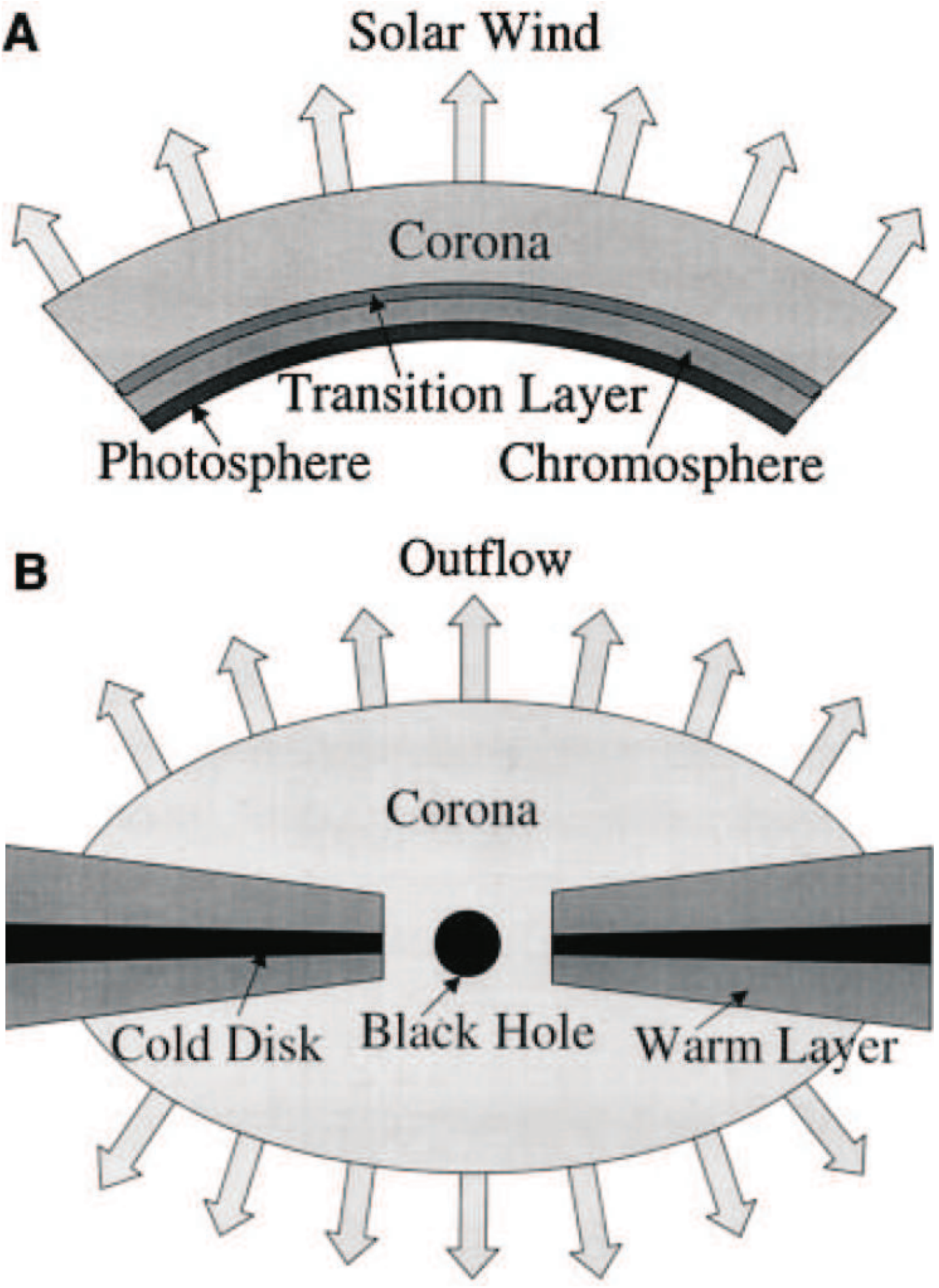}} \caption{\label{three_layer} Schematic diagrams of the solar atmosphere (A) and accretion
disk structure (B). (Figure~2 in Ref.\cite{Zhang2000}).}
\end{figure}

It is worth noticing that the purported coronae in both XRBs and AGNs have strikingly similar properties, e.g.,
they all have electron temperatures of the order of hundreds keV, in spite that the temperatures, inner disk
radii and variability time scales of their cold accretion disks all scale with their BH masses (and accretion
rates) as predicted in the SSD model. This means that their coronae are scale independent. It is perhaps not
coincidental that the electrons' velocities in a corona are approximately the same as the Keplerian orbital
velocities of the inner disk, which are also roughly the same as the launching velocities of jets. Of course
these velocities are also the varialized velocities of the central BHs. It is plausible that turbulent small
scale magnetic fields lifts the plasma in the ``warm layer" shown in Figure~\ref{three_layer} to form the
corona, which is then launched into the jets by the rotating large scale magnetic fields through either the BP
or BZ mechanism; during these processes the magnetic fields are mostly responsible for changing the directions
of motion of plasmas by the Lorentz force. If the corona is clumped as discussed in Section~\ref{corona}
 \cite{Wang2012}, then the plasma channeled into the jets should be clumpy and thus the
ejected jets should be episodic and have knotted structures, in agreement with observations; this scenario is
illustrated in Figure~\ref{clumpy_jet}. Alternatively episodic ejections may also occur in the disk, similar to
the coronal mass ejection on the Sun \cite{Yuan2009a}.

\begin{figure}
\center{
\includegraphics[angle=0,scale=0.40]{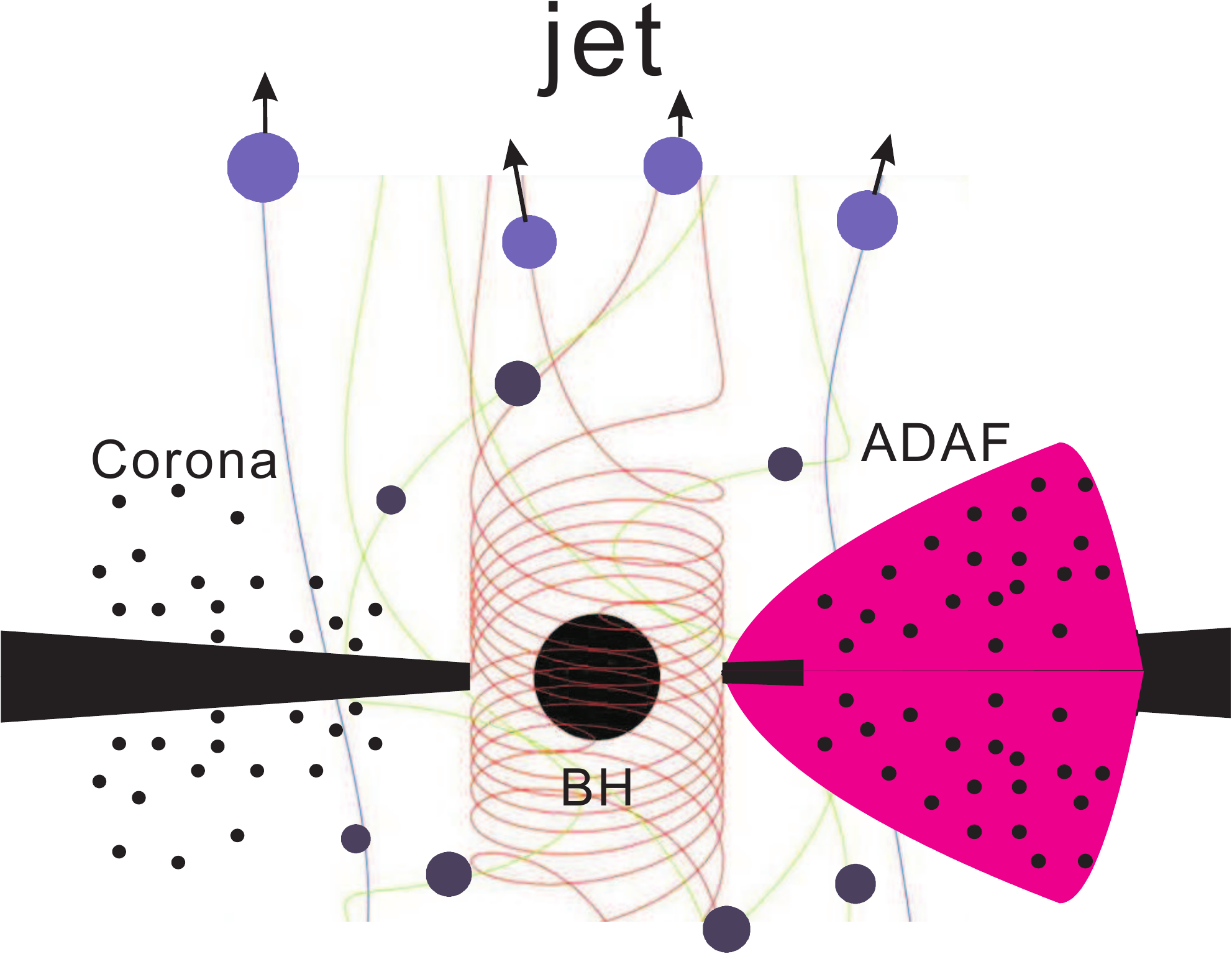}} \caption{\label{clumpy_jet} The clumps in the clumpy corona/ADAF of a
BHXB are channeled into the rotating and wound-up magnetic field lines and then ejected from the system. The
launched jets should be episodic and have knotted structures.}
\end{figure}

Interestingly, the soft X-ray excess (SXE) frequently observed in AGNs can be interpreted as the warm layer
found in XRBs, since the inferred plasmas parameters, with electron temperature of 0.1 to 0.3 keV and Compton
scattering optical depth of around 10, are similar to that of XRBs and universal among AGNs with different BH
masses, but seem to be related only with $\dot m$ \cite{Czerny2003,Ai2011,Done2012}. Therefore their warm layers
are also scale independent. Indeed, magnetic flares, perhaps caused by magnetic reconnections in the cold disk,
can produce the warm layers in both XRBs and AGNs \cite{Nayakshin1997,Nayakshin2001}.

\section{Further developments on state transitions}\label{transitions}
To further understand the mechanism of the spectral state transitions, it is important to study those spectral
state transitions in individual BHXBs during different outbursts. The advantages are obvious -- uncertainties in
our estimates of black hole mass, the binary properties such as the orbital period, or the source distance will
not play any role in producing the observed diverse transition properties in individual BHXBs, which can only be
driven by the accretion process under different initial conditions. The sources which firstly allowed such a
study were the BH transients GX~339--4 and XTE~J1550--564, the NS transient Aquila X--1, and the flaring NS
low-mass XRB (LMXB) 4U~1705--44, in which a remarkable correlation between the luminosity of the hard-to-soft
transition and the peak luminosity of the following soft state was found \cite{Yu2004,Yu2007,Yu2007a}. More
recently, a comprehensive study of the hard-to-soft spectral state transitions detected in all the bright XRBs
in a period of about five years with simultaneous X-ray monitoring observations with the RXTE/ASM and the
Swift/BAT confirmed the correlation between the hard-to-soft transition and the peak luminosity of the following
soft state \cite{Yu2009,Tang2011}, as shown in the upper panel of Figure~\ref{yu_2} \cite{Yu2009}. More
important was the discovery of the correlation between the transition luminosity and the rate-of-change of the
luminosity during the rising phase of an outburst of transients or a flare of persistent sources, as shown in
the lower panel of Figure~\ref{yu_2} \cite{Yu2009}.

The above correlation implies that in most cases it is the rate-of-increase of the mass accretion rate, rather
than the mass accretion rate itself, determines the hard-to-soft spectral transition; this is depicted in
Figure~\ref{yu_3} \cite{Yu2009}. In addition, the discovery of the relation between the hard X-ray peak flux and
the waiting time of transient outburst in the BH transient GX~339--4 (as shown in Figure~\ref{yu_1}) supports
that the total mass in the disk determines the peak soft state luminosity of the following outburst
\cite{Yu2007}. Therefore spectral states should be understood in the non-stationary accretion regime, which
would be described by both $\dot m$ and $\ddot m$, as well as the total mass in the disk before an outburst.

\begin{figure}
\center{
\includegraphics[angle=0,scale=0.55]{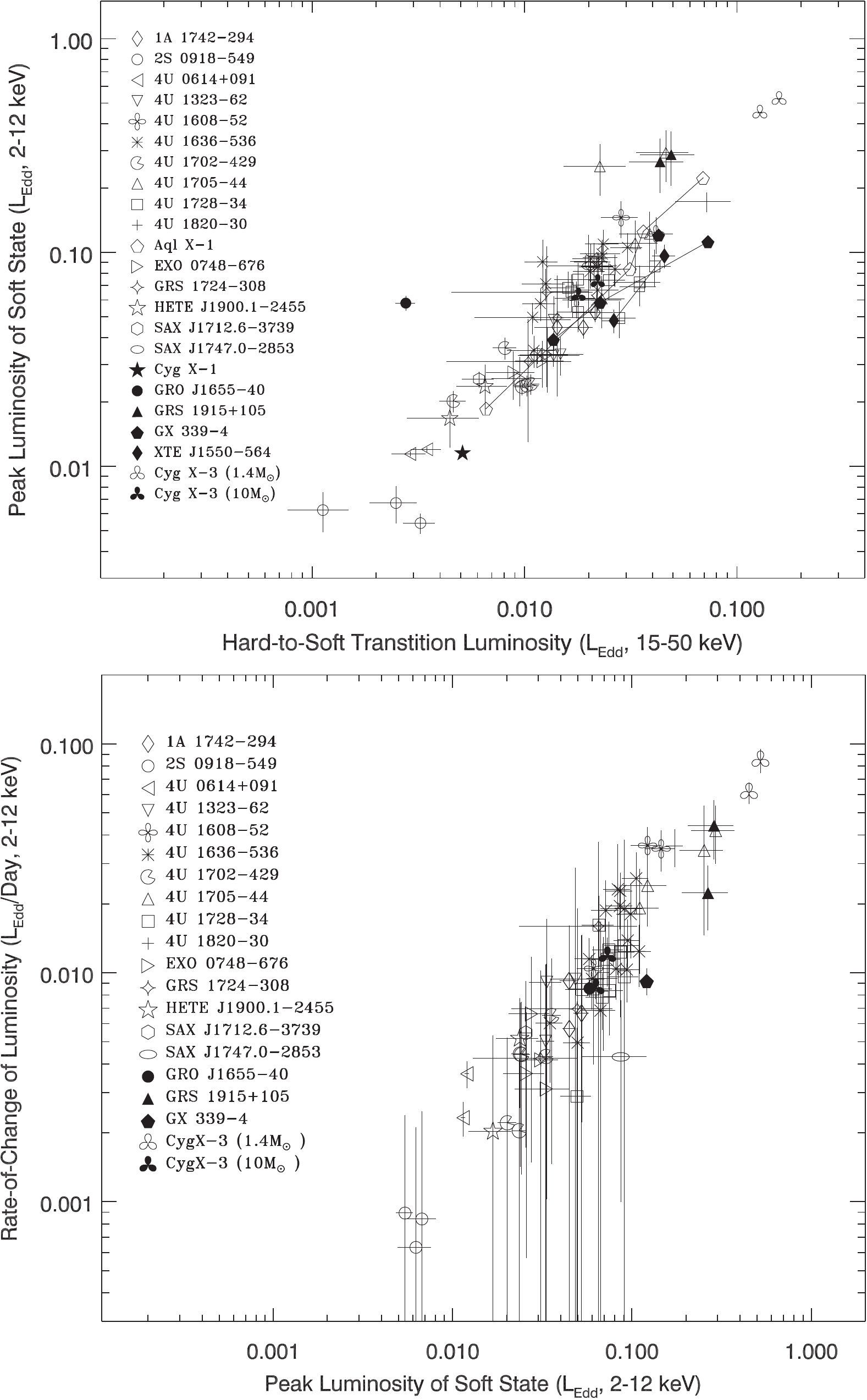}}
\caption{\label{yu_2} Upper panel: correlation between the transition luminosity (15--50 keV) and the peak
luminosity of the following soft state (2--12 keV) in Eddington units. Lower panel: correlation between the peak
luminosity of the soft state and the maximum rate-of-increase of the X-ray luminosity around the hard-to-soft
transition. (Adapted from Figures~24 and 27 in Ref.\cite{Yu2009})}
\end{figure}

\begin{figure}
\center{
\includegraphics[angle=0,scale=0.4]{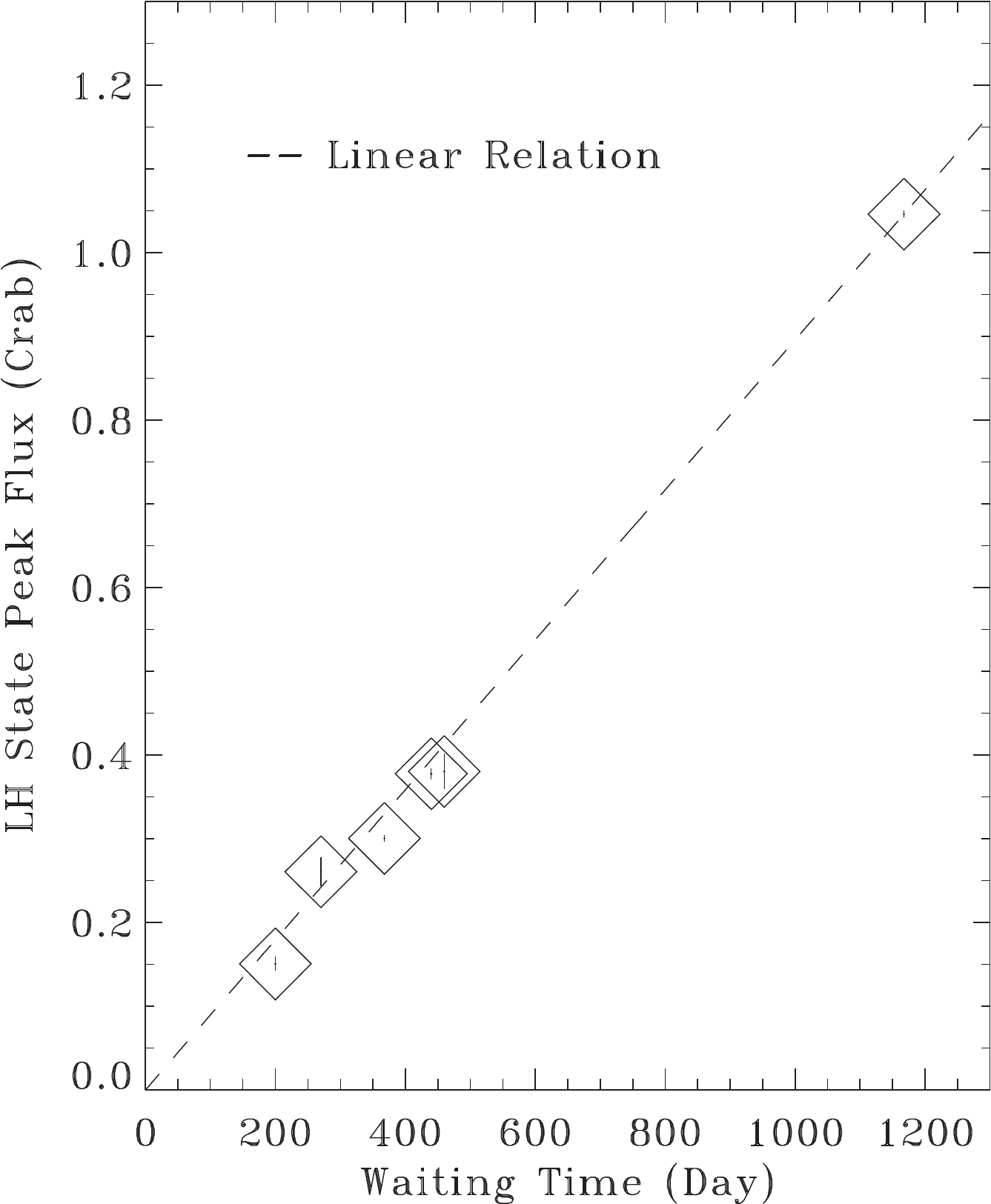}}
\caption{\label{yu_3} Correlation between the peak fluxes of the initial low/hard states in the outbursts of GX
339--4 and the time since the latest low/hard state peak in the previous outburst. The dashed line passes the
origin and the data point of maximal peak flux, showing an example of a linear relation. (Adapted from Figure~3
in Ref.\cite{Yu2007})}
\end{figure}

\begin{figure}
\center{
\includegraphics[angle=0,scale=0.4]{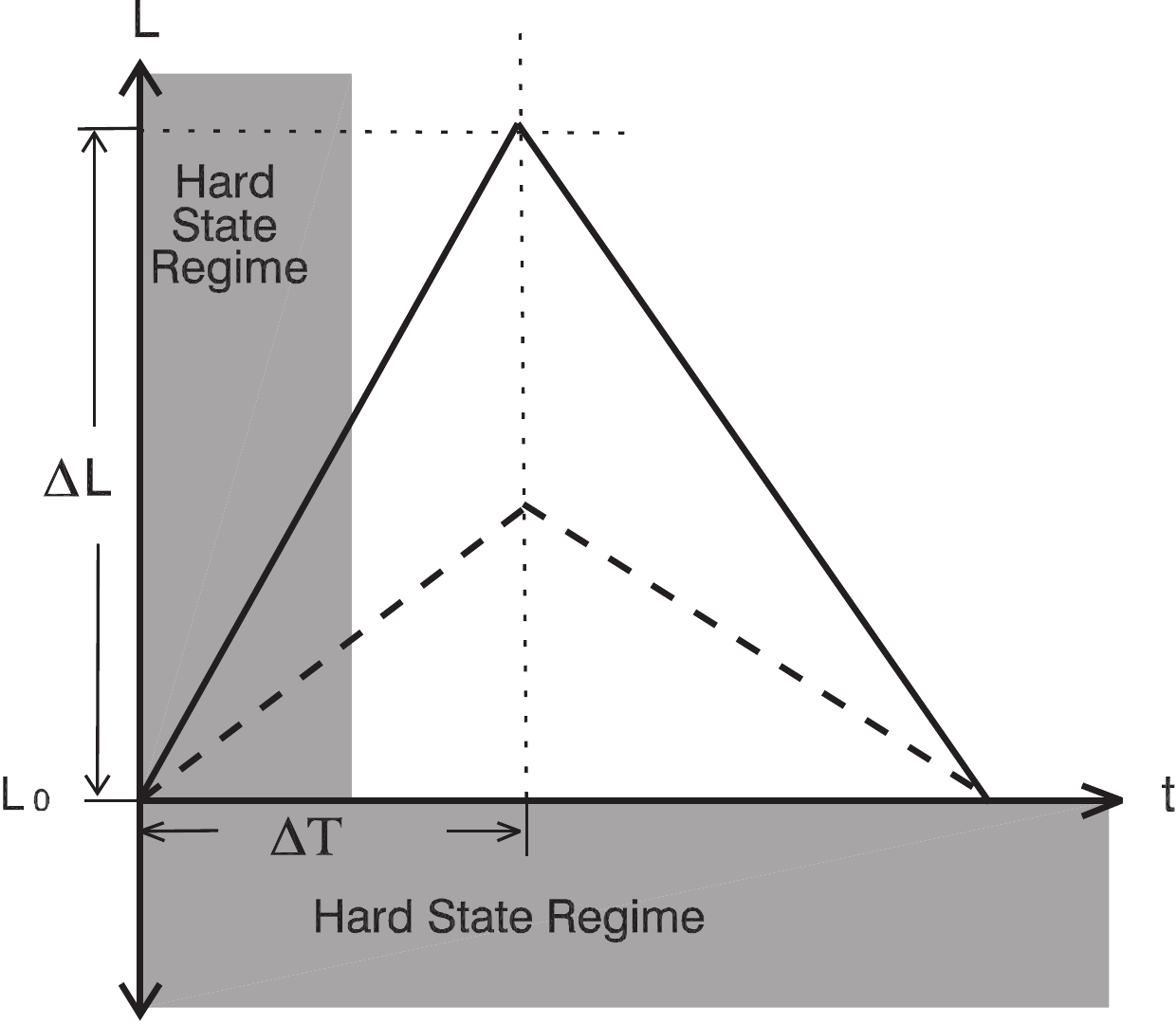}}
\caption{\label{yu_1} A schematic picture of the regimes of the hard state. Two assumed transient outbursts of
different peak luminosities are shown. When a source is under stationary accretion, spectral transitions between
the hard state and the soft state occurs at a nearly a constant luminosity $L_{0}$. When a source is undergoing
an outburst or flare, the hard-to-soft transition occurs at a luminosity above $L_{0}$. The additional
luminosity roughly proportional to $\frac{\Delta L}{\Delta T}$. The soft-to-hard transitions are expected to
occur around $L_{0}$. (Figures~28 in Ref.\cite{Yu2009})}
\end{figure}

\section{Further developments on thermal stability of SSD}\label{disk}

The SSD model predicts that when the accretion rate is over a small fraction of the Eddington rate, which
corresponds to $L\gtrsim 0.06 L_{\rm Edd}$, the inner region of the disk is radiation-pressure-dominated and
then both secularly \cite{Lightman1974} thermally unstable \cite{Shakura1976,Piran1978}. However, observations
of the high/soft state of black hole X-ray binaries with luminosity well within this regime ($0.01L_{\rm
Edd}\lesssim L\lesssim 0.5L_{\rm Edd}$) indicate that the disk has very little variability, i.e., quite stable
\cite{Gierlinski2004}. It has been well established that the accretion flow in this state is described by the
SSD model \cite{Zdziarski2004,Done2007}. Radiation magnetohydrodynamic simulations of a vertically stratified
shearing box have confirmed the absence of the thermal instability \cite{Hirose2009}. Recently, the thermal
stability is revisited by linear analysis, by taking into account the role of magnetic field in the accretion
flow \cite{Zheng2011}. By assuming that the field responds negatively to a positive temperature perturbation, it
was found that the threshold of accretion rate above which the disk becomes thermally unstable increases
significantly, compared with the case of not considering the role of magnetic field. This accounts for the
stability of the observed sources with high luminosities, as shown in Figure~\ref{xiamen_1}. If the magnetic
pressure is less than about 24\% of the total pressure, then this model can explain the ``heart-beat"
limit-cycle instability observed in GRS 1915+105 at its highest luminosity; this peculiar source holds the
highest accretion rate (or luminosity) among BHXBs.

\begin{figure}
\center{
\includegraphics[angle=0,scale=0.5]{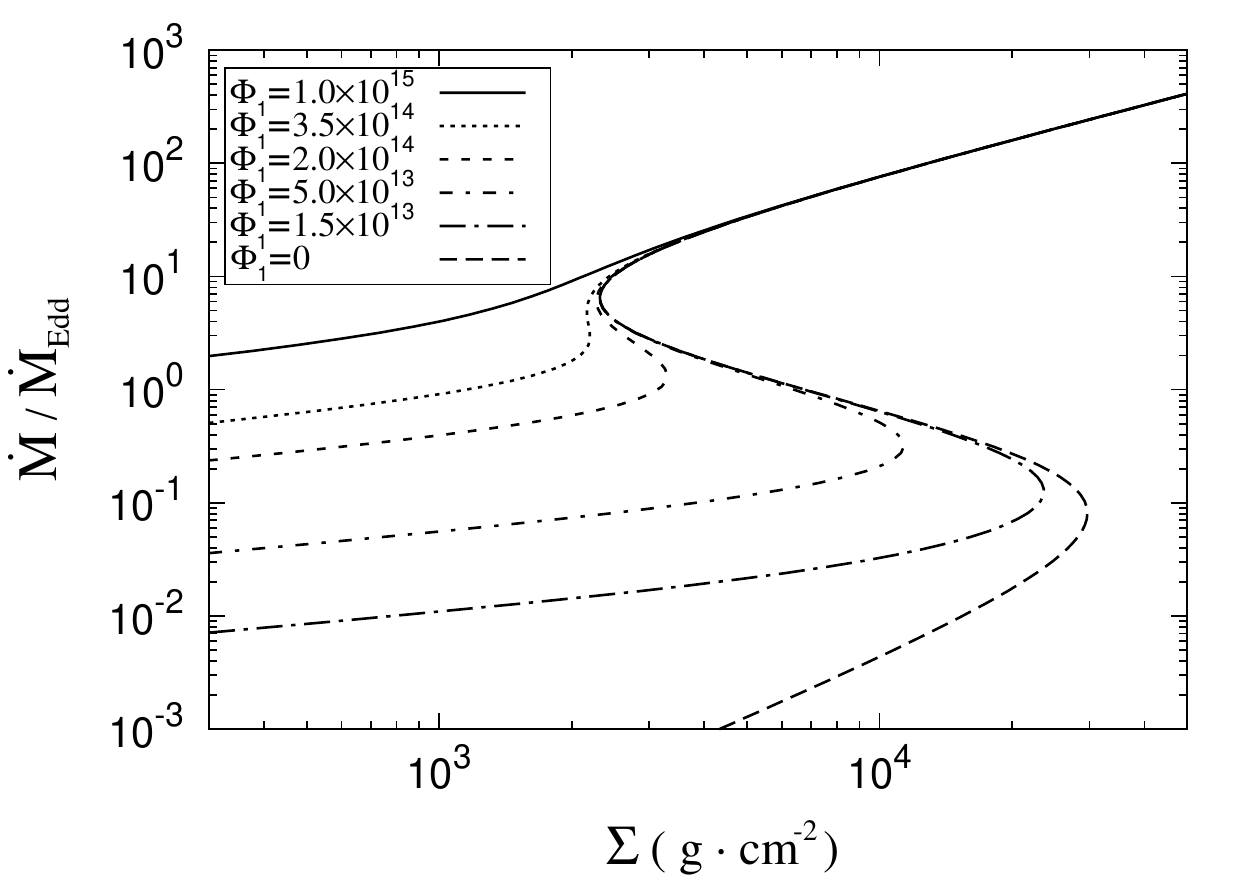}}
\caption{\label{xiamen_1} The thermal equilibrium curves of a thin disk at $10 r_{\rm g}$ for different
$\Phi_1$, which is defined as $\Phi_1 \equiv B_\varphi H$ and is in unit of Gs$\cdot$cm. Other model parameters
are BH mass $M_{\rm BH}=10M_{\odot}$ and viscosity parameter $\alpha=0.1$~. (Figure~2 in Ref.\cite{Zheng2011})}
\end{figure}

Observations of GRS 1915+105 showed that $t_{\rm high}$ (the duration of the outburst phase) is comparable to
$t_{\rm low}$ (duration of the quiescent phase) and $L_{\rm high}$ is 3 to 20 times larger than $L_{\rm low}$
\cite{Belloni1997}. However, numerical calculations showed that $t_{\rm high}$ is less than 5 percent of $t_{\rm
low}$, and $L_{\rm high}$ is around two orders of magnitude larger than $L_{\rm low}$. Some efforts have been
made to improve the theory in order to explain observations either by some artificial viscosity prescription
\cite{Nayakshin2000} or by additional assumption of the energy exchange between the disk and corona
\cite{Janiuk2002}. Taking into account the stress evolution process, it was found that the growth rate of
thermally unstable modes can decrease significantly owing to the stress delay, which may help to understand the
``heart-beat" limit-cycle variability of GRS 1915+105 \cite{Lin2011}. The limit-cycle properties are found to be
dominated by the mass-supply rate (accretion rate at the outer boundary) and the value of the $\alpha$-viscosity
parameter in the SSD model that assumes that the viscous torque is proportional to the total pressure
\cite{Xue2011}. It was also found that only the maximal outburst luminosity (in Eddington units) is positively
correlated with the spin of a BH, providing another way to probe BH spin \cite{Xue2011}; this is mainly due to
the smaller inner disk radius and thus higher radiative efficiency for larger $a_*$ as shown in
Figures~\ref{r_isco}, \ref{eta} and \ref{eta_r}.

\section{Unification and Outlook}
Despite of the many progresses made over the last decades on the study of BHXBs and microquasars, there are still many outstanding and unresolved issues, which can be pierced by two big pictures. One big picture is related to the astrophysics of BHXBs and microquasars, which is centered on understanding the accretion-outflow (wind or jet) connections at different accretion rates for different types of astrophysical systems involving BHs. The other big picture is related to the fundamental physics of BHXBs and microquasars, which is centered on identifying astrophysical BHs and testing theories of strong gravity. These two pictures are actually also entangled together, because a strong gravity theory, such as GR, is needed to describe the astrophysical aspects of these systems, and we need to understand the astrophysics of BHXBs and microquasars before we can start to test any strong gravity theory\cite{Abramowicz2013}.

\begin{figure}
\centering
\includegraphics[angle=0,scale=0.5]{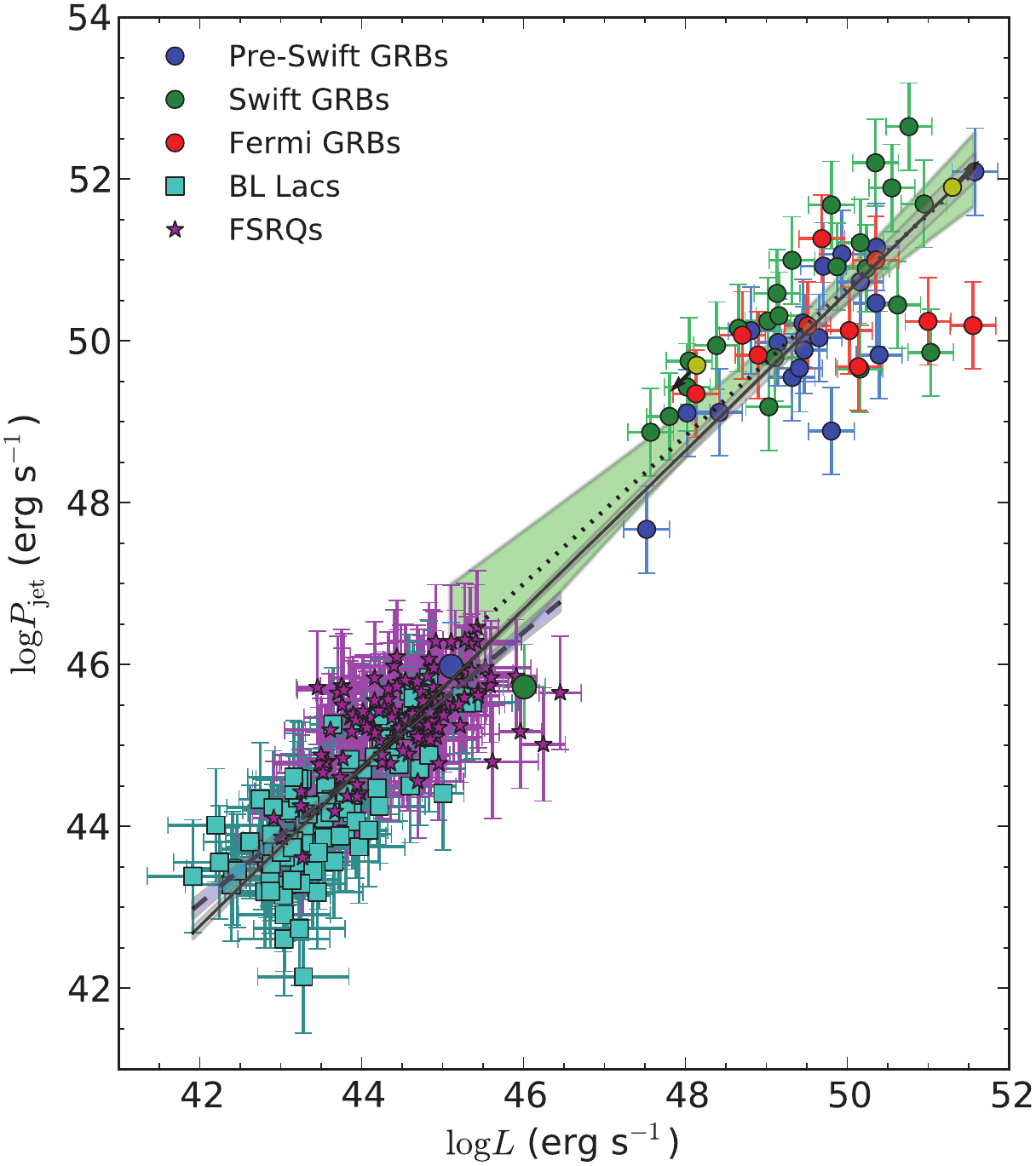}  % debeam.py
\caption{The relation between the collimation-corrected gamma-ray luminosity and the
kinetic power for AGNs and GRBs. The shaded regions display the $2\sigma$ confidence band of the fits. The
blazar and GRB best-fit models (dashed and dotted lines, respectively) follow correlations which are consistent,
within the uncertainties, with the best-fit model obtained from the joint data set (solid line).(Figure~3 in Ref.\cite{Nemmen2012})} \label{AGN_GRB}
\end{figure}

Figure~\ref{AGN_GRB} shows the relation between the collimation-corrected gamma-ray luminosity and the
kinetic power for AGNs and gamma-ray bursts (GRBs)\cite{Nemmen2012}. For at least some of GRBs, a super-Eddington accreting compact stellar object, probably a BH with around $10M_{\odot}$, is believed to be responsible for powering the highly relativistic jets, which produces intense gamma-rays through violent collisions of blobs in the jets\cite{Zhang2007c}. The two types of AGNs shown here, flat-spectrum radio quasars (FSRQs) and BL Lacs, have all been observed to have mildly relativistic jets. Their central engines are supermassive BHs with around $10^{7-9}M_{\odot}$ with accretion rates just below (for FSRQs) or far below (for BL Lacs) Eddington rate\cite{Zhang2012b}. The good correlation over about 10 orders of magnitudes between these very different systems with very different accretion rates suggests that there must be some common mechanisms responsible for the accretion-outflow (wind/jet) connections for all accreting BHs\cite{Zhang2007d}.

\begin{figure*}[!t]
\centering
\hbox{
{\includegraphics[angle=0,scale=0.6]{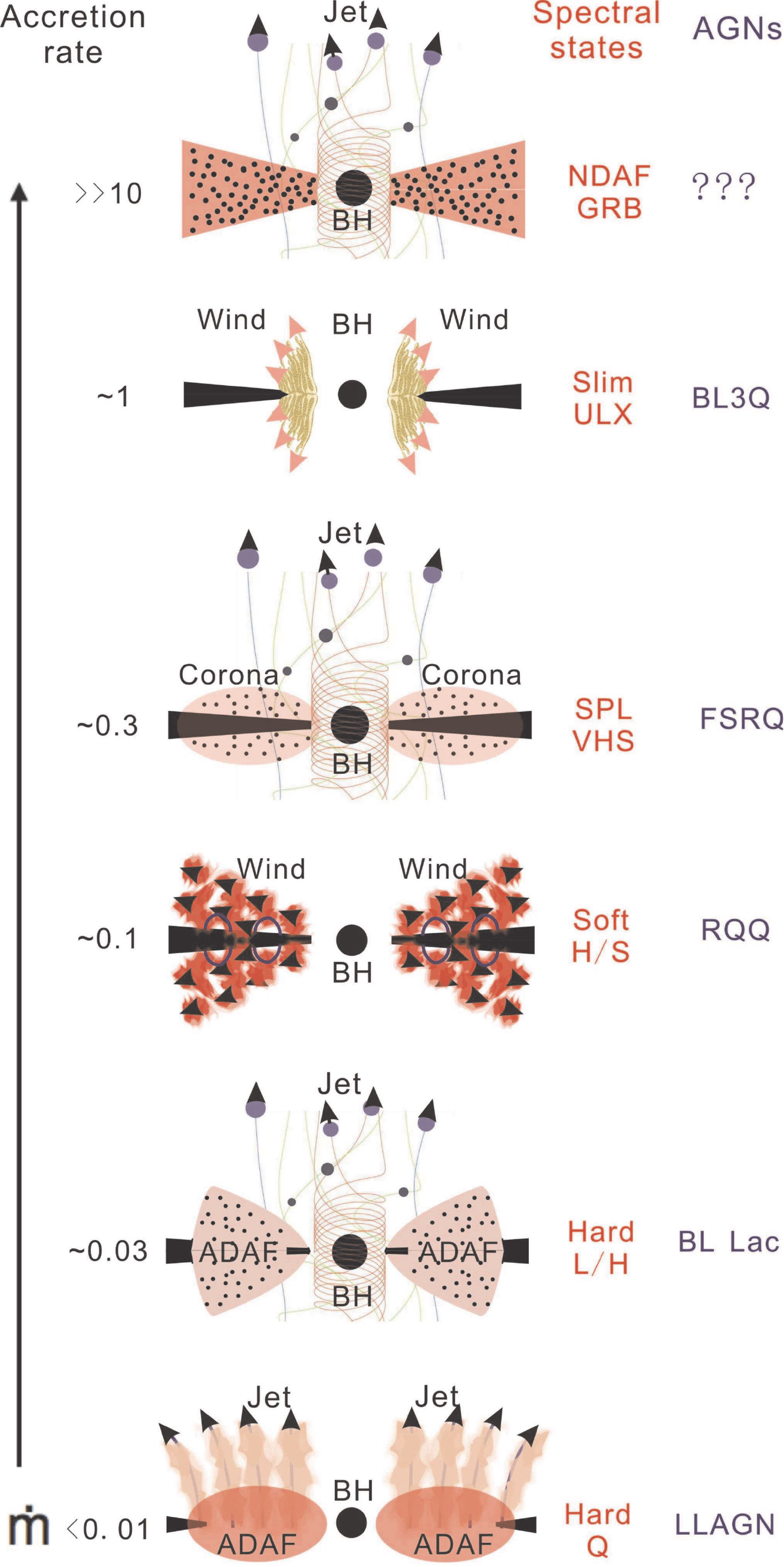}} {\begin{minipage}[0]{9.0cm}
\vspace{-15.8cm}
{\large \bf Some necessary elaborations and points: }\\
\vspace{5mm}
\begin{itemize}
\item $\dot m \gg 10$: the two-phase clumpy thick disk is in the form of neutrino dominated accretion flow (NDAF), which can extend to ISCO due to the extremely low opacity of neutrinos. The large scale magnetic field lines are wound-up by the spinning BH and rotating disk. The discontinuous jet is produced by the BZ mechanism. Most likely no such ultra-Eddington AGN exists.
\item $\dot m \sim 1$: the thin disk is truncated by radiation pressure, which drives near-spherical winds out. BL3Q is predicted as the early phase of a quasar, when the radiation pressure drives all gas in the radiation cone out so that no broad-line region (BLR) can be formed\cite{Liu2011a}. Actually all gas mixed with dust in the radiation cone can be blown out at substantially below Eddington rate, since the dust has a very high opacity to emissions from visible to UV. There BL3Q should be generic in the early phase of a quasar's active cycle, if the quasar activity is triggered by enhanced mass inflow consisting of a mixture of gas and dust.
\item $\dot m \sim 0.3$: the thin disk extends to around ISCO and the two-phase corona is clumpy, so the jet is produced mostly via the BZ mechanism and is discontinuous. The dusty torus and BLR for in a FSRQ are not shown.
\item $\dot m \sim 0.1$: the thin disk extends to around ISCO and near-equatorial wind is thermally driven out; small scale and turbulent magnetic fields may be responsible for launching the plasma out of the disk via magnetic reconnections. The dusty torus and BLR in a RQQ are not shown.
\item $\dot m \sim 0.03$: the thin disk is made of an inner and outer part; the inner disk extends to around ISCO. The two-phase ADAF is clumpy, so the jet produced via a mixture of the BP and BZ mechanisms is discontinuous. The dusty torus and BLR in a BL Lac are not shown.
\item $\dot m < 0.01$: ``Q" refers to the very low luminosity quiescent state that normally displays a hard power-law spectrum. The large scale magnetic fields rotating with the truncated thin disk and thick ADAF channel the continuous plasma to form collimated and continuous jets via the BP mechanism. There is neither dusty torus nor BLR in LLAGNs \cite{Liu2011a}.
\end{itemize}
\end{minipage}}}
\caption{Unification scheme of accretion-outflow connections of accreting BHs for different ranges of accretion rates, corresponding to different spectral states or different types of astrophysical systems. From very low to very high accretion rates in units of the Eddington rate $\dot m$, these states are: hard/quiescent state, hard/low state, thermal-dominated soft state, steep power-law state, slim disk, and NDAF disk; different types of AGNs with similar accretion-outflow structures are also labeled for comparison. The ubiquitous dusty tori and BLRs are absent in all BHXBs and microquasars. The lack of dusty tori can be easily understood since there is no dust supply in them. It is then plausible that the lack of BLRs is the consequence of no dusty tori, suggesting that a BLR may be formed out of the evaporated inner dusty torus by the anisotropic radiation of the accretion disk in an AGN\cite{Liu2011a}.} \label{unification}
\end{figure*}

Putting together all related phenomenologies and some theoretical modeling of BHXBs and microquasars discussed in this article, an unification scheme is illustrated in Figure~\ref{unification} with the major elements in the accretion-outflow connections in different types of astrophysical systems harboring both stellar mass BHs and supermassive BHs with accretion rates over several orders of magnitudes. The types of accreting stellar mass BHs include BHXBs and microquasars in different spectral states, as well as ultra-luminous X-ray sources (ULXs) (some of which are most likely ultra-Eddington accreting BHs\cite{Feng2011}), and GRBs. The types of accreting supermassive mass BHs include low-luminosity AGNs (LLAGNs), BL Lac objects, normal radio quiet quasars (RQQs), FSRQs and broad-line-less luminous quasars (BL3Qs) \cite{Liu2011a} (I create the acronym ``BL3Q" here just for fun).

This scheme is not a theoretical model, or even toy model yet. It, however, can be used as one possible chain to pierce many observed phenomenologies together, providing a possible frame work for further theoretical developments on accretion-outflow connections. Future observations will scrutinize this unification scheme and revise it inevitably. Of particular importance is the reliable determinations of the mass and spin of accreting BHs. The BH mass allows us to determine the accretion rate in units of Eddington rate, i.e., $\dot m$, which is a key parameter to the unification scheme of accretion-outflow connections. The BH spin is of course another key parameter, because it can determine the radiative efficiency of the disk and jet power in the BZ mechanism.

Once we have well determined BH parameters and a consistent and predictable theory describing the observed accretion-outflow connections, we are then ready to study the fundamental physics of BHs with BHXBs and microquasars, such as the properties of event horizon, space-time around Schwarzschild and Kerr BHs, BH spin energy (and mass) extraction. For example, both the broad iron line and CF fitting methods can be used to measure a BH's spin. However the two methods are equivalent only if the metric is accurately described with Kerr metric. Different spin measurements for a BH with the two different methods would invalidate Kerr metric, provided that we understand the accretion disk physics thoroughly. On the other hand, a correct metric is of course required to describe accurately the accretion flow around BHs. This is one example of the entanglement or interplay between astrophysics and fundamental physics of accreting BHs. The stake is high, but the job is difficult.

 {\it Acknowledgment.} I appreciate inputs from Profs. Lijun Gou, Weimin Gu, Lixin Li, Bifang Liu, Dingxiong Wang, Wenfei Yu, Feng Yuan, and Shu Zhang. The editors of this book are thanked for inviting me to write this article, as well as their patience, persistency, proof reading it, and offering comments and suggestions to improve it in the end.
 This work is partially supported with funding by the 973 Program of China under grant
2009CB824800, by the National Natural Science Foundation of China under grant Nos. 11133002 and 10725313, and by
the Qianren start-up grant 292012312D1117210.
\bibliographystyle{unsrt}
\bibliography{black_hole_binaries}{}% Produces the bibliography via BibTeX.

\end{document}